\documentclass[]{interact}

\usepackage[utf8]{inputenc}
\usepackage{amssymb}
\usepackage{amsmath}
\usepackage{bm}
\usepackage{amsthm}
\usepackage{mathrsfs}
\usepackage{url}
\usepackage{dsfont}
\usepackage{wasysym}
\usepackage{comment}
\usepackage[english]{babel}
\usepackage{graphicx}
\usepackage{graphics}
\usepackage{color}
\usepackage{tcolorbox}
\usepackage{subcaption}
\usepackage{multirow}
\usepackage{tabularx}
\usepackage{lmodern}
\usepackage{lipsum}
\usepackage{tikz}
\usetikzlibrary{calc,trees,positioning,arrows.meta,chains,shapes.geometric, chains,decorations.pathreplacing,decorations.pathmorphing,shapes,  matrix,shapes.symbols}
\tikzset{>={Latex[width=2mm, length=1.2mm]}}
\usepackage{rotating}
\usepackage{longtable}
\allowdisplaybreaks
\usepackage{lscape}
\usepackage{fancyhdr}
\usepackage{acronym}
\usepackage{enumerate}

\begin{document}

\title{Residual lifetime prediction for heterogeneous degradation data by Bayesian semi-parametric method}

\author{
\name{Barin Karmakar\textsuperscript{a}\thanks{Corresponding author: Barin Karmakar. Email: barinsaheb1996@gmail.com} and Biswabrata Pradhan\textsuperscript{a}}
\affil{\textsuperscript{a}SQC \& OR Unit, Indian Statistical Institute,\\
203, B.T. Road, Kolkata, PIN-700108, India.}
}

\maketitle

\begin{abstract}
Degradation data are considered for assessing reliability in highly reliable systems. The usual assumption is that degradation units come from a homogeneous population. But in presence of high variability in the manufacturing process, this assumption is not true in general; that is different sub-populations are involved in the study. Predicting residual lifetime of a functioning unit is a major challenge in the degradation modeling, especially in heterogeneous environment. To account for heterogeneous degradation data, we have proposed a Bayesian semi-parametric approach to relax the conventional modeling assumptions. We model the degradation path using the Dirichlet process mixture of normal distributions. Based on the samples obtained from posterior distribution of model parameters we obtain residual lifetime distribution for individual unit. Transformation based MCMC technique is used for simulating values from the derived residual lifetime distribution for prediction of residual lifetime. A simulation study is undertaken to check performance of the proposed semi-parametric model compared with parametric model. Fatigue Crack Size data is analyzed to illustrate the proposed methodology.


\end{abstract}


\section{Introduction}

For some highly reliable engineering products, it is difficult to estimate reliability due to the fact that products take too long time to fail. In this context, degradation measures taken over time are used to estimate reliability. The evolution of the degradation measures may be observed using sensor technology through a procedure known as Condition Monitoring (See Nelson \cite{NELSON}). Some examples of degradation data include vibration signals for monitoring excessive wear induced in rotating machinery, acoustic emissions for monitoring crack propagation, temperature changes and oil debris for engine lubrication, decrease of brightness of light bulbs, etc. Inferences on lifetime distribution and residual lifetime of a product can be done by modeling the underlying degradation mechanism that represents the evolution of degradation resulting failure. There have been a number of works on degradation modeling, see for example, Lu and Meeker \cite{Lu}, Padgett and Tomlinson \cite{TOMLINSON}, Muller and Zhang \cite{Muller} and Park and Padgett \cite{Padgett}. Degradation data provide more information than the bare lifetimes and hence precise inferences can be made as discussed in Lu and Meeker \cite{Lu}. The modeling of degradation data can be considered by different processes. The commonly used degradation processes are the  Wiener, Inverse-Gaussian, and Gamma processes (See Lawless and Crowder \cite{Lawless}). Lu and Meeker introduced  general path model \cite{Lu} to model degradation data. Ye and Xie \cite{Ye} have done a comparative study between general path model and stochastic processes for modeling degradation data.

\vspace{2mm}

Estimation of residual lifetime distribution of systems operating in the field plays a key role in implementing condition-based maintenance decision making (see Jardine and Banjevic \cite{Banjevic}).  Different degradation models are considered to derive the distribution function of residual lifetime. Gebraeel et al. (\cite{Gebraeel2006}, \cite{Gebraeel2005}) have introduced  a Bayesian degradation path model for predicting residual life time of units. Zhou et al. \cite{Zhou} have proposed an empirical Bayes approach to update the stochastic parameters of the degradation model for predicting soft-failure of a functioning device. Liu et al. \cite{Jiang} have considered modified Weiner process to predict the remaining useful lifetime; and Si et al. \cite{SiWang} have introduced a Wiener-process-based degradation model with a recursive filter algorithm for remaining useful lifetime prediction. Recently, Qin et al. \cite{LinChai} proposed a sparse variational Bayesian technique for remaining useful life prediction. Xu et al. \cite{XuA1} developed a robust Bayesian framework to estimate degradation state for remaining useful lifetime prediction in the presence of outliers. Brumm et al. \cite{Brumm} considered a joint model of degradation signals and time‐to‐event data for the prediction of remaining useful life. Qin and Shen \cite{Qin} introduced functional variance process for degradation modeling and remaining useful lifetime prediction. In practice it is often seen that different units degrade at different rates which motivates to incorporate heterogeneity in degradation modeling ( for example see, Wang et al. \cite{WangXian}, Zhai et al. \cite{Zhai}). There can be several reasons for heterogeneity, such as unobservable  factors, unobservable usage patterns ( see Ye et al. \cite{YeHong}). A key assumption in most of the degradation models is that the degradation rate is homogeneous, see for example, Lu and Meeker \cite{Lu} and Robinson and Crowder \cite{Robinson}. However, in practice, the assumption of homogeneity is not appropriate, particularly in presence of high variability in the manufacturing process, where the population consists of a number of homogeneous sub-populations. Ye et al. \cite{Chen} have considered heterogeneous degradation rates to estimate the distribution of remaining useful life. Xu et al. \cite{XuHu} considered product to product heterogeneity to analyze fatigue-crack data. Wen et al. \cite{Wen} considered Wiener process to predict remaining lifetime in presence of heterogeneous population. Xu et al. \cite{XuA2} considered a recursive Bayesian method for remaining useful life prediction for gamma degradation process with heterogeneous effects under conjugate priors.
\vspace{2mm}

Different mixture models are used for modeling heterogeneous failure time data. Mixture of gamma distributions and mixture of normal distributions (Kontar et al. \cite{Kontar}) are commonly used for mixture models. In the context of degradation data, Yuan and Ji \cite{Yuan} analyzed Laser emitters data and noticed that some units degrade faster than other units indicating the fact that they come from different sub-populations. They have considered a finite mixture of normal distributions for random effects in the linear degradation model. Model selection criteria like Akaike information criterion (AIC) and Bayesian information criterion (BIC) are used to estimate the number of mixture components. One of the major difficulty in using finite mixture models is defining the true number of components in the mixture, which denotes different sub-population under study. Limiting true number of mixture components may lead to wrong estimate of parameters. To overcome the above limitation of pre-specifying the number of sub-populations, Santos and Loschi \cite{Santos} considered Bayesian nonparametric models for degradation data.

\vspace{2mm}

Dirichlet process introduced by Fergusan \cite{Ferguson1973} is one of the most popular Bayesian nonparametric methods. Dirichlet process prior puts probability 1 to the set of discrete probability measures. To get rid of this problem Dirichlet process mixture (DPM) of normal distributions is used in the literature (See Escobar and West \cite{Escobar}), when the true distribution is continuous. Lo \cite{Lo} introduced Monte Carlo simulation based methods for density estimation using Dirichlet process mixtures. Escobar (\cite{Escobar1988}, \cite{Escobar}) and  MacEachern \cite{MacEachern} developed Gibbs sampling methods for Dirichlet process in a normal mixture model. But these techniques tended to produce limited posterior inference for example the Markov chain produced by these techniques tends to mix very slowly. To overcome this problem, Ishwaran and Zarepour (\cite{Ishwaran2000}, \cite{Ishwaran2001}) introduced Gibbs sampling methods for the approximate Dirichlet process which consider a truncation approximation of Dirichlet process. In the context of modeling heterogeneous degradation data, Bayesian non-parametric method have gained popularity in recent times, see for example, Santos and Loschi \cite{Santos}, Li et al. \cite{Meng} and Cheng and Yuan \cite{Cheng}. Recently Li et al. \cite{LiWang} used non-parametric techniques to model heterogeneous degradation data. Nguyen et al. \cite{Nguyen} used Bayesian non-parametric model for remaining useful life prediction of individual units with sparse degradation data, but they consider Dirichlet process prior for distribution of model parameters which neglects the fact that the true distribution of the parameters for random effect in the hierarchical model may be continuous. Bayesian semi-parametric techniques are widely recognized for mixed effect models due to its flexibility to model the multi-modal, skewed nature of random effect distribution ( see Ishwaran and Takahara \cite{Ishwaran3},  Zhang and Davidian \cite{Davidian}). There is lack of research on modeling degradation data using Bayesian semi-parametric techniques to study the distributions of residual lifetimes of individual units. Also, there are not much literature on comparison of Bayesian parametric and semi-parametric method in the context of degradation modeling, which is essential to understand the validity of model assumptions. Furthermore, the distribution of residual lifetime is required to properly characterized, which is essential for prediction.

\vspace{2mm}

In this article, we introduce a simulation based prediction method for residual lifetime of a functioning unit, when the unit comes from a heterogeneous population. The main contributions of this article are summarized as follows. We consider a Bayesian semi-parametric technique, where the degradation path is modeled using general path model and the distribution of random effect is modeled by Dirichlet process mixture of normal distribution, and parametric prior is considered for other model parameters. The proposed semi-parametric construction deals with the problem when the true random effect distribution is continuous. We estimate residual lifetime distribution of a functioning unit conditioned on the observed degradation measures of other units and degradation observation of the concerned unit. Samples generated from posterior distributions of parameters by Gibbs sampling method. We show that the residual lifetime distribution depends on the posterior samples of parameters and it does not have closed form expression. So an approximate residual lifetime distribution is derived. We use transformation based MCMC technique introduced by Dutta and Bhattacharya \cite{Dutta} to simulate samples from the approximated distribution. Finally, we use the simulated samples for predicting residual life time of a unit. The accuracy of predicted results produced by the proposed Bayesian semi-parametric method is compared with Bayesian parametric method, where a unimodal parametric distribution is considered for the random effect and other parameters. 

\vspace{2mm}

The rest of this paper is organized as follows. Degradation model is discussed in Section 2. Residual lifetime distribution is obtained based on proposed degradation model in Section 3. We discuss method of generating observations from posterior distribution in Section 4. Prediction evaluation criteria to check the performance of the proposed models are given in Section 5. A simulation study is undertaken to asses the performance and efficiency of the proposed Bayesian semi-parametric method in Section 6. A real-life data on fatigue crack size is analyzed to illustrate the proposed methodology in Section 7. We conclude this article in Section 7.

\section{Degradation model}

Consider the situation where degradation measurements are taken for different units at some fixed time points and the units are from a heterogeneous population where degradation rate vary significantly among the units. Appropriate modeling of degradation data is required in this heterogeneous situation.

\subsection{Bayesian semi-parametric degradation model}

Let $Y_{t}$ be the random variable denoting the degradation measurement of a unit at the $t$th time point. Consider the general path model introduced by Lu and Meeker \cite{Lu}, where $Y_{t}$ is assumed to have the following representation.


\begin{equation} \label{eq1}
Y_{t} = \eta(t; \bm{\alpha}, \bm{\beta}) + \epsilon \hspace{4mm} 
\end{equation}

where $\eta( \cdot ; \bm{\alpha}, \bm{\beta})$ denotes the true degradation path, $\bm{\alpha} = (\alpha_1, \alpha_2,\dots,\alpha_{k_1})^{t}$ is a $k_1$ × 1 vector of parameters that are common to all units, known as fixed effect and $\bm{\beta}= (\beta_{1}, \beta_{2},\dots,\beta_{{k_2}})^{t}$ is
a $k_2$ × 1 vector of random effects for representing individual unit characteristics. Further it is assumed that the measurement error at the $t$th time point is $\epsilon$ and $\epsilon \sim \mathcal{N}(0,\sigma_{\epsilon}^2)$. It is assumed that the measurement errors for distinct time points are identically distributed and independent to each other. We also assume that random effect $\bm{\beta}$ is independent of $\epsilon$. Suppose we have $n$ sample units randomly selected from the population and degradation measurements are taken at time points $t_{i1},t_{i2},\dots,t_{in_i}$ for $i$th individual where $i =1,\dots,n$. Let $Y_{ij}$ be random variable denoting the degradation measurement of $i$th unit at time $t_{ij}$. Using equation $(1)$, $Y_{ij}$ is represented as follows.

\vspace{-4mm}

\begin{equation} \label{eq2}
Y_{ij} = \eta(t_{ij}; \bm{\alpha}, \bm{\beta}_{i}) + \epsilon_{ij}, \hspace{4mm} i =1,\dots,n, j =1,\dots,n_i,
\end{equation}

For each sample unit $i$, the degradation path is defined to be an observed sequence of degradation readings $\textbf{y}_i$ over time $\textbf{t}_i$, where $\textbf{y}_i = ( y_{i1},\dots,y_{i{n_i}})$,  $\textbf{t}_i = ( t_{i1},\dots,t_{in_i})$ and $\bm{\beta}_{i}$ is the random effect for unit $i$. Linear and nonlinear structures for true degradation path in general path models are considered in different contexts. For example Lu and Meeker \cite{Lu} have considered a non-linear model to estimate reliability of a unit. They considered that a unit fails when the true path $\eta(\cdot)$ crosses some predetermined threshold value, say $D$. On the other hand, Robinson and Crowder \cite{Robinson}, Zhou et al. \cite{Zhou}, Gebraeel et al. \cite{Gebraeel2005}, assumed that failure of a unit occurs if its observed degradation measurement reaches a predetermined threshold $D$. This threshold value may be fixed or may vary with time.

\vspace{2mm}

One of the important issues in using the general path models is the specification of the distribution of the random effects. The random-effects $\bm{\beta}$ represent unit wise effect and can be interpreted as degradation rates. Most of the existing literature assumes that the units under test originate from homogeneous populations, hence a unimodal distribution is assumed for the random effects for modeling degradation path. But in practice, it may happen that units come from heterogeneous population consisting of different homogeneous sub-populations. In this context, Yuan and Ji \cite{Yuan} have used finite mixture distribution for random effects to incorporate heterogeneity. Bayesian method is considered to account for uncertainty of number of mixture components. But in finite mixture modeling, one of the major problems is that the model assumes a finite mixture of $K$ component regardless of sample size $n$ and that ignores the fact that $K$ can also grow as the sample size $n$ increases. It may happen that a unit is degrading significantly different from all other units and a mixture model with fixed number of mixing components will not able to recognize this unit as it is coming from distinct sub-population other than $K$ many components chosen for modeling.  This may lead to biased parameter estimates which in turn effects the estimation of reliability and residual lifetime distribution. A flexible approach to deal with this problem is to consider a unknown distribution function for random effects (see Dunson \cite{Dunson}). A Bayesian approach to this problem is done by considering unknown distribution function $F$ as a random quantity, and hence construction of a prior for $F$ is necessary. To deal with this situation Bayesian nonparametric techniques can be considered ( see Santos and Loschi \cite{Santos}), where the number of mixture components slowly grows with sample sizes.  This will also help to avoid misspecification of the distribution. In this article, we consider an unknown distribution function for random effects to overcome the above-mentioned problems.

\vspace{2mm}

Let us consider $\textbf{Y}_{train}$ = $\{Y_{ij}\}_{i=1,\dots,n}^{j=1,\dots,n_i}$ as the collection of random variables that denotes the degradation measures of $n$ units. We consider these $n$ units as training set. Suppose that there is a new unit with degradation observations at some time points $t_{1},\dots,t_{k}$, where $t_{1}<\dots<t_{k}$. Let us consider the new unit as $(n+1)$th unit, where $\textbf{Y}_{n+1} = \textbf{Y}_{new} = \{Y_{new, t_{1}},\dots,Y_{new, t_{k}}\}$ is the collection of random variables for degradation measurements of the new unit. Let us consider $\textbf{y}_{new} = \{y_{new, t_{1}},\dots,y_{new, t_{k}}\}$  as the observed degradation measures of the new unit and $\textbf{t}_{n+1} = \{t_{1},\dots,t_{k}\}$. We consider $\bm{Y} = [ \textbf{Y}_{train}, \textbf{Y}_{new}]$, which is the collection of random variables that represents the degradation measures of the units in the training set and the new unit. We utilize the entire degradation information contained in the $(n+1)$ units to model the degradation path which helps in deriving the residual lifetime distribution of the new unit. In this work we use mixed effect model introduced by Lu and Meeker \cite{Lu} for degradation modeling with one fixed effect $\alpha$ and random effect $\beta_i$, $i=1,\dots,(n+1)$. Let us consider $\beta_{(n+1)}$ be the random effect on this new unit, which we represent as $\beta_{new}$. We assume that the random effect $\beta_i|F  \overset{\mathrm{iid}} \sim F$, $i=1,\dots,(n+1)$, where $F$ is unspecified distribution and has a density function. The proposed mixed effect model is given as:

\begin{equation}
\begin{aligned}[b]
Y_{ij} & =  \eta(t_{ij}; \bm{\alpha}, \bm{\beta}_{i}) + \epsilon_{ij}, \hspace{2mm} i =1,\dots,(n+1), j =1,\dots,n_i\\
  & \beta_i|F  \overset{\mathrm{iid}} \sim F, \hspace{2mm} \epsilon_{ij} \overset{\mathrm{iid}} \sim \mathcal{N}(0,\sigma_{\epsilon}^2), \hspace{2mm} \epsilon_{ij} \perp \beta_i \hspace{2mm} \text{for all $i,j$} 
\end{aligned}
\label{eq:3}
\end{equation}

\vspace{2mm}

It is important to consider a proper structure for $ \eta(t_{ij}; \bm{\alpha}, \bm{\beta}_{i})$, given in equation (3). Considering a linear degradation model can often help in deriving closed form expression for posterior distribution, but one can also consider non-linear models instead of linear model. The traditional approach is to compare those models using model selection criteria like Bayesian Information Criterion (BIC), Conditional Predictive Ordinate (CPO), Deviance Information Criterion (DIC) etc. and choose the best model. In this article, we conduct analysis based on the linear degradation model where, $\eta(t; \bm{\alpha}, \bm{\beta}) = \alpha + \beta t$ $\forall \hspace{1mm} t>0$. Next we consider selection of prior for $F$. One of the commonly used priors for $F$ is the Dirichlet process (DP) prior, which is a probability measure defined on the space of distribution functions, introduced by Fergusan \cite{Ferguson1973}. Dirichlet process prior is almost surely (a.s) discrete hence it would be inappropriate to use this since we assumed that density function exits with respect to $F$. In order to overcome this problem, we consider Dirichlet process mixture (DPM) of continuous distributions (See Escobar \cite{Escobar}, Lo \cite{Lo}). Also we consider a  parametric prior $\pi_1(\cdot)$ on the fixed effect parameter $\alpha$ and the $\pi_2(\cdot)$ on variance of measurement error $\sigma_{\epsilon}^2$ and they are independent to each other. In the proposed method, the degradation path is modeled using general path model and distribution of random effect is modeled through Bayesian non-parametric method, while parametric prior is considered for other model parameters. So essentially we consider a Bayesian semi-parametric technique for our analysis.

\vspace{2mm}

In a Dirichlet process mixture modeling, the distribution function of random effect is represented by mixture over some simple parametric distribution functions where the mixing distribution is given by a Dirichlet process prior. We consider that $\beta_i|\theta_i \sim f(\cdot)$, $i=1,\dots,(n+1)$, where $f(\cdot)$ is a continuous distribution on $\mathbb{R}$ with a probability density function and $\theta_i$'s are unobserved random elements. Suppose that $\theta_i|G \overset{\mathrm{iid}}{\sim} G$ and G has a Dirichlet process prior denoted by $G \sim DP( \gamma, G_0)$, where $\theta_i$ $\in$  $\Theta$, $i=1,\dots,(n+1)$ and $G_0$ is the center or baseline probability measure on the measurable space ($\Theta$,$\mathscr{B}$), where $\mathscr{B}$ is the corresponding Borel $\sigma$-algebra and $\gamma \in \mathbb{R}^{+}$, considered as concentration parameter. Under these assumptions, for $i = 1,\dots,(n+1)$, and $j = 1,\dots,n_i$, the proposed degradation model can be hierarchically represented as :

\begin{equation}
\begin{aligned}[b]
& Y_{ij}|\alpha, \beta_i, \sigma_{\epsilon}^2 \overset{\mathrm{ind}}\sim \mathcal{N}(\alpha + \beta_i t_{ij}, \sigma_{\epsilon}^2)\\
     &   {\beta}_i|\theta_i \overset{\mathrm{ind}}\sim f(\beta_i|\theta_i)\\
     &  \theta_i|G \overset{\mathrm{iid}} \sim G\\
    &   G \sim DP(\gamma, G_0), \alpha \sim \pi_1(\alpha), \sigma_{\epsilon}^2 \sim \pi_2(\sigma_{\epsilon}^2)
\end{aligned}
\label{eq:4}
\end{equation}

The hierarchical representation given by model (4), explains the semi-parametric nature of the proposed model. We also assume that the random variable for degradation measurement at $t$th time, $Y_{it}$ are conditionally independent with $\beta_1,\dots,\beta_{i-1},\beta_{i+1},...,\beta_{n+1}$ given $\alpha, \beta_i$ and $\sigma_{\epsilon}^2$, $i = 1,\dots,(n+1)$, and $ \forall \hspace{1mm} t>0$. The parameter $\gamma$ controls the concentration of the prior for G about $G_0$. It can be proved that, for any measurable subset $A$ of $\Theta$, $E[G(A)]= {G_0}(A)$ and $Var[G(A)]= \frac{{G_0}(A)[1-{G_0}(A)]}{1+\gamma}$. Observe that $\forall A \in \mathscr{B}$, $G(A)$ is highly concentrated about $G_0(A)$ for large values of $\gamma$. On the other hand, as $\gamma$ tends to zero the expected shape of G is different from the one assumed by the baseline probability measure $G_0$. Since $G$ is a discrete probability measure with probability 1, so there will be repetition among the drawn values from this distribution, in other words if we draw $m$ values say $\theta_1,\dots,\theta_m$ from $G$, then there will be $k_m$ unique values of $\theta_i$'s where $k_m \leq m$. One can observe that clustering is induced among $\beta_i$'s, $i=1,\dots,(n+1)$ due to the fact that there are similar values among the unobserved random element $\theta_i$'s. So there is a positive probability that given $\theta_i$'s the distribution of random effect $\beta_i$'s are same for distinct $i$'s. The expected number of clusters or, equivalently, expected number of distinct $\theta_i$ among the $m$ drawn sample values is given by $E[k_m] = \sum_{i=1}^{m}\frac{\gamma }{\gamma + i -1}$. If $m \rightarrow \infty $, it follows that $E[k_m] \approx \gamma  \log(\frac{\gamma  + m}{\gamma })$. Our next goal is to choose a suitable expression for $f(\cdot), \pi_1(\cdot)$ and $\pi_2(\cdot)$ and construct the hierarchy of the model.

\subsection{Dirichlet process mixture of normal distribution for degradation model}

As discussed in Section 2.1, we assume that the mixing probability measure $G$ has a Dirichlet process prior denoted by $G \sim DP( \gamma, G_0)$. Sethuraman \cite{Sethuraman} showed that $G$ can be represented as, $G(\cdot) = \sum\limits_{h=1}^{\infty} p_h \delta_{m_h}(\cdot)$, where $m_h$ are random samples from $G_0$, $0 \leq p_h \leq 1$ $\forall$ $h$ and $p_1 = V_1$, $p_h = (1 -V_1)(1 -V_2)\dots(1 -V_{h-1})V_h$ $\forall$ $h \geq 2$. Ishawarn and Zarepour \cite{Ishwaran2000} represented this sum of infinitely many terms by sum of finitely many $N$ terms. This method approximate Dirichlet process prior by $\mathbb{G}_N$, where $\mathbb{G}_N(\cdot) = \sum\limits_{h=1}^{N}p_h \delta_{Z_h}(\cdot)$ and $\delta_{Z}(\cdot)$ denotes a discrete measure concentrated at $Z$. Note that $\mathbb{G}_N$ is a random probability measure, and as $N \rightarrow \infty$, it converges almost surely to a Dirichlet process with baseline probability measure $G_0$ and Dirichlet mass parameter $\gamma $, denoted by $DP(\gamma, G_0)$. Let us consider, $\textbf{Z}=(Z_1,\dots,Z_N)$, and $Z_1,\dots,Z_{N}$ are i.i.d random variables with absolutely continuous distribution function $G_0$ and also they are independent of $\textbf{p} = (p_1,\dots,p_N)$. A probability measure simulated from the prior $\mathbb{G}_N$ is defined by choosing its random weights $p_1,\dots,p_N$ by the stick-breaking construction, where $p_1 = V_1$, $p_h = (1 -V_1)(1 -V_2)\dots (1 -V_{h-1})V_h, \hspace{0.8mm} h = 2,\dots, N$, with $V_1, V_2,\dots,V_{N-1}$ are i.i.d $Beta(1, \gamma)$ random variables, setting $V_N = 1$ ensures that $\sum\limits_{h=1}^{N}p_h = 1$. 

Ishawaran and Zarepour \cite{Ishwaran2000} utilized this approximated Dirichlet process $\mathbb{G}_N$ and introduced block Gibbs sampler to estimate posterior $\mathbb{G}_N^{*}$. They have constructed an efficient MCMC method, which represents the nonparametric hierarchical model completely in terms of random variables. We use this technique to represent the hierarchical structure of the random effect distribution, given by model (4). Let us consider,  $K_1,\dots,K_{n+1}$ be the classification variables of the random effects $\beta_1,\dots,\beta_{n+1}$ and $K_i$'s are conditionally independent random variables given $\textbf{p}$ which identifies $Z_k$ with each associated $\theta_i$, particularly $\theta_i = Z_{K_i}$, $i=1,\dots,(n+1)$. The clustering nature of the hidden variables $\theta_i$'s are described by the classification variables $K_1,\dots,K_{n+1}$. Let us consider $\pi_3(\cdot)$ and $\pi_4(\cdot)$ be the prior distributions for $\textbf{p}$ and $\textbf{Z}$ respectively. It follows that the hierarchy for distribution of $\beta_i$. $i=1,\dots,(n+1)$ as presented in equation (4) can be rewritten as (See Ishawaran and Zarepour \cite{Ishwaran2000})

\begin{equation}
\begin{aligned}[b]
& \beta_i| \textbf{Z} ,K_i \overset{\mathrm{ind}} \sim f(\beta_i| Z_{K_i})\\
    & K_i|\textbf{p} \overset{\mathrm{iid}} \sim \sum\limits_{h=1}^{N}p_h \delta_h(\cdot)\\
    & \textbf{p} \sim \pi_3(\textbf{p}), \textbf{Z} \sim \pi_4(\textbf{Z})
\end{aligned}
\label{eq:5}
\end{equation}

The expression given by model (5) implies that a random effects on distinct units can have distribution with identical parameter values which is induced due to discrete nature of classification random variables. This representation helps to the fact that any degradation trajectory that is significantly different from the others can be classified as a distinct cluster. In this work we take the truncation parameter $N$ as the total number of units on which the proposed degradation model is built. We consider a normal distribution as a prior for the fixed effect parameter $\alpha$ and hierarchical priors are considered for its parameters. To complete the prior specification of $\mathbb{G}_N(\cdot)$, we consider $Z_h = (\mu_h, \sigma_h^2)$, where $\mu_h$ and $\sigma_h^2$ are mean and variance of conditional distribution of random effect given $\bm{\mu}, \bm{\sigma}^2,K_i = h$ which is normal and denoted as $\mathcal{N}(\cdot)$ in this work. Normal prior is considered for $\mu_h$'s and gamma prior for $(\sigma_{h}^2)^{-1}$, $h=1,\dots,N$ which we denote as $\mathcal{G}(\cdot)$. We consider Gamma distribution with the form $\mathcal{G}(a_0,b_0)$ so that mean is $\frac{a_0}{b_0}$. Gamma prior is considered for inverse of variance of error measurements $\sigma_{\epsilon}^2$. The hierarchical structure of the proposed model is as follows,

\begin{equation}
\begin{aligned}[b]
& \hspace{25mm} Y_{ij}|\alpha, \beta_i, \sigma_{\epsilon}^2 \overset{\mathrm{ind}} \sim \mathcal{N}( \alpha + \beta_i t_{ij}, \sigma_{\epsilon}^2), \hspace{2mm} i =1,\dots,n+1, j =1,\dots,n_i\\
    & \hspace{40mm} \alpha \sim \mathcal{N}( \mu_{\alpha}, \sigma_{\alpha}^2) \\
    & \hspace{35mm} \beta_i| \bm{\mu}, \bm{\sigma}^2,K_i \overset{\mathrm{ind}} \sim \mathcal{N}(\mu_{K_{i}},{\sigma}_{K_{i}}^2), \hspace{2mm} i =1,\dots,n+1 \\
    & \hspace{35mm} K_i|p \overset{\mathrm{iid}} \sim \sum\limits_{h=1}^{N}p_h \delta_h(\cdot), \hspace{2mm} i =1,\dots,n+1 \\
    & \hspace{25mm} p_1 = V_1, \hspace{2mm} p_h=V_h \prod\limits_{j=1}^{h-1} (1- V_{j}), \hspace{2mm} h=2,\dots,N, \hspace{2mm} V_l|\gamma \overset{\mathrm{i.i.d}} \sim Beta(1,\gamma),\\ & \hspace{25mm} \hspace{2mm} l=1,\dots,N-1 \hspace{2mm} \& \hspace{2mm} V_N=1 \\
    & \hspace{40mm} \mu_h|\sigma_z^2 \sim \mathcal{N}(m_{\mu},\sigma_z^2), \hspace{2mm} h =1,\dots,N\\
    & \hspace{40mm} ({\sigma_z^2})^{-1} \sim \mathcal{G}(\tau_1,\tau_2) \\
    & \hspace{40mm} ({\sigma_h^2})^{-1} \sim \mathcal{G}(a,b), \hspace{2mm} h =1,\dots,N \\
    & \hspace{40mm} (\sigma_{\epsilon}^2)^{-1} \sim \mathcal{G}(a_{\epsilon},b_{\epsilon})\\
    & \hspace{42mm} \gamma \sim \mathcal{G}(\eta_1, \eta_2)  
\end{aligned}
\label{eq:6}
\end{equation}

In this work, we use model (6) to derive the posterior distribution of parameters. Observe that, classification variables of random effects $K_i$'s, $i=1,\dots,(n+1)$ are conditionally independent random variables given $\textbf{p}$, where $\textbf{p}$ is constructed by stick-breaking construction. Note that $K_{n+1}$ is the classification variable corresponding to random effect of the new unit. The posterior distribution of the parameters is given by
\begin{equation} \label{eq7}
\pi(\bm{\beta},\alpha, \bm{\mu},\sigma_z^2,\textbf{K},\textbf{p},\bm{\sigma}^2,\sigma_{\epsilon}^2,
    \gamma|\bm{Y} = \bm{y}),
\end{equation}

where $\bm{\mu}= (\mu_1,\dots,\mu_N)$, $\bm{\sigma}^2= (\sigma_1^2,\dots,\sigma_N^2)$, $\bm{\beta} = (\beta_1,\dots,\beta_{n+1})$, $\textbf{K}= (K_1,\dots, K_{n+1})$ and $\textbf{p}= (p_1,\dots, p_{N})$. In the next section we consider derivation of residual lifetime distribution of a unit conditioned on the observed degradation measurements. We show that this distribution depends on the samples from posterior distribution of parameters and hence it is required to generate samples from the posterior distribution given by expression (7). We discuss in details on the prior specification in Section 4.

\section{ Residual lifetime distribution} 

Our goal is to derive distribution function of residual lifetime of a new unit for which we have degradation measurements at time points $t_{1},\dots,t_{k}$. In this work, we consider that failure of a unit occurs when the random variable for observed degradation measurement crosses a predetermined threshold value $D$. Suppose the degradation observations of the new unit has not crossed threshold value $D$ till time $t_k$, that is $\underset{1 \leq a \leq k}{max} \{y_{new, t_a}\}$ $\leq D$. Let $T_{new}$ be a random variable which is defined as the first time the degradation measurement reaches the threshold value $D$ subtracted from $t_k$, which is given as follows in equation (8). 

\vspace{-4mm}

\begin{equation}
\begin{aligned}[b]
  T_{new} = &   \underset{s_1 \in (t_k, \infty)} {\text{inf}} \{s_1 - t_k : Y_{new, s_1} \geq D \}\\
    = & \underset{s \in (0, \infty)} {\text{inf}} \{s: Y_{new, s + t_k} \geq D\}, \hspace{5mm} s = s_1 - t_k,  
\end{aligned}
\label{eq:7}
\end{equation}

The residual lifetime of the new unit functioning at time $t_k$ is defined as the first time the degradation measures crosses the threshold value $D$ given that the degradation measures at time point $t_1,\dots,t_k$ are less than $D$. So, given the collection of random variables for degradation observations $\textbf{Y}_{new}$, it is enough to to derive the conditional distribution function of random variable $T_{new}$ given $\textbf{Y}_{new}=\textbf{y}_{new}$. We make the following assumptions for deriving the residual lifetime distribution of a functioning unit.

\begin{enumerate}
    \item The random variable for residual lifetime has support on $(0,\infty)$.
\vspace{1mm}
    \item Once the degradation path exceeds the predefined threshold $D$ for first time say at time point $t^{*}$ it will never crosses back the threshold after $t^{*}$, that is there does not exist $y \geq 0$ such that $P(Y_{t^{*} + y} \leq D) > 0$.
\end{enumerate}

The distribution function of $T_{new}$ given $\textbf{Y}_{new}=\textbf{y}_{new}$ is obtained as.
\vspace{-2mm}
\begin{equation}
\begin{aligned}[b]
  F_{T_{new}|\textbf{Y}_{new}}(t)  = & P[T_{new} \leq t| \textbf{Y}_{new}=\textbf{y}_{new}]\\
    = & 1 - P[T_{new} > t| \textbf{Y}_{new}=\textbf{y}_{new}]\\
    = &  1 - P[ \underset{s \in (0, \infty)} {\text{inf}} (s: Y_{new, s + t_k} \geq D) > t| \textbf{Y}_{new}=\textbf{y}_{new}]\\
    = & 1 - P[Y_{new, t + t_k} < D| \textbf{Y}_{new}=\textbf{y}_{new}], \hspace{2mm} \text{by} \hspace{2mm} \text{assumption} \hspace{0.8mm} 2\\
    = & P[Y_{new, t + t_k} \geq D| \textbf{Y}_{new}=\textbf{y}_{new}]. 
\end{aligned}
\label{eq:8}
\end{equation}

The degradation model as defined by model (6), implies that the distribution function of residual lifetime of the  new unit depends on the parameters $\alpha, \beta_{new}$ and $\sigma_{\epsilon}^2$. So, given these parameters and individual degradation observations, the distribution function of $T_{new}$ is as follows.
\vspace{-2mm}
\begin{equation*}
\begin{aligned}[b]
  F_{T_{new}|\textbf{Y}_{new},\alpha, \beta_{new},\sigma_{\epsilon}^2}(t)  = & P[T_{new} \leq t| \textbf{Y}_{new}=\textbf{y}_{new},\alpha = \alpha^{*}, \beta_{new}= \beta_{new}^{*},\sigma_{\epsilon}^2 = \sigma_{\epsilon}^{2*}]\\
    = & 1 - P[T_{new} > t| \textbf{Y}_{new}=\textbf{y}_{new},\alpha = \alpha^{*}, \beta_{new}= \beta_{new}^{*},\sigma_{\epsilon}^2 = \sigma_{\epsilon}^{2*} ]\\
    = & P[Y_{new, t + t_k} \geq D|\alpha = \alpha^{*}, \beta_{new}= \beta_{new}^{*},\sigma_{\epsilon}^2 = \sigma_{\epsilon}^{2*}]. \\
     & \hspace{2mm} [\text{using (6)}]\\
    = & \Phi\Big[ \frac{\alpha^{*} + \beta_{new}^{*}(t+t_k) - D}{\sigma_{\epsilon}^{*}}\Big].  \hspace{51.2mm} (10)
\end{aligned}
\end{equation*}

Observe that the expression in equation (10) is a distribution function when $\beta_{new}^{*} > 0$. If we use equation (9) for computing the distribution function of residual lifetime of the new unit, only individual degradation observations of this new unit are utilized, whereas it would be more appropriate if one can use the degradation observations of $(n+1)$ units. In order to tackle this issue, we compute distribution function of the random variable $T_{new}$ given $\bm{Y}=\bm{y}$, which is denoted as $F_{T_{new}|\bm{Y}}(t)$ where $\bm{y}$ is the observed value of the random vector $\bm{Y}$ and $\bm{Y} = [ \textbf{Y}_{train}, \textbf{Y}_{new}]$. So it is enough to derive $P[Y_{new, t + t_k} \geq D|\bm{Y}= \bm{y}]$, since the event $[ T_{new} \leq t]$ equals the event $[ Y_{new, t + t_k} \geq D]$ under the assumption 2. Thus, it is also possible to derive residual lifetime distribution of each unit of the training set by considering the individual unit as the unit for which residual lifetime is to be predicted and rest of the units as training set. So, if there are $n$ units in training set, then we will fix one unit as a new unit and use $(n-1)$ units as training data. This process can be done for all $n$ units.

\vspace{2mm}

Sometimes degradation measurements of a unit are available after its degradation path crosses the threshold value. In our proposed method these information are also used to compute residual lifetime distribution. The residual lifetime distribution of a new unit conditioned on $\bm{Y} = \bm{y}$, is given by,

\vspace{-2mm}
\begin{equation*}
\begin{aligned}[b]
  F_{T_{new}|\bm{Y}}(t)  = & P[T_{new} \leq t|\bm{Y}= \bm{y}]\\
    = & P[Y_{new, t + t_k} \geq D|\bm{Y} = \bm{y}]\\
    = & \int_{\bm{\gamma_1}} P[Y_{new, t + t_k} \geq D,\bm{\gamma_1} |\bm{Y}= \bm{y}] d\bm{\gamma_1}, \hspace{1mm}  \bm{\gamma_1} = (\alpha, \beta_1,\dots,\beta_n,\beta_{new},\sigma_{\epsilon}^2)\\
    = & \int_{\bm{\gamma_1}} P[Y_{new, t + t_k} \geq D,|\bm{\gamma_1}, \bm{Y}= \bm{y}] \pi(\bm{\gamma_1} |\bm{Y}= \bm{y} )d\bm{\gamma_1}, \hspace{1mm}, \hspace{1mm}  d\bm{\gamma_1} = d\alpha d\beta_1 \dots d\beta_{new} d\sigma_{\epsilon}^2 \\ 
    = & \int_{\bm{\gamma_1}} P[Y_{new, t + t_k} \geq D|\bm{\gamma_1}] \pi(\bm{\gamma_1} |\bm{Y}= \bm{y} )d\bm{\gamma_1}\\
     = & E_{\bm{\gamma_1}|\bm{Y}= \bm{y} }\big[P[Y_{new, t + t_k} > D|\bm{\gamma_1}]\big]. \hspace{65mm} (11)
\end{aligned}
\end{equation*}


It may be noted that to obtain analytic expression of $F_{T_{new}|\bm{Y}}(t)$, represented in equation (11), a large number of multivariate integrals need to be computed. To avoid this issue, we compute this expression given in equation (11) approximately based on observations generated from the posterior distribution of $\bm{\gamma_1}$. Let $\alpha^{(i)}$, $\beta_{1}^{(i)}$, \dots, $\beta_{n}^{(i)}$, $\beta_{new}^{(i)}$ and $(\sigma_{\epsilon}^2)^{(i)}$, $i=1,\dots,N_{1}$ are the samples generated from posterior distribution of $\bm{\gamma_1}$ and $N_{1}$ is number of posterior samples. Let us consider $\epsilon_{new}$ be the random variable for measurement error for the degradation measurement at the time point $(t+t_k)$. Next, we derive the the approximated distribution function denoted as $F_{T_{new}|\bm{Y}}^{approx}(t)$, which is given as follows.

\vspace{-2mm}

\begin{eqnarray*}
   F_{T_{new}|\bm{Y}}^{approx}(t) & = &  \frac{1}{N_{1}} \sum_{i=1}^{N_{1}}  P[Y_{new, t + t_k} \geq D|\bm{\gamma_1} = \bm{\gamma_1}^{(i)}]\\
   & = & \frac{1}{N_{1}} \sum_{i=1}^{N_{1}}  P[Y_{new, t + t_k} \geq D|\alpha= \alpha^{(i)},\beta_{1} = \beta_{1}^{(i)},\dots,\beta_{n} = \beta_{n}^{(i)},\beta_{new} = \beta_{new}^{(i)},\sigma_{\epsilon}^2= (\sigma_{\epsilon}^2)^{(i)}]\\
   & = & \frac{1}{N_{1}} \sum_{i=1}^{N_{1}}  P[Y_{new, t + t_k} \geq D|\alpha= \alpha^{(i)},\beta_{new} = \beta_{new}^{(i)},\sigma_{\epsilon}^2= (\sigma_{\epsilon}^2)^{(i)}] \\
   & = & \frac{1}{N_{1}} \sum_{i=1}^{N_{1}}  P[\alpha^{(i)} + \beta_{new}^{(i)}(t+t_k) + \epsilon_{new} \geq D|\alpha= \alpha^{(i)}, \beta_{new} = \beta_{new}^{(i)},\sigma_{\epsilon}^2= (\sigma_{\epsilon}^2)^{(i)}]\\
   & = & \frac{1}{N_{1}} \sum_{i=1}^{N_{1}}  P[ \epsilon_{new} \geq D - \alpha^{(i)} - \beta_{new}^{(i)}(t+t_k)|\alpha= \alpha^{(i)},\beta_{new} = \beta_{new}^{(i)},\sigma_{\epsilon}^2= (\sigma_{\epsilon}^2)^{(i)}]\\
   & = & \frac{1}{N_{1}} \sum_{i=1}^{N_{1}}  P[ \epsilon_{{new}} \geq D - \alpha^{(i)} - \beta_{new}^{(i)}(t+t_k)|\sigma_{\epsilon}^2= (\sigma_{\epsilon}^2)^{(i)}] \hspace{2mm}\\
   & = & \frac{1}{N_{1}} \sum_{i=1}^{N_{1}}  \Phi\Big[ \frac{\alpha^{(i)} + \beta_{new}^{(i)}(t+t_k) - D}{\sigma_{\epsilon}^{(i)}}\Big], \hspace{51.2mm} (12)\\
\end{eqnarray*}

\begin{equation}\tag{13}
\begin{aligned}[b]
   f_{T_{new}|\bm{Y}}^{approx} (t)= {\frac{1}{N_1} \sum_{i=1}^{N_1}  \phi\Big[ \frac{\alpha^{(i)} + \beta_{new}^{(i)}(t + t_k) - D}{\sigma_{\epsilon}^{(i)}}\Big] \times \frac{\beta_{new}^{(i)}}{\sigma_{\epsilon}^{(i)}}}.
\end{aligned}
\end{equation}


where $\Phi(\cdot)$, $\phi(\cdot)$ are respectively cdf, pdf of standard normal random variable. So we approximate the residual lifetime distribution by the expression derived in equation (12). For the approximated distribution we have,

\vspace{-3mm}

\begin{equation}\tag{14}
\begin{aligned}[b]
   F_{T_{new}|\bm{Y}}^{approx} (0)  = & P[T_{new} \leq 0| \bm{Y}= \bm{y}]\\
    = & \frac{1}{N_1} \sum_{i=1}^{N_1}  \Phi\Big[ \frac{\alpha^{(i)} + \beta_{new}^{(i)}t_k - D}{\sigma_{\epsilon}^{(i)}}\Big] > 0.
\end{aligned}
\end{equation}

\vspace{-2mm}

Equation (14) implies that the random variable representing the residual lifetime of a new unit given the entire degradation data can take negative values with respect to the approximated distribution. Similarly, the same phenomena follows for the residual lifetime distribution given by equation (10). But we assumed that the support of $T_{new}$ is  $(0,\infty)$. Because of this, we consider to derive the residual lifetime distribution function for the new unit with an additional constraint $T_{new} > 0$ which we denote as $F_{T_{new}|\bm{Y},T_{new} >0}(t)$ given $\bm{Y}= \bm{y}$ with an additional constraint $T_{new} > 0$ which we denote as $F_{T_{new}|\bm{Y},T_{new} >0}(t)$ and $F_{T_{new}|\bm{Y}= \bm{y},T_{new} >0}^{approx}(t)$ is the corresponding approximated distribution function. This expression of this distribution function is given as below:

\begin{equation}\tag{15}
\begin{aligned}[b]
   F_{T_{new}|T_{new} > 0, \alpha, \beta_{new},\sigma_{\epsilon}^2}(t)  = & P[T_{new} \leq t|T_{new} > 0, \alpha = \alpha^{*}, \beta_{new}= \beta_{new}^{*},\sigma_{\epsilon}^2 = \sigma_{\epsilon}^{2*}] \\
    = &  \frac{P{(0<T_{new} \leq t|\alpha = \alpha^{*}, \beta_{new}= \beta_{new}^{*},\sigma_{\epsilon}^2 = \sigma_{\epsilon}^{2*})}}{1 - P{(T_{new} \leq 0|\alpha = \alpha^{*}, \beta_{new}= \beta_{new}^{*},\sigma_{\epsilon}^2 = \sigma_{\epsilon}^{2*})}}.\\
    = & \frac{\Phi\Big[ \frac{\alpha^{*} + \beta_{new}^{*}(t + t_k) - D}{\sigma_{\epsilon}^{*}}\Big] - \Phi\Big[ \frac{\alpha^{*} + \beta_{new}^{*}t_k - D}{\sigma_{\epsilon}^{*}}\Big]}{ 1 -\Phi\Big[ \frac{\alpha^{*} + \beta_{new}^{*}t_k - D}{\sigma_{\epsilon}^{*}}\Big]}
\end{aligned}
\end{equation}

\vspace{-2mm}

\begin{equation}\tag{16}
\begin{aligned}[b]
   F_{T_{new}|\bm{Y},T_{new} > 0}^{approx}(t)  = & \frac{\frac{1}{N_1} \sum_{i=1}^{N_1}  \Phi\Big[ \frac{\alpha^{(i)} + \beta_{new}^{(i)}(t + t_k) - D}{\sigma_{\epsilon}^{(i)}}\Big] - \frac{1}{N_1} \sum_{i=1}^{N_1}  \Phi\Big[ \frac{\alpha^{(i)} + \beta_{new}^{(i)}t_k - D}{\sigma_{\epsilon}^{(i)}}\Big]}{ 1 - \frac{1}{N_1} \sum_{i=1}^{N_1}  \Phi\Big[ \frac{\alpha^{(i)} + \beta_{new}^{(i)}t_k - D}{\sigma_{\epsilon}^{(i)}}\Big]}.
\end{aligned}
\end{equation}

The conditional pdf of $T_{new}$ given $\bm{Y}= \bm{y}$ and $T_{new} > 0$ with respect to the approximated distribution function given by equation (16), is:

\begin{equation}\tag{17}
\begin{aligned}[b]
     f_{T_{new}|\bm{Y},T_{new} >0}^{approx} (t)= \frac{\frac{1}{N_1} \sum_{i=1}^{N_1}  \phi\Big[ \frac{\alpha^{(i)} + \beta_{new}^{(i)}(t + t_k) - D}{\sigma_{\epsilon}^{(i)}}\Big] \times \frac{\beta_{new}^{(i)}}{\sigma_{\epsilon}^{(i)}}}{ 1 - \frac{1}{N_1} \sum_{i=1}^{N_1}  \Phi\Big[ \frac{\alpha^{(i)} + \beta_{new}^{(i)}t_k - D}{\sigma_{\epsilon}^{(i)}}\Big]}.
\end{aligned}
\end{equation}

\vspace{2mm}

Observe that if $\beta_{new}^{(i)} >0$ $\forall$ $i$, then $F_{T_{new}|\bm{Y}}^{approx}(t)$, $F_{T_{new}|\bm{Y},T_{new} > 0}^{approx}(t)$, given by equations (12) and (16) is a cumulative distribution function and $f_{T_{new}|\bm{Y},T_{new} >0}^{approx}(t)$, $f_{T_{new}|\bm{Y}}^{approx}(t)$ given by equation (13) and (17) is a probability density function. Similarly, the expression derived in equation (15) is a distribution function when $\beta_{new}^{*} >0$. Kolmogorov-Smirnov (KS) distance can be used to measure the closeness of the approximated distribution and the true distribution. Observe that the approximated pdf of residual lifetime depends on the samples from the posterior distribution of $\alpha$, $\beta_{new}$ and $\sigma_{\epsilon}^2$. In this procedure entire observed degradation measures are utilized to derive residual lifetime distribution of a new unit. The proposed method has an advantage in the sense that, as soon as new degradation signals are available, we update the posterior distribution of required parameters and sample from the updated posterior distribution and use the samples to estimate residual lifetime distribution. Next we consider how to generate samples from posterior distribution of required parameters using the proposed Bayesian semi-parametric degradation model.

\section{Generating observations from posterior distribution}

Here we consider simulations from the posterior distribution. To check the impact of prior distribution on the posterior estimates for the proposed semi-parametric model, we consider 5 different prior combinations for parameters, which are given in Table 1. Similarly, we also consider different prior combinations for the parametric model which are presented in Table 2.

\renewcommand{\arraystretch}{1.5}
\begin{table}[ht!]
\centering
    \caption{Prior Specifications for Bayesian semi-parametric model}
\begin{tabular}{|c| p{11cm}|} 
 \hline
 Scenario &  \hspace{40mm} Prior Distribution \\
 \hline
1 & $\alpha \sim \mathcal{N}(0, 1000)$, $\mu_h \sim \mathcal{N}(0, \sigma_z^2)$, $({\sigma_h^2})^{-1} \sim \mathcal{G}(1,0.01)$, $h=1,\dots,N$, $({\sigma_z^2})^{-1} \sim \mathcal{G}(0.01,0.01)$, $({\sigma_{\epsilon}^2})^{-1} \sim \mathcal{G}(0.01,0.01)$, $\gamma \sim \mathcal{G}(0.01, 0.01)$.\\
\hline
 2 & $\alpha \sim \mathcal{N}(0, 1000)$, $\mu_h \sim \mathcal{N}(m_{\mu}, \sigma_z^2)$, $({\sigma_h^2})^{-1} \sim \mathcal{G}(a,b)$, $h=1,\dots,N$, $({\sigma_z^2})^{-1} \sim \mathcal{G}(0.01,0.01), ({\sigma_{\epsilon}^2})^{-1} \sim \mathcal{G}(0.01,0.01)$, $\gamma \sim \mathcal{G}(2, 2)$. \\
 \hline
 3 & $\alpha \sim \mathcal{N}(0, 1000)$, $\mu_h \sim \mathcal{N}(m_{\mu}, \sigma_z^2)$, $({\sigma_h^2})^{-1} \sim \mathcal{G}(a,b)$, $h=1,\dots,N$, $({\sigma_z^2})^{-1} \sim \mathcal{G}(0.01,0.01) , ({\sigma_{\epsilon}^2})^{-1} \sim \mathcal{G}(0.01,0.01)$, $\gamma \sim \mathcal{G}(0.01, 0.01)$. \\
 \hline
 4 & $\alpha \sim \mathcal{N}(0, 1000)$, $\mu_h \sim \mathcal{N}(m_{\mu}, \sigma_z^2)$, $\sigma_h \sim \text{Half-Cauchy}(A_1)$, $h=1,\dots,N$, ${\sigma_z} \sim \text{Half-Cauchy}(A_2), ({\sigma_{\epsilon}^2})^{-1} \sim \mathcal{G}(0.01,0.01)$, $\gamma \sim \mathcal{G}(2, 2)$. \\
 \hline
 5 & $\alpha \sim \mathcal{N}(0, 1000)$, $\mu_h \sim \mathcal{N}(m_{\mu}, \sigma_z^2)$, $\sigma_h \sim \text{Half-Cauchy}(A_1)$, $h=1,\dots,N$, $\sigma_z \sim \text{Half-Cauchy}(A_2), ({\sigma_{\epsilon}^2})^{-1} \sim \mathcal{G}(0.01,0.01)$, $\gamma \sim \mathcal{G}(0.01, 0.01)$. \\ 
 \hline
\end{tabular}
\end{table}

\begin{table}[ht!]
\centering
    \caption{Prior Specifications for Bayesian parametric model}
\begin{tabular}{|c| p{11cm}|}
 \hline
 Scenario &  \hspace{40mm} Prior Distribution \\ 
 \hline
1 & $\alpha \sim \mathcal{N}(0, 1000)$, $\mu_{\beta} \sim \mathcal{N}(0, \sigma_z^2)$, $({\sigma_{\beta}^2})^{-1} \sim \mathcal{G}(0.01,0.01)$, $({\sigma_z^2})^{-1} \sim \mathcal{G}(0.01,0.01), ({\sigma_{\epsilon}^2})^{-1} \sim \mathcal{G}(0.01,0.01)$. \\
\hline
 2 & $\alpha \sim \mathcal{N}(0, 1000)$, $\mu_{\beta} \sim \mathcal{N}(m_{\mu}, \sigma_z)$, $({\sigma_z^2})^{-1} \sim \mathcal{G}(0.01,0.01)$, $({\sigma_{\beta}^2})^{-1} \sim \mathcal{G}(a,b), ({\sigma_{\epsilon}^2})^{-1} \sim \mathcal{G}(0.01,0.01)$.  \\
 \hline
 3 & $\alpha \sim \mathcal{N}(0, 1000)$, $\mu_{\beta} \sim \mathcal{N}(m_{\mu}, \sigma_z^2)$, $\sigma_{\beta} \sim \text{Half-Cauchy}(A_1)$, $\sigma_z \sim \text{Half-Cauchy}(A_2), ({\sigma_{\epsilon}^2})^{-1} \sim \mathcal{G}(0.01,0.01)$. \\
 \hline
\end{tabular}
\end{table}

For the semi-parametric model, we consider vague priors for fixed effect $\alpha$, inverse of error variance $\sigma_{\epsilon}^2$ and concentration parameter $\gamma$ of Dirichlet process in the first scenario. In scenario 2, we consider informative priors for mean $\mu_h$ and inverse of variance $(\sigma_h^2)^{-1}$. Santos and Loschi \cite{Santos} used observed data for choosing the values of hyperparameters for prior distributions of $\mu_h$ and $(\sigma_h^2)^{-1}$. Motivated by their ideas, we choose $m_{\mu}$, $a$ and $b$ as follows. We fit a simple linear regression model assuming $\eta(t_{ij}; \bm{\alpha}, \bm{\beta}_{i}) = \alpha + \beta_i t_{ij}$, for the degradation path of each unit. Assuming that there are enough observations to estimate $\beta_i$ for each $i$, let $\beta_i^ {*}$ be its least square estimate of for $i$th unit, $i=1,\dots, (n+1)$. Now we consider $m_{\mu}$ and $v_{\beta}$ as the sample average and variance of the $(n+1)$ many $\beta_i^ {*}$. For the hyper-parameters of prior distribution of $(\sigma_h^2)^{-1}$, we consider $ a= \sqrt{v_{\beta}}, b={\sqrt[3]{v_{\beta}}}$. Ishwaran and James \cite{Ishwaran2002} used a informative prior for $\gamma$ for normal mean mixture model. In this article, we consider informative prior $\mathcal{G}(2, 2)$ for $\gamma$ for prior 2, whereas vague prior $\mathcal{G}(0.01, 0.01)$ is considered for $\gamma$ in scenario 3. A vague gamma prior is considered for $({\sigma_z^2})^{-1}$ for the first three scenarios, where the hyperpriors are taken as $(0.01, 0.01)$. In a hierarchical model, Gelman (\cite{Gelman}) introduced half-Cauchy prior for the standard deviation parameter which is denoted by Half-Cauchy(A). We consider half-cauchy distribution for the standard deviation $(\sigma_h)^{-1}$ and $(\sigma_z)^{-1}$ with $A_1=A_2=25$. Prior 4 and 5 is constructed with these half-cauchy distribution for $(\sigma_h)^{-1}$ and $(\sigma_z)^{-1}$  along with consideration of informative and vague prior for the concentration parameter $\gamma$ respectively. For each scenario we consider $({\sigma_{\epsilon}^2})^{-1} \sim \mathcal{G}(0.01,0.01)$.

For the Bayesian parametric model, we consider three different prior scenarios. For Bayesian parametric method, we consider $\beta_i \sim N(\mu_{\beta}, \sigma_{\beta}^2)$, and $\mu_{\beta} \sim \mathcal{N}(0, \sigma_z^2)$ and $(\sigma_{\epsilon}^2)^{-1} \sim G(a_{\epsilon}, b_{\epsilon})$. Firstly, we consider a non-informative prior for $\sigma_{\beta}^2$ and $\sigma_z^2$. For the second scenario we take informative prior for both mean and variance parameter of random effects. The hyperpriors $m_{\mu},a$ and b are chosen according to the procedure described by prior 2 for semi-parametric model. For prior 3 we consider half-Cauchy prior for $(\sigma_h)^{-1}$ and $(\sigma_z)^{-1}$ with $A_1=A_2=25$. Similar to the semi-parametric method, we consider a vague prior for $({\sigma_{\epsilon}^2})^{-1} \sim \mathcal{G}(a_{\epsilon},b_{\epsilon})$, with $(a_{\epsilon} ,b_{\epsilon})= (0.001, 0.001)$ for each scenario. Furthermore, it is also assumed that parameters are independent with respect to prior distributions for both parametric and semi-parametric model.

\vspace{2mm}

\subsection{Conditional distributions for Gibbs sampling}

To simulate observations from the posterior distribution given by (7), we use the hierarchical structure of the model described by the expression in(6). To generate observations from posterior distribution, we use Gibbs sampling method. It may be noted that due to Dirichlet process prior there are ties among the classification variables. Let us consider that ${{K_1}^{*},\dots,{K_m}^{*}}$ denotes $m \leq (n +1)$ unique values of $\textbf{K}$, where $\textbf{K} = (K_1,\dots,K_{n+1})$. In this article, we denote a vector $\textbf{Q}_{(-i)}$ which contains all $Q_j$'s where $j \neq i$. In each iteration of the Gibbs sampler we simulate from the following conditional distributions :

\begin{enumerate}

\item Conditional distribution of $\alpha$ given $(\bm{\beta}, \bm{\mu},\sigma_z^2,\textbf{K},\textbf{p},\bm{\sigma}^2,\sigma_{\epsilon}^2,\gamma,\bm{Y})$ is 
    
    \begin{center}
    $\mathcal{N}\Big(\frac{\sigma_{\alpha}^2 \sum\limits_{i=1}^{(n+1)} \sum\limits_{j=1}^{n_i} (y_{ij} -\beta_i t_{ij}) + \sigma_{\epsilon}^2 \mu_{\alpha}}{\sigma_{\alpha}^2 \sum\limits_{i=1}^{n +1}n_{i} + \sigma_{\epsilon}^2} , \frac{\sigma_{\alpha}^2 \sigma_{\epsilon}^2}{\sigma_{\alpha}^2 \sum\limits_{i=1}^{n + 1}n_{i} + \sigma_{\epsilon}^2}\Big)$.\\
\end{center} 

    \item Conditional distribution of $\beta_i$ given $(\bm{\beta}_{(-i)},\alpha, \bm{\mu},\sigma_z^2,\textbf{K},\textbf{p},\bm{\sigma}^2,\sigma_{\epsilon}^2,
    \gamma,\bm{Y})$ is
    
    \begin{center}
    $ \mathcal{N}\Big(\frac{\sigma_{K_i}^2 \sum\limits_{j=1}^{n_i}t_{ij} (y_{ij} - \alpha) + \sigma_{\epsilon}^2 \mu_{K_i}}{\sigma_{K_i}^2 \sum\limits_{j=1}^{n_i}t_{ij}^2 + \sigma_{\epsilon}^2} , \frac{\sigma_{K_i}^2 \sigma_{\epsilon}^2}{\sigma_{K_i}^2 \sum\limits_{j=1}^{n_i}t_{ij}^2 + \sigma_{\epsilon}^2}\Big), \hspace{5mm}  i=1,\dots,n+1$.\\
\end{center} 

\item Conditional distribution of $(\sigma_{\epsilon}^2)^{-1}$ given $(\bm{\beta},\alpha, \bm{\mu},\sigma_z^2,\textbf{K},\textbf{p},\bm{\sigma}^2,
    \gamma,\bm{Y})$ is,

\begin{center}
    $ \mathcal{G}\Big(a_{\epsilon} + \frac{\sum\limits_{i=1}^{(n+1)}n_i}{2}, b_{\epsilon} + \frac{\sum\limits_{i=1}^{(n+1)}\sum\limits_{j=1}^{n_i}(y_{ij}- \alpha - \beta_i t_{ij})^2}{2}\Big)$.
\end{center}

\item Conditional distribution of $K_i$ given $(\bm{\beta},\alpha, \bm{\mu},\sigma_z^2,\textbf{K}_{(-i)},\textbf{p},\bm{\sigma}^2,\sigma_{\epsilon}^2,
    \gamma,\bm{Y})$ is, 

\begin{center}
    $ \sum\limits_{h=1}^{N}p_{h,i} \delta_k(\cdot)$, \hspace{5mm} $i=1,\dots,n+1$,\\
    \vspace{0.9mm}
    where $(p_{1,i},\dots,p_{N,i}) \propto (\frac{p_1}{\sigma_1}\exp(-\frac{(\beta_i - \mu_1)^2}{2\sigma_1^2}),\dots,\frac{p_N}{\sigma_N}\exp(-\frac{(\beta_i - \mu_N)^2}{2\sigma_N^2}))$.
\end{center}

\item Conditional distribution of $\textbf{p}$ given $(\bm{\beta}, \bm{\mu},\sigma_z^2,\textbf{K},\bm{\sigma}^2,\sigma_{\epsilon}^2,\gamma,\bm{Y})$ is represented in terms of distribution of $V_h^*$'s, $h = 1,\dots, (N - 1)$, where $p_1 = V_1^*$ and $p_h = (1 -V_1^*)(1 -V_2^*)\dots(1 -V_{h-1}^*)V_h^* \hspace{1mm}, h = 2,\dots, (N - 1)$ and $p_N = 1-\sum\limits_{h=1}^{N-1}p_h$ and the $V_h^*$ are independent and $V_h^* \sim Beta( 1+ r_h, \gamma + \sum\limits_{l=k+1}^{N} r_l)$, where $r_h = \sum\limits_{i=1}^{n} I_{K_i=h} $, $h=1,\dots,N$.

\vspace{2mm}
\item Conditional distribution of ${\mu}_h$ given $(\bm{\beta},\alpha, \bm{\mu}_{(-h)},\sigma_z^2,\textbf{K},\textbf{p},\bm{\sigma}^2,\sigma_{\epsilon}^2,\gamma,\bm{Y}= \bm{y})$, for each $h \in [{K_1}^{*},\dots,{K_m}^{*}]$ is $N(\mu_h^*,{\sigma_h^2}^*)$, where $\mu_h^* = {\sigma_h^2}^* \times  \big(\sum\limits_{i: K_i = h}X_i/{\sigma_h^2} + m_{\mu}/\sigma_z^2 \big) $, and ${\sigma_h^2}^* = \frac{{\sigma_h^2} {\sigma_z^2}}{n_h {\sigma_z^2} + {\sigma_h^2}}$, where $n_h$ is the number of times $K_h^*$ occurs in $\textbf{K}$. For $h \in \textbf{K} - [{K_1}^{*},\dots,{K_m}^{*}]$ independently simulate from $N(m_{\mu},{\sigma_z^2})$.

\vspace{2mm}
\item a) Conditional distribution of $(\sigma_z^2)^{-1}$ given $(\bm{\beta},\bm{\mu},\textbf{K},\textbf{p},\bm{\sigma}^2,\sigma_{\epsilon}^2,\gamma,\bm{Y})$ is, $\mathcal{G}( \tau_1 + \frac{N}{2}, \tau_2 + \sum\limits_{h=1}^{N} (\mu_h - m_{\mu})^2)$, for gamma prior on $(\sigma_z^2)^{-1}$. 

\vspace{2mm}

\hspace{-5.5mm} b) Conditional distribution of $\sigma_z$ given $(\bm{\beta}, \bm{\mu},\textbf{K},\textbf{p},\bm{\sigma}^2,\sigma_{\epsilon}^2,\gamma,\bm{Y})$ is denoted as $f(\sigma_z|.)$, where $f(\sigma_z|.) \propto (\frac{1}{\sigma_z})^{N} \exp\Big[ -\frac{1}{2\sigma_z^2}\sum\limits_{h=1}^{N} (\mu_h - m_{\mu})^2\Big] \times \Big( 1 + \frac{\sigma_z^2}{A_2}\Big)^{-1}$ for half-cauchy prior for $\sigma_z$.

\vspace{2mm}
\item 

a) Conditional distribution of $(\sigma_h^2)^{-1}$ given $(\bm{\beta},\alpha, \bm{\mu},\sigma_z^2,\textbf{K},\textbf{p},\bm{\sigma}^2_{(-h)},\sigma_{\epsilon}^2,\gamma,\bm{Y})$, for each $h \in [{K_1}^{*},\dots,{K_m}^{*}]$ is,  is $\mathcal{G}( a + \frac{r_h}{2}, b + \sum\limits_{i:K_i=h} (\beta_i - \mu_h)^2)$ and for $h \in \textbf{K} - [{K_1}^{*},\dots,{K_m}^{*}]$ independently simulate from $\mathcal{G}(a,b)$ for Gamma prior of $(\sigma_h^2)^{-1}$. 

\vspace{4mm}

\hspace{-5.5mm} b) Conditional distribution of $\sigma_h$ given $(\bm{\beta},\alpha,\bm{\mu},\sigma_z^2,\textbf{K},\textbf{p},\bm{\sigma}^2_{(-h)},\sigma_{\epsilon}^2,\gamma,\bm{Y})$, for each $h \in [{K_1}^{*},\dots,{K_m}^{*}]$ is denoted as $f(\sigma_h|.)$, where $f(\sigma_h|.) \propto (\frac{1}{\sigma_h})^{r_h} \exp\Big[ -\frac{1}{2\sigma_h^2}\sum\limits_{i: K_i = h} (\beta_i - \mu_h)^2\Big] \times \Big( 1 + \frac{\sigma_h^2}{A_1}\Big)^{-1}$ and for $h \in \textbf{K} - [{K_1}^{*},\dots,{K_m}^{*}]$ independently simulate from $\text{Half-Cauchy}(A_1)$ for half-cauchy prior of $\sigma_h$.

\vspace{2mm}
\item Conditional distribution of concentration parameter $\gamma$ given $(\bm{\beta}, \alpha, \bm{\mu},\sigma_z^2,\textbf{K},\textbf{p},\bm{\sigma}^2,\sigma_{\epsilon}^2,\bm{Y})$ is, 

\begin{center}
    $\mathcal{G}( N + \eta_1 - 1, \eta_2 - \sum\limits_{k=1}^{N-1}\log(1-V_k^*))$.
\end{center}

\end{enumerate} 

\vspace{2mm}

For simulating observations from the conditional distribution given by 7-b) and 8-b) we use Metropolis-Hastings algorithm, where the proposal distribution is taken as gamma distribution with shape parameter 1 and rate parameter as the last iterated value. So, suppose that we start with initial value $\sigma_z^{0}$, then for the $(i+1)$th step in Metropolis-Hastings, we draw values from $\mathcal{G}(1,\sigma_z^{i})$, where $\sigma_z^{i}$ is the value obtained in the $i$th step. For semi-parametric method we take truncation number $N$ as the number of units in the study for each case. We simulate total 50000 MCMC samples from the posterior distributions for each scenario and discard initial 5000 samples as burn-in. We consider only the observations with lag sizes 50, to decrease the autocorrelation. The convergence of the generated Markov chains is validated by auto-correlation plot and trace plot. We compute the residual lifetime distribution of a unit based on the generated observations from posterior distribution of parameters. To make the residual lifetime distribution valid as discussed in section 3, if there is any negative value generated for the conditional distribution of the random effects by the Gibbs sampling method, we discard them and take only the samples in $(0,\infty)$. Next we discuss on generating observations from the derived residual lifetime distribution given by equation (12) and (16).

\subsection{ Transformation based Markov chain Monte Carlo (TMCMC)}

Once we estimate the probability density function of residual lifetime distribution, our next goal is to predict residual lifetime for a unit functioning at some point $t_k$. One way to predict is simulating observation from the estimated distribution and a specific sample quantile of that generated samples can be used as predicted residual lifetime. MCMC techniques like Metropolis-Hastings (MH) helps to simulate observation from a given target distribution. However, several challenges arise for implementation of this algorithm. For example, a very large number of iterations (of the order of millions) are usually necessary for simulating observations from the target distribution. Another key issue for implementing MH algorithm is that it often leads to poor acceptance rates. To overcome these issues, we use a MCMC technique called Transformation based Markov chain Monte Carlo (TMCMC) algorithm introduced by Dutta and Bhattacharya \cite{Dutta} for simulating samples from the estimated residual lifetime distribution. This algorithm use simple deterministic bijective transformations of a low-dimensional random variable from some arbitrary distribution to update parameters simultaneously within a block. This algorithm produces a rapidly mixing Markov chain and a better acceptance rate compared to the standard MH algorithms ( see \cite{Dutta1}, \cite{Dutta2}). For univariate case, TMCMC can be reduced to a MH algorithm with a specific proposal distribution, though in higher dimensions the proposal does not admit a mixture form and TMCMC cannot be a special case of the MH algorithm. 


Consider $S$ be the state space and suppose $T : S \times S' \rightarrow S$ for some $S'$ (may be a subset of $S$) is a differentiable transformation. TMCMC technique is based on constructing forward and backward transformations $T$ which are to be defined in such a way that the detailed balance and irreducibility hold for the Markov chain generated by this technique. Consider a transformation, where the current state is $x$, then the  forward move is proposed by $x' = x \epsilon$, where $\epsilon \in (0,1)$ is a simulated value from some arbitrary density of the form $ g(\epsilon)I_{\epsilon}(0,1)$. The backward transformation $\frac{x'}{\epsilon}$ is applied for moving back to $x$ from $x'$. In general, for given $\epsilon$ and the current state $x$, forward transformation is denoted by $T(x,\epsilon)$, and the backward transformation by $T^{b}(x,\epsilon)$. One can observe that, the regions covered by the two transformations are disjoint. An important advantage associated with this algorithm is that whatever be the choice of the density $g(\epsilon)$, it cancels in the acceptance ratio of the proposed TMCMC algorithm. For more details see Dutta and Bhattacharya \cite{Dutta}.


The transformation $T$ is considered to be a differentiable transformation and the corresponding Jacobian is constructed as $J(x, \epsilon) = |\frac{\partial (T(x,\epsilon),\epsilon)}{\partial (x,\epsilon)}|$ which is non-zero almost everywhere. In this article, we simulate from the residual lifetime having density function given in equation (13) and (17). Under $f_{T_{new}|\bm{Y}}^{approx} (t)$, residual lifetime can take negative values whereas, residual lifetime has support on $(0, \infty)$ with respect to $f_{T_{new}|\bm{Y},T_{new} >0}^{approx}(t)$. We use the following transformations,

\begin{enumerate}[a)]
    \item For simulating observations with respect to pdf given in equation (13), we take $S = S' = (-\infty,\infty)$ and $T(x,\epsilon) = x + \epsilon$. For all $x \in S$, $T^{b}(x,\epsilon) = x - \epsilon$; $A = (0,\infty)$, with Jacobian is 1.

    \vspace{2mm}
    
    \item For simulating observations with respect to pdf given in equation (17), we consider $S = S' = (0,\infty)$ and $T(x,\epsilon) = x\epsilon$. For all $x \in S$, $T^{b}(x,\epsilon) = \frac{x}{\epsilon}$; $A = (0,1)$, with Jacobian is $\epsilon$.
\end{enumerate}  

Suppose $g$ is a density on $A$ where $A$ is a subset of $S'$ such that $T(x,A)$ and $T^{b}(x,A)$ are disjoint and $0<p<1$, $\alpha(\cdot)$ be the acceptance ratio and $\pi(\cdot)$ be the target distribution. Consider an initial value $x_0$ and $x_t$ be the value that the chain takes at $t$th iteration, where $t \geq 1$. Then the MCMC algorithm based on transformation is constructed as follows.

\vspace{5mm}

\begin{enumerate}
    \item Generate $ \epsilon \sim g(\cdot)$ and $u \sim U(0,1)$ independently.
    
    \vspace{2mm}
    \item 
    if $u<p$,
    $ x' = T(x_t, \epsilon) $ and $\alpha( x_t, \epsilon) = min(1, \frac{1-p}{p}\frac{\pi(x')}{\pi(x)} J(x,\epsilon))$ \\

    \hspace{-5mm} else if $p<u<1$, $x' = T^{b}(x_t, \epsilon) $ and $\alpha( x_t, \epsilon) = min(1, \frac{p}{1-p}\frac{\pi(x')}{\pi(x)} \frac{1}{J(x,\epsilon)})$

    \vspace{2mm}
    \item set, 
         $x_{t+1}=\left\{
    \begin{array}{ll}
      x', \hspace{2mm} \text{with probability} \hspace{2mm} \alpha( x_t, \epsilon).\\
      x_t, \hspace{2mm} \text{with probability} \hspace{2mm} 1- \alpha( x_t, \epsilon).
    \end{array}
  \right.$

  \vspace{2mm}

  \item Repeat Steps 1-3 $N_2$ times.
     
\end{enumerate}

We have simulated 10000 MCMC iterates after discarding the initial 1000 iterations as burn-in. We take samples from lag size 10 from the rest of the samples and use the median of this as our predicted value of residual lifetime. For this procedure, we take initial value as $x_0 = 1$ and $p=\frac{1}{2}$ for each case.

\section{Prediction evaluation criteria}

Suppose there are $n_1$ many units for which residual lifetime is to be predicted. Consider the predicted life as ${T_{ip}}$ and actual life as $T_{ia}$ for the $i$th individual unit. The prediction accuracy for $i$th individual unit is assessed by the root mean squared error (RMSE) and mean absolute error (MAE) which are given as,

\begin{enumerate}
    \item RMSE = $ \sqrt{\frac{1}{n_1} \sum\limits_{i=1}^{n_1}(T_{ip} - T_{ia})^2}$ 

    \vspace{2mm}

    \item MAE = $\frac{1}{n_1} \sum\limits_{i=1}^{n_1}|T_{ip} - T_{ia}|$
  
\end{enumerate}

We evaluate both the proposed Bayesian semi-parametric method and parametric method with respect to these error criteria. Lower values of RMSE and MAE indicates better prediction.


\section{Simulation study}

In this section we perform different simulation experiments to demonstrate the performance of the proposed Bayesian semi-parametric degradation modeling approach. We consider three component mixture of gamma and Weibull distributions for random effects for first two experiments. For the last three experiments, we consider two, three and five component mixture of normal distributions. For each case, we consider true fixed effect $\alpha$ as 2. The threshold value $D$ and the distribution of error measurement and random effect used for the study are given in Table 3. For each case, we generate degradation measures for $n= 10,30, 50$ units. Degradation observations are generated in the time interval $[0,1]$ at equally spaced times of length $\frac{1}{10}$ and $\frac{1}{30}$ respectively. We stop generating observations if degradation measure crosses the respective threshold value $D$ before the maximum time or till the last time point. So, maximum number of observations, denoted as $m = \max\{n_i\}$ per unit is 11 and 31 respectively for $i=1,\dots,n$. Suppose, for some unit degradation measures crosses threshold before the maximum time which is 1 in this case, we stop observing data. We generate observations assuming the true degradation model as, $Y_{ij} = \alpha + \beta_i t_{ij} + \epsilon_{ij}, \hspace{2mm} j =1,\dots,n_i, i =1,\dots,n$. 

\begin{table}[ht!]
\centering
    \caption{Scenarios for Simulation Study}
\begin{tabular}{|c| p{11cm}|}
 \hline
 Case &  \hspace{20mm} Distributions and Threshold value \\ 
 \hline
1 &  $\beta_i \overset{\mathrm{i.i.d}}{\sim} 0.4\mathcal{G}(40,20) + 0.25\mathcal{G}(70,20) + 0.35\mathcal{G}(100,20)$, $\epsilon_{ij} \overset{\mathrm{i.i.d}}{\sim} \mathcal{N}(0, 0.7)$, $D=11$.\\
\hline
 2 & $\beta_i \overset{\mathrm{i.i.d}}{\sim} 0.4\text{Weibull}(20,2) + 0.25\text{Weibull}(40,4) + 0.35\text{Weibull}(80,8)$, $\epsilon_{ij} \overset{\mathrm{i.i.d}}{\sim} \mathcal{N}(0, 0.7)$ , $D=12$.  \\
 \hline
 3 & $\beta_i \overset{\mathrm{i.i.d}}{\sim} 0.6\mathcal{N}(3,0.1) + 0.4\mathcal{N}(6,0.2)$, $\epsilon_{ij} \overset{\mathrm{i.i.d}}{\sim} \mathcal{N}(0, 0.5)$, $D=9$. \\
 \hline
 4 & $\beta_i \overset{\mathrm{i.i.d}}{\sim} 0.4\mathcal{N}(2,0.1) + 0.3\mathcal{N}(4,0.15) + 0.3\mathcal{N}(6,0.12)$, $\epsilon_{ij} \overset{\mathrm{i.i.d}}{\sim} \mathcal{N}(0, 0.6)$, $D=9$. \\
 \hline
 5 & $\beta_i \overset{\mathrm{i.i.d}}{\sim} 0.3\mathcal{N}(2,0.1) + 0.2\mathcal{N}(4,0.15) + 0.2\mathcal{N}(6,0.12) + 0.15\mathcal{N}(8,0.15) + 0.15\mathcal{N}(10,0.1)$, $\epsilon_{ij} \overset{\mathrm{i.i.d}}{\sim} \mathcal{N}(0, 0.7)$, $D=13$. \\
 \hline
\end{tabular}
\end{table}

For gamma mixture, we consider parametrization of the form $\mathcal{G}(a_1,b_1)$ where shape parameter is $a_1$ and rate parameter is $b_1$ and for Weibull mixture, where each component is denoted as $\text{Weibull}(a_2,b_2)$ has shape parameter $a_2$ and scale parameter $b_2$. For each combination of $n$ and $m$, we conduct the analysis with respect to each priors described in Tables 1 and 2. The goal of this article is to predict residual lifetime of each unit, so it is required to compare the predicted value with the true value. To obtain the true value of residual lifetime, we run the simulation procedure to generate degradation path from the above-mentioned models. We generate observations using the sample interval of length 0.001 and obtain the time when the degradation path crosses the predefined respective threshold for the first time. We consider this as true value to check the prediction accuracy of the proposed methods. We predict residual lifetime of each unit at the last time point when the degradation measurement has not crossed threshold value $D$. We evaluate the prediction accuracy of the residual life using the error criteria introduced in Section 4.2. We compare both distribution functions introduced in equations (12) and (16) for residual lifetime prediction. Suppose $M_1$ denotes the method for residual lifetime prediction with constraint $T_{new}>0$ and $M_2$ without the constraint. The proposed Bayesian semi-parametric method is compared with a Bayesian parametric method. We present analysis for residual lifetime distribution based on prior 2 for each case. For each case, Kolmogorov-Smirnov (KS) statistic is calculated using the generated samples with respect to the residual lifetime distribution given in equation (15) to measure the distance between true and approximated distribution. The estimate for parameters and residual lifetime obtained by both the parametric and semi-parametric methods are rounded off to three decimal places for each cases. The degradation path simulated from the different cases are represented in Figure 1.  

\vspace{0.5cm}

\begin{figure}[htbp]
    \centering
   
    \begin{subfigure}[b]{0.3\textwidth}
        \centering
        \includegraphics[width=\linewidth]{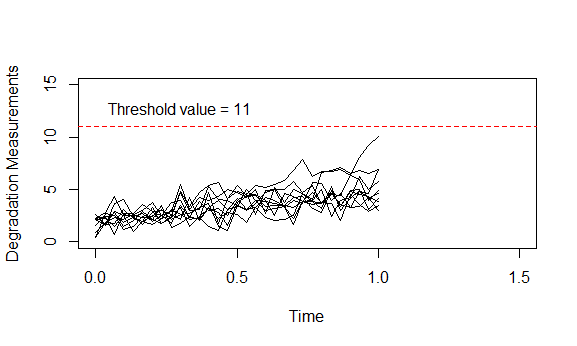}
        \caption{Case 1}
        \label{fig:1}
    \end{subfigure}
    \hspace{-1mm}
    \begin{subfigure}[b]{0.3\textwidth}
        \centering
        \includegraphics[width=\linewidth]{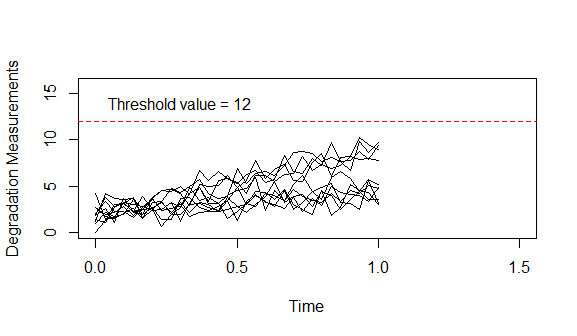}
        \caption{Case 2}
        \label{fig:2}
    \end{subfigure}
    \hspace{-1mm}
    \begin{subfigure}[b]{0.3\textwidth}
        \centering
        \includegraphics[width=\linewidth]{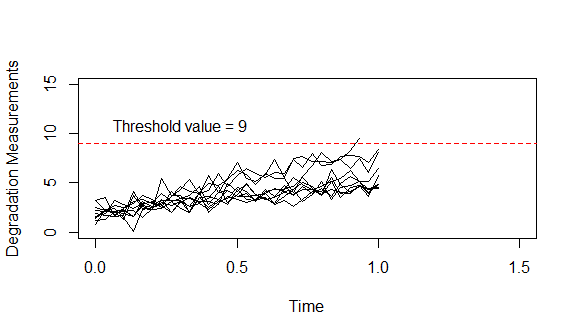}
        \caption{Case 3}
        \label{fig:3}
    \end{subfigure}

    \vspace{0.5cm} 

    \begin{subfigure}[b]{0.3\textwidth}
        \centering
        \includegraphics[width=\linewidth]{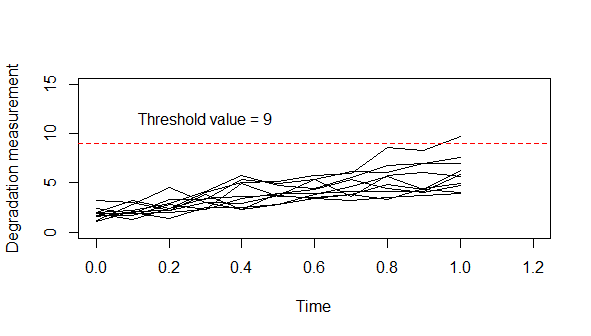}
        \caption{Case 4}
        \label{fig:1}
    \end{subfigure}
    \hspace{-1mm}
    \begin{subfigure}[b]{0.3\textwidth}
        \centering
        \includegraphics[width=\linewidth]{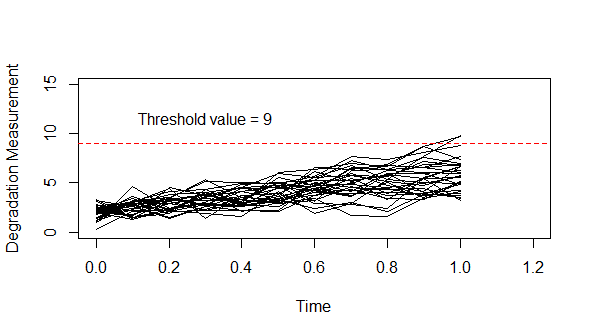}
        \caption{Case 5}
        \label{fig:1}
    \end{subfigure}
    
    \caption{Degradation paths generated with $n=10$ and $m=31$.}
\end{figure}


\subsection{Case 1}

For the first experiment we consider a 3-component mixture of gamma distribution for random effect where mean of first component is 2, second component is 3.5 and third component is 5. So we expect that some units degrade significantly faster than others, which is also evident from Figure 1(a). Here we present the analysis for $n=10$ and $m=31$ with respect to prior 2 and 5 for semi-parametric model and prior 2 and 3 for parametric model. The mean, standard deviation (s.d) and 95\% Credible interval for random effects for each of 10 units, fixed effect $\alpha$ and variance of error $\sigma_{\epsilon}^2$ computed based on generated posterior samples under prior 2 are presented in Table 4, whereas posterior estimates with respect to prior 3 of parametric model and prior 5 for semi-parametric model are given in Table 5. The trace plot and autocorrelation plot for the generated mcmc samples of $\alpha$, $\beta_i$, $i=1,2,3$ and $\sigma_{\epsilon}^2$ for semi-parametric and parametric method are provided in Figure 15 in Appendix B.

\renewcommand{\arraystretch}{1}
\begin{table}[ht!]
    \centering
    \caption{Posterior mean, standard deviation and 95\% Credible Interval of random effects ($\beta_i$, $i=1,\dots,10$ ), fixed effect ($\alpha$) and variance of measurement error ($\sigma_{\epsilon}^2$) based on Prior 2 for both parametric and semi-parametric method}. 

    \scalebox{0.95}{\begin{tabular}{|c|c|c|c|c|c|c|}
\hline
\multirow{3}{*}{Parameter} & \multicolumn{3}{c|}{Parametric method} & 
    \multicolumn{3}{c|}{Semi-parametric method} \\
\cline{2-7}
 & mean & s.d & 95\% Credible interval & mean & s.d & 95\% Credible interval  \\
\hline
 $\beta_1$  &  2.345  &    0.282  & (1.820, 2.888) &     2.339  &   0.291 & (1.749, 2.923) \\
 \hline
 $\beta_2$      &  6.166  & 0.279    & (5.628, 6.716)  &  6.157   &  0.289  & (5.602, 6.764) \\
 \hline
$\beta_3$    &   2.461  &   0.288   & (1.906, 2.998) &    2.465    &   0.276 & (1.941, 2.986) \\
\hline
 $\beta_4$   &  4.751     & 0.277  & (4.229, 5.286)  &    4.748     &  0.279 & (4.241, 5.309)\\
 \hline
$\beta_5$  &  3.403     &  0.284  & (2.838, 3.982) & 3.399  & 0.266  & (2.890, 3.923) \\
\hline
$\beta_6$  &  1.860  &  0.282   & (1.309, 2.404) &  1.856 &  0.277 & (1.335, 2.445) \\
\hline
 $\beta_7$  &     1.993   &    0.271    & (1.472, 2.512) &     1.996   &   0.279  & (1.452, 2.595)\\
 \hline
 $\beta_8$      &  3.542   &  0.287   &  (2.975, 4.087) &  3.531  &  0.274&  (2.992, 4.093) \\
 \hline
$\beta_9$    &  3.204  &   0.266 & (2.672, 3.730) &    3.203  &    0.281 & (2.659, 3.770) \\
\hline
 $\beta_{10}$   &  2.458    &   0.269    & (1.925, 2.985) &   2.454  & 0.280  & (1.902, 3.000)\\
 \hline 
$\alpha$  & 1.951    &   0.089 & (1.783, 2.124) &  1.952  &  0.09 & (1.773, 2.121) \\
\hline
$\sigma_{\epsilon}^2$  &  0.653 &  0.072 & (0.527, 0.801) &   0.649  &  0.071 & (0.526, 0.797)\\
\hline
\end{tabular}}
\vspace{2mm}
\end{table}

\begin{table}[ht!]
    \centering
    \caption{Posterior mean, standard deviation and 95\% Credible Interval of random effects ($\beta_i$, $i=1,\dots,10$), fixed effect ($\alpha$) and variance of measurement error ($\sigma_{\epsilon}^2$) based on Prior 3 for parametric and prior 5 for semi-parametric method}.

    \scalebox{0.95}{\begin{tabular}{|c|c|c|c|c|c|c|}
\hline
\multirow{3}{*}{Parameter} & \multicolumn{3}{c|}{Parametric method} & 
    \multicolumn{3}{c|}{Semi-parametric method} \\
\cline{2-7}
 & mean & s.d & 95\% Credible interval & mean & s.d & 95\% Credible interval  \\
\hline
$\beta_1$ & 2.326 & 0.283 & (1.743, 2.917) & 2.335 & 0.286 & (1.792, 2.906) \\
\hline
$\beta_2$ & 6.189 & 0.281 & (5.639, 6.759) & 6.179 & 0.277 & (5.664, 6.752) \\
\hline
$\beta_3$  & 2.453 & 0.277 & (1.936, 2.986) & 2.454 & 0.282 & (1.927, 3.005) \\
\hline
 $\beta_4$ & 4.781 & 0.267 & (4.233, 5.285) & 4.773 & 0.281 & (4.224, 5.352) \\
 \hline
$\beta_5$ & 3.422 & 0.271 & (2.896, 3.964) & 3.421 & 0.278 & (2.843, 3.961) \\
\hline
$\beta_6$  & 1.866 & 0.278 & (1.311, 2.437) & 1.846 & 0.281 & (1.301, 2.443) \\
\hline
$\beta_7$ & 1.981 & 0.269 & (1.471, 2.511) & 1.988 & 0.275 & (1.453, 2.505) \\
\hline
$\beta_8$ & 3.544 & 0.268 & (3.021, 4.074) & 3.556 & 0.261 & (3.067, 4.058) \\
\hline
$\beta_9$ & 3.206 & 0.267 & (2.667, 3.730) & 3.209 & 0.287 & (2.636, 3.766)\\
\hline
$\beta_{10}$ & 2.467 & 0.278 & (1.903, 3.005) & 2.453 & 0.283 & (1.908, 2.993) \\
\hline
$\alpha$  &   1.944  &  0.084 & (1.767, 2.100) & 1.946  &  0.085 & (1.780, 2.114) \\
\hline
$\sigma_{\epsilon}^2$  &  0.658  &  0.073 & (0.533, 0.807) &  0.656  &  0.069 & (0.533, 0.798)\\
\hline
\end{tabular}}
\vspace{2mm}
\end{table}


It is evident from Tables 4 and 5, that both parametric and proposed semi-parametric method produces credible interval for fixed effect and error variance, that contains the true value of the parameters. We also find in our analysis that as number of observations per unit is increased, the estimates are closer to true value. The units that are slowly degrading are characterized by random effects with low posterior means and significantly higher posterior means corroborate to the fact that the unit is degrading faster. We observe that for each unit, the posterior mean of random effects is very close for both the methods. It is found that the posterior estimates are not significantly different with respect to the prior choices as the number of observations per unit increases, so we consider prior 2 to compute residual lifetime distribution function for both the model. For each unit we apply the above-mentioned TMCMC technique to simulate observations from the target residual lifetime distribution.

\begin{table}[hbt!]
    \centering
    \caption{Estimated median and 95\% predictive interval of residual lifetime using a Bayesian parametric method and proposed Bayesian semi-parametric method with $n= 10, m=31$}.
    \scalebox{0.95}{\begin{tabular}{|c|c|c|c|c|c|}
\hline
\multirow{2}{*}{Unit} & \multirow{2}{*}{True value} & \multicolumn{2}{c|}{Parametric} & 
    \multicolumn{2}{c|}{Semi-parametric} \\
\cline{3-6}
 & & median & 95\% Predictive interval  & median & 95\% Predictive interval  \\
\hline
1 & 2.175 & 2.866 & (1.883, 3.991) & 2.834 & (1.788, 4.119) \\
\hline
2 & 0.355 & 0.473 & (0.216, 0.780) & 0.450 & (0.150, 0.715) \\
\hline
3 & 3.156 & 2.664 & (1.775, 3.709) & 2.700 & (1.698, 3.808) \\
\hline
4 & 0.375 & 0.899 & (0.514, 1.277) & 0.922 & (0.528, 1.278) \\
\hline
5 & 1.260 & 1.673 & (1.081, 2.274) & 1.637 & (1.110, 2.278) \\
\hline
6 & 2.917 & 3.859 & (2.409, 5.969) & 3.882 & (2.343, 5.522) \\
\hline
7 & 2.992 & 3.521 & (2.190, 5.052) & 3.533 & (2.288, 5.070) \\
\hline
8 & 1.033 & 1.558 & (0.989, 2.130) & 1.540 & (1.026, 2.194) \\
\hline
9 & 1.502 & 1.850 & (1.193, 2.539) & 1.821 & (1.148, 2.505) \\
\hline
10 & 2.513 & 2.686 & (1.748, 3.908) & 2.687 & (1.741, 3.668)\\
\hline
\end{tabular}}
\end{table}

\begin{table}[hbt!]
    \centering
    \caption{KS distance between true and approximated residual lifetime distribution for Bayesian parametric method and proposed Bayesian semi-parametric method}.
    \scalebox{0.95}{\begin{tabular}{|c|c|c|c|c|}
\hline
\multirow{2}{*}{Unit} & \multicolumn{2}{c|}{$m=11$} & 
    \multicolumn{2}{c|}{$m=31$} \\
\cline{2-5}

 &  Parametric & Semi-parametric &  Parametric & Semi-parametric \\
\hline
1 & 0.2257 & 0.2671 & 0.4121 & 0.4217 \\
\hline
2 & 0.3654 & 0.3389 & 0.3944 & 0.4621 \\ 
\hline
3 & 0.3395 & 0.3145 & 0.8358 & 0.8239 \\
\hline
4 & 0.2820 & 0.2741 & 0.2592 & 0.2996 \\ 
\hline
5 & 0.4021 & 0.3912 & 0.1565 & 0.2107 \\ 
\hline
6 & 0.6803 & 0.6706 & 0.3096 & 0.2949 \\ 
\hline
7 & 0.7745 & 0.8118 & 0.5214 & 0.5324 \\ 
\hline
8 & 0.5524 & 0.5154 & 0.0686 & 0.0822 \\ 
\hline
9 & 0.3663 & 0.3789 & 0.4403 & 0.4678 \\ 
\hline
10 & 0.2559 & 0.2325 & 0.6409 & 0.6283 \\ 
\hline
\end{tabular}}
\end{table}

\begin{table}[hbt!]
    \centering
    \caption{Prediction accuracy for Parametric and Semi-Parametric method}.
    \scalebox{0.95}{\begin{tabular}{|c|p{1cm}|p{1cm}|p{1cm}|p{1cm}|p{1cm}|p{1cm}|p{1cm}|p{1cm}|}
\hline
\multirow{3}{*}{Scenario} & \multicolumn{4}{c|}{$M_1$} & \multicolumn{4}{c|}{$M_2$} \\ \cline{2-9}
 & \multicolumn{2}{c|}{Parametric} & \multicolumn{2}{c|}{Semi-parametric} & \multicolumn{2}{c|}{Parametric} & \multicolumn{2}{c|}{Semi-parametric} \\ \cline{2-9}
 &  RMSE & MAE & RMSE & MAE &  RMSE & MAE &  RMSE & MAE \\
\hline
$n = 10, m=11$ & 1.710 & 1.028 & 1.831 & 1.059 & 1.710 & 1.028 & 1.762 & 1.045 \\
\hline
$n= 10, m=31$ & 0.527 & 0.690 & 0.520 & 0.681 & 0.536 & 0.693 & 0.541 & 0.698\\ 
\hline
$n= 30, m=11$ & 2.448 & 1.174 & 2.653 & 1.191 & 2.447 & 1.173 & 2.490 & 1.174 \\
\hline
$n= 30, m=31$ & 1.051 & 0.918 & 1.056 & 0.921 & 1.052 & 0.922 & 1.053 & 0.920 \\ 
\hline
$n =50, m=11$ & 0.796 & 0.747 & 1.761 & 1.155 & 0.772 & 0.749 & 1.778 & 1.159 \\ 
\hline
$n= 50, m=31$ & 1.040 & 0.899 & 1.036 & 0.898 & 1.033 & 0.897 & 1.046 & 0.901 \\ 
\hline
\end{tabular}}
\end{table}

We generate samples according to the discussion done in Section 4.2 and they are used for prediction. Predictive intervals are constructed using the 95\% highest probability density region of the generated samples. The histograms of the generated residual lifetimes for units 1-3 along with the trace plot of and autocorrelation plot of generated samples for semi-parametric and parametric method are provided in Figure 3 and 4 in the Appendix A to this article. It is evident from Table 6 that for most of the units both parametric and semi-parametric method produce intervals containing the true value, although predictive interval produced for unit 4 does not contain true value for both the parametric and semi-parametric method. The reason for this is both unimodal and mixture of normal distributions for random effect is unable to model gamma mixtures properly.

Kolmogorov-Smirnov (KS) distance obtained for $m=11, 31$ and $n=10$ is presented in Table 7. We observe that for the slowly degrading units for example unit 3, 6 semi-parametric method produced smaller KS value compared to parametric method. The reason for this is that the mixture of normal is able to estimate the associated random effect generated from the mixing component with smallest mean, whereas the unimodal distribution for parametric method is unable to estimate this. On the other side, KS distance for parametric method is lesser for faster degrading units, due to the fact that the mean of the posterior distribution of random effects is shifted towards the higher value for this method and hence it produces more accurate estimates for the units coming from the component with highest mean. We evaluate overall prediction accuracy using RMSE and MAE for each combination of $n$ and $m$. It is evident from Table 8 that, as the number of observations per unit increases for each $n= 10,30$ and 50, both RMSE and MAE decreases under $M_1$ and $M_2$ for most of the cases. Both the proposed semi-parametric and parametric method for $M_1$ and $M_2$, is nearly equal as both the observations and sample size increases.

\subsection{Case 2}

For the Case 2, We consider a 3-component mixture of Weibull distribution for random effect. The degradation observation generated for this case is given in Figure 1(b) for $n=10, m =31$. The mean, standard deviation (s.d) and 95\% Credible interval of random effects for each of 10 units, fixed effect $\alpha$ and variance of error $\sigma_{\epsilon}^2$ computed based on generated posterior samples with respect to the same prior combinations as discussed for Case 1 are presented in Tables 9 and 10 respectively. The trace plot and autocorrelation plot for the generated mcmc samples of $\alpha$, $\beta_i$, $i=1,2,3$ and $\sigma_{\epsilon}^2$ for semi-parametric and parametric method are provided in Figure 16 in Appendix B.

\renewcommand{\arraystretch}{1}
\begin{table}[ht!]
    \centering
    \caption{Posterior mean, standard deviation and 95\% Credible Interval of random effects ($\beta_i$, $i=1,\dots,10$ ), fixed effect ($\alpha$) and variance of measurement error ($\sigma_{\epsilon}^2$) based on Prior 2 for both parametric and semi-parametric method}.

    \scalebox{0.95}{\begin{tabular}{|c|c|c|c|c|c|c|}
\hline
\multirow{3}{*}{Parameter} & \multicolumn{3}{c|}{Parametric method} & 
    \multicolumn{3}{c|}{Semi-parametric method} \\
\cline{2-7}
 & mean & s.d & 95\% Credible interval & mean & s.d & 95\% Credible interval  \\
\hline
$\beta_1$ & 7.220 & 0.350 & (6.514, 7.907) & 7.196 & 0.325 & (6.595, 7.887) \\ \hline
$\beta_2$ & 3.760 & 0.339 & (3.105, 4.415) & 3.732 & 0.339 & (3.067, 4.456) \\ \hline
$\beta_3$ & 7.157 & 0.329 & (6.518, 7.791) & 7.165 & 0.335 & (6.544, 7.862) \\ \hline
$\beta_4$ & 2.020 & 0.341 & (1.374, 2.710) & 2.047 & 0.348 & (1.394, 2.701) \\ \hline
$\beta_5$ & 6.157 & 0.334 & (5.514, 6.832) & 6.148 & 0.346 & (5.474,  6.832) \\ \hline
$\beta_6$ & 2.404 & 0.339 & (1.794, 3.090) & 2.373 & 0.349 & (1.710, 3.089) \\ \hline
$\beta_7$ & 6.943 & 0.337 & (6.266, 7.582) & 6.950 & 0.329 & (6.361, 7.608) \\ \hline
$\beta_8$ & 2.016 & 0.335 & (1.361, 2.660) & 2.024 & 0.342 & (1.370, 2.666) \\ \hline
$\beta_9$ & 1.927 & 0.325 & (1.313, 2.581) & 1.917 & 0.335 & (1.249, 2.609) \\ \hline
$\beta_{10}$ & 2.026 & 0.342 & (1.357, 2.733) & 2.002 & 0.343 & (1.348, 2.709) \\ \hline
$\alpha$ & 2.062 & 0.105 & (1.857, 2.259) & 2.069 & 0.107 & (1.845, 2.265) \\ \hline
$\sigma_\epsilon^2$ & 0.989 & 0.113 & (0.792, 1.225) & 0.983 & 0.116 & (0.783,  1.252) \\ 
\hline
\end{tabular}}
\vspace{2mm}
\end{table}

\begin{table}[ht!]
    \centering
    \caption{Posterior mean, standard deviation and 95\% Credible Interval of random effects ($\beta_i$, $i=1,\dots,10$), fixed effect ($\alpha$) and variance of measurement error ($\sigma_{\epsilon}^2$) based on Prior 3 for parametric and prior 5 for semi-parametric method}.

    \scalebox{0.95}{\begin{tabular}{|c|c|c|c|c|c|c|}
\hline
\multirow{3}{*}{Parameter} & \multicolumn{3}{c|}{Parametric method} & 
    \multicolumn{3}{c|}{Semi-parametric method} \\
\cline{2-7}
 & mean & s.d & 95\% Credible interval & mean & s.d & 95\% Credible interval  \\
\hline
$\beta_1$ & 7.237 & 0.337 & (6.543, 7.888) & 7.236 & 0.338 & (6.655, 7.952)\\
\hline
$\beta_2$ & 3.745 & 0.340 & (3.075, 4.378) & 3.749 & 0.344 & (3.062, 4.450) \\
\hline
$\beta_3$ & 7.164 & 0.351 & (6.539, 7.848) & 7.172 & 0.349 & (6.460, 7.846) \\
\hline
$\beta_4$ & 2.026 & 0.355 & (1.338, 2.754) & 2.033 & 0.342 & (1.382, 2.729) \\
\hline
$\beta_5$ & 6.171 & 0.342 & (5.528, 6.838) & 6.152 & 0.342 & (5.499,  6.806) \\
\hline
$\beta_6$ & 2.390 & 0.342 & (1.735, 3.075) & 2.389 & 0.355 & (1.675, 3.127) \\
\hline
$\beta_7$ & 6.979 & 0.336 & (6.303, 7.644) & 6.975 & 0.346 & (6.318, 7.659) \\
\hline
$\beta_8$ & 1.995 & 0.325 & (1.345, 2.628) & 2.007 & 0.352 & (1.318, 2.742) \\
\hline
$\beta_9$ & 1.906 & 0.352 & (1.188, 2.551) & 1.906 & 0.337 & (1.255, 2.558) \\
\hline
$\beta_{10}$ & 2.009 & 0.334 & (1.371, 2.663) & 2.017 & 0.347 & (1.359, 2.708) \\
\hline
$\alpha$ & 2.065 & 0.106 & (1.851, 2.268) & 2.064 & 0.110 & (1.833, 2.259) \\
\hline
$\sigma_\epsilon^2 $ & 0.992 & 0.112 & (0.802, 1.236) & 0.988 & 0.119 & (0.772, 1.242) \\
\hline
\end{tabular}}
\vspace{2mm}
\end{table}

\vspace{2mm}

It is evident from Tables 9 and 10, that 95\%  credible interval produced by both parametric and proposed semi-parametric do not contain true value for error variance under different prior combinations. On the other hand 95\% credible intervals for both the method contains the true value of the fixed effect parameter. We observe that for each unit, the posterior mean of random effects is very close for both the methods. The posterior estimates are not significantly influenced by prior distributions as the number of observations per unit increases. We consider prior 2 to compute residual lifetime distribution function for both the model.

\begin{table}[hbt!]
    \centering
    \caption{Estimated median and 95\% predictive interval of residual lifetime using a Bayesian parametric method and proposed Bayesian semi-parametric method with $m=31$}.
    \scalebox{0.95}{\begin{tabular}{|c|c|c|c|c|c|}
\hline
\multirow{2}{*}{Unit} & \multirow{2}{*}{True value} & \multicolumn{2}{c|}{Parametric} & 
    \multicolumn{2}{c|}{Semi-parametric} \\
\cline{3-6}
 & & median & 95\% Predictive interval  & median & 95\% Predictive interval  \\
\hline
1  & 0.194 & 0.378 & (0.100, 0.643) & 0.381 & (0.109, 0.678) \\
\hline
2  & 1.095 & 1.657 & (0.984, 2.289) & 1.663 & (1.026, 2.349) \\
\hline
3  & 0.188 & 0.391 & (0.099, 0.657) & 0.391 & (0.125, 0.677) \\
\hline
4  & 2.987 & 3.970 & (2.350, 6.543) & 3.846 & (2.134, 6.024) \\
\hline
5  & 0.286 & 0.630 & (0.315, 1.023) & 0.606 & (0.252, 0.934) \\
\hline
6  & 2.556 & 3.174 & (1.832, 4.619) & 3.189 & (1.730, 4.488) \\
\hline
7  & 0.171 & 0.433 & (0.159, 0.759) & 0.432 & (0.172, 0.716) \\
\hline
8  & 2.862 & 3.965 & (2.225, 6.019) & 3.950 & (2.359, 5.796) \\
\hline
9  & 2.717 & 4.183 & (2.439, 6.174) & 4.171 & (2.394, 6.284) \\
\hline
10 & 3.554 & 3.871 & (2.205, 6.123) & 4.063 & (2.347, 6.236) \\
\hline
\end{tabular}}
\end{table}

\begin{table}[hbt!]
    \centering
    \caption{KS distance between true and approximated residual lifetime distribution for Bayesian parametric method and proposed Bayesian semi-parametric method}.
    \scalebox{0.95}{\begin{tabular}{|c|c|c|c|c|}
\hline
\multirow{2}{*}{Unit} & \multicolumn{2}{c|}{$m=11$} & 
    \multicolumn{2}{c|}{$m=31$} \\
\cline{2-5}

 &  Parametric & Semi-parametric &  Parametric & Semi-parametric \\
\hline
1  & 0.166 & 0.810 & 0.291 & 0.283 \\
\hline
2  & 0.997 & 0.622 & 0.135 & 0.160 \\
\hline
3  & 0.980 & 0.720 & 0.098 & 0.122 \\
\hline
4  & 1.000 & 0.661 & 0.227 & 0.284 \\
\hline
5  & 1.000 & 0.577 & 0.294 & 0.254 \\
\hline
6  & 0.989 & 0.841 & 0.487 & 0.473 \\
\hline
7  & 0.334 & 0.607 & 0.154 & 0.163 \\
\hline
8  & 1.000 & 0.677 & 0.219 & 0.217 \\
\hline
9  & 0.325 & 0.551 & 0.244 & 0.266 \\
\hline
10 & 0.999 & 0.827 & 0.532 & 0.463 \\
\hline
\end{tabular}}
\end{table}

\begin{table}[hbt!]
    \centering
    \caption{Prediction accuracy for Parametric and Semi-Parametric method}.
    \scalebox{0.95}{\begin{tabular}{|c|p{1cm}|p{1cm}|p{1cm}|p{1cm}|p{1cm}|p{1cm}|p{1cm}|p{1cm}|}
\hline
\multirow{3}{*}{Scenario} & \multicolumn{4}{c|}{$M_1$} & \multicolumn{4}{c|}{$M_2$} \\ \cline{2-9}
 & \multicolumn{2}{c|}{Parametric} & \multicolumn{2}{c|}{Semi-parametric} & \multicolumn{2}{c|}{Parametric} & \multicolumn{2}{c|}{Semi-parametric} \\ \cline{2-9}
 &  RMSE & MAE & RMSE & MAE &  RMSE & MAE &  RMSE & MAE \\
\hline
$n = 10, m=11$ & 2.258 & 1.306 & 0.477 & 0.572 & 2.202 & 1.299 & 0.497 & 0.582 \\
\hline
$n= 10, m=31$ & 0.734 & 0.777 & 0.726 & 0.780 & 0.729 & 0.779 & 0.749 & 0.792 \\
\hline
$n= 30, m=11$ & 0.865 & 0.793 & 0.951 & 0.837 & 0.872 & 0.797 & 0.941 & 0.837 \\
\hline
$n= 30, m=31$ & 0.872 & 0.794 & 0.879 & 0.799 & 0.871 & 0.798 & 0.888 & 0.803 \\
\hline
$n =50, m=11$ & 0.925 & 0.791 & 1.131 & 0.884 & 0.934 & 0.796 & 1.123 & 0.883 \\
\hline
$n= 50, m=31$ & 0.858 & 0.822 & 0.846 & 0.821 & 0.852 & 0.821 & 0.856 & 0.824 \\
\hline
\end{tabular}}
\end{table}

Predictive intervals are constructed using the 95\% highest probability density region of the generated samples from residual lifetime distribution for individual unit. The estimates for residual lifetime along with the true residual lifetime values for each unit are presented in Table 11. The histograms of the generated residual lifetimes for units 1-3 along with the trace plot of and autocorrelation plot of generated samples for semi-parametric and parametric method are provided in Figure 5-6, given in the Appendix A to this article. It is evident from Table 11 that for most of the units both parametric and semi-parametric methods produce intervals containing the true value. Predictive interval produced by parametric method fails to contain true value for unit 5 whereas the semi-parametric method fails to contain true value for unit 7. We find that both parametric and semi-parametric method fails to accurately estimate the tail of the true distribution in both case 1 and 2, due to the reason that the each of the 3-component mixture in both the cases are skewed, but the methods relies on either mixture of normals or unimodal normal distributions.

Kolmogorov-Smirnov (KS) distance obtained for $m=11, 31$ and $n=10$ is given in Table 12. We observe that for the slowly degrading units for example unit 4, 5 KS statistic is very large for parametric method compared to the semi-parametric method when $m=10$. The reason for this is that the mixture of normal is able to model the heterogeneous nature of the true distribution, whereas the unimodal distribution for parametric method is unable to estimate this when there are not enough observations per unit. On the other side, KS distance for parametric method is lesser for faster degrading units for $m=10$. The same pattern also evident for Case 1, which indicates that the parametric method performs well for the units coming from mixing component with highest mean. As number of observations per individual increases value of KS statistic decreases for both the method. To assess the overall prediction accuracy, RMSE and MAE are calculated for each combination of $n$ and $m$. Similar to Case 1, It is evident from Table 13 that, as the number of observations per unit increases for $n= 30, 50$, both RMSE and MAE decreases under $M_1$ and $M_2$. Both the proposed semi-parametric and parametric method for $M_1$ and $M_2$, is nearly equal as both the observations and sample size increases. For both case 1 and 2 method $M_2$ can produce non-negative values which can decrease the overall RMSE or MAE, but method $M_1$ will produce positive values which is appropriate for residual life according to our assumption. In both scenarios for case 1 and 2, it is observed that the estimated median of residual lifetime is higher than true value. The reason for this is the residual lifetime distribution functions derived to simulate observations involves terms which are average of many distribution functions, as a result this induces a much longer upper tail and hence higher values are simulated.

\subsection{Case 3}

We consider a 2-component mixture of normal distribution for random effect for Case 3, where mean of first component is 3 and the second component is 6. Degradation observation generated for this case is given in Figure 1(c). The mean, standard deviation (s.d) and 95\% Credible interval of random effects for each of 10 units, fixed effect $\alpha$ and variance of error $\sigma_{\epsilon}^2$ with respect to the different prior combinations for both the parametric and semi-parametric model are given in Tables 14 and 15. The trace plot and autocorrelation plot for the generated mcmc samples of $\alpha$, $\beta_i$, $i=1,2,3$ and $\sigma_{\epsilon}^2$ for semi-parametric and parametric method are provided in Figure 17 in Appendix B.

\renewcommand{\arraystretch}{1}
\begin{table}[ht!]
    \centering
    \caption{Posterior mean, standard deviation and 95\% Credible Interval of random effects ($\beta_i$, $i=1,\dots,10$ ), fixed effect ($\alpha$) and variance of measurement error ($\sigma_{\epsilon}^2$) based on Prior 2}.

    \scalebox{0.95}{\begin{tabular}{|c|c|c|c|c|c|c|}
\hline
\multirow{3}{*}{Parameter} & \multicolumn{3}{c|}{Parametric method} & 
    \multicolumn{3}{c|}{Semi-parametric method} \\
\cline{2-7}
 & mean & s.d & 95\% Credible interval & mean & s.d & 95\% Credible interval  \\
\hline
$\beta_1$ & 2.483 & 0.294 & (1.906, 3.088) & 2.494 & 0.296 & (1.925, 3.079) \\
\hline
$\beta_2$ & 6.737 & 0.324 & (6.150, 7.400) & 6.752 & 0.312 & (6.125, 7.347) \\
\hline
$\beta_3$ & 2.841 & 0.288 & (2.277, 3.408) & 2.875 & 0.287 & (2.313, 3.421) \\
\hline
$\beta_4$ & 3.014 & 0.282 & (2.466, 3.575) & 3.029 & 0.293 & (2.464, 3.608) \\
\hline
$\beta_5$ & 3.589 & 0.282 & (3.041, 4.126) & 3.604 & 0.284 & (3.055, 4.170) \\
\hline
$\beta_6$ & 5.635 & 0.290 & (5.100, 6.234) & 5.651 & 0.289 & (5.103, 6.219) \\
\hline
$\beta_7$ & 3.101 & 0.291 & (2.542, 3.690) & 3.118 & 0.280 & (2.584, 3.668) \\
\hline
$\beta_8$ & 2.874 & 0.281 & (2.329, 3.400) & 2.900 & 0.299 & (2.342, 3.484) \\
\hline
$\beta_9$ & 3.165 & 0.287 & (2.634, 3.737) & 3.163 & 0.291 & (2.622, 3.747) \\
\hline
$\beta_{10}$ & 6.300 & 0.301 & (5.713, 6.871) & 6.303 & 0.301 & (5.684, 6.863) \\
\hline
$\alpha$ & 2.033 & 0.090 & (1.859, 2.201) & 2.026 & 0.089 & (1.853, 2.203) \\
\hline
$\sigma_\epsilon^2$ & 0.728 & 0.091 & (0.572, 0.917) & 0.729 & 0.087 & (0.571, 0.911) \\
\hline
\end{tabular}}
\vspace{2mm}
\end{table}

\begin{table}[ht!]
    \centering
    \caption{Posterior mean, standard deviation and 95\% Credible Interval of random effects ($\beta_i$, $i=1,\dots,10$), fixed effect ($\alpha$) and variance of measurement error ($\sigma_{\epsilon}^2$) based on Prior 3 for parametric and prior 5 for semi-parametric method}. 

    \scalebox{0.95}{\begin{tabular}{|c|c|c|c|c|c|c|}
\hline
\multirow{3}{*}{Parameter} & \multicolumn{3}{c|}{Parametric method} & 
    \multicolumn{3}{c|}{Semi-parametric method} \\
\cline{2-7}
 & mean & s.d & 95\% Credible interval & mean & s.d & 95\% Credible interval  \\
\hline
$\beta_1$ & 2.488 & 0.293 & (1.928, 3.096) & 2.463 & 0.271 & (1.964, 3.022) \\
\hline
$\beta_2$ & 6.752 & 0.335 & (6.117, 7.417) & 6.764 & 0.337 & (6.100, 7.417) \\
\hline
$\beta_3$ & 2.835 & 0.287 & (2.275, 3.438) & 2.843 & 0.297 & (2.268, 3.442) \\
\hline
$\beta_4$ & 2.995 & 0.293 & (2.453, 3.570) & 2.996 & 0.293 & (2.420, 3.604) \\
\hline
$\beta_5$ & 3.588 & 0.304 & (3.014, 4.191) & 3.588 & 0.298 & (3.026, 4.171) \\
\hline
$\beta_6$ & 5.637 & 0.296 & (5.050, 6.214) & 5.644 & 0.281 & (5.082, 6.195) \\
\hline
$\beta_7$ & 3.101 & 0.290 & (2.557, 3.653) & 3.091 & 0.288 & (2.537, 3.656) \\
\hline
$\beta_8$ & 2.883 & 0.307 & (2.295, 3.467) & 2.896 & 0.299 & (2.298, 3.467) \\
\hline
$\beta_9$ & 3.150 & 0.299 & (2.574, 3.768) & 3.160 & 0.290 & (2.632, 3.721) \\
\hline
$\beta_{10}$ & 6.304 & 0.292 & (5.708, 6.880) & 6.307 & 0.281 & (5.776, 6.858) \\
\hline
$\alpha$ & 2.034 & 0.095 & (1.837, 2.215) & 2.032 & 0.093 & (1.837, 2.201) \\
\hline
$\sigma_\epsilon^2$ & 0.727 & 0.092 & (0.568, 0.924) & 0.724 & 0.091 & (0.556, 0.909) \\
\hline
\end{tabular}}
\vspace{2mm}
\end{table}


Observe that 95\%  credible interval produced by both parametric and proposed semi-parametric contain true fixed effect value under each prior combinations. The estimates for fixed effect parameter are closer to true value as the number of observations per unit increases, whereas the posterior mean for error variance seems to higher than true value for both models. We observe that for fast degrading units, the difference between posterior means of random effect for parametric and semi-parametric method decreases as number of observations per unit increases. We consider prior 2 to compute residual lifetime distribution function for both the model.

\begin{table}[hbt!]
    \centering
    \caption{Estimated median and 95\% predictive interval of residual lifetime using a Bayesian parametric method and proposed Bayesian semi-parametric method with $m=31$}.
    \scalebox{0.95}{\begin{tabular}{|c|c|c|c|c|c|}
\hline
\multirow{2}{*}{Unit} & \multirow{2}{*}{True value} & \multicolumn{2}{c|}{Parametric} & 
    \multicolumn{2}{c|}{Semi-parametric} \\
\cline{3-6}
 & & median & 95\% Predictive interval  & median & 95\% Predictive interval  \\
\hline
1  & 0.780 & 1.844 & (0.965, 2.866) & 1.772 & (0.929, 2.766) \\
\hline
2  & 0.017 & 0.158 & (0.004, 0.367) & 0.165 & (0.002, 0.379) \\
\hline
3  & 1.017 & 1.409 & (0.734, 2.267) & 1.421 & (0.808, 2.276) \\
\hline
4  & 0.656 & 1.274 & (0.643, 1.978) & 1.292 & (0.604, 1.987) \\
\hline
5  & 0.484 & 0.947 & (0.469, 1.533) & 0.936 & (0.429, 1.463) \\
\hline
6  & 0.043 & 0.257 & (0.008, 0.517) & 0.237 & (0.004, 0.500) \\
\hline
7  & 0.798 & 1.234 & (0.664, 2.038) & 1.217 & (0.574, 1.918) \\
\hline
8  & 1.048 & 1.421 & (0.703, 2.161) & 1.403 & (0.715, 2.180) \\
\hline
9  & 0.497 & 1.212 & (0.618, 1.904) & 1.215 & (0.674, 1.955) \\
\hline
10 & 0.010 & 0.158 & (0.001, 0.361) & 0.146 & (0.004, 0.367) \\
\hline
\end{tabular}}
\end{table}

\begin{table}[hbt!]
    \centering
    \caption{KS distance between true and approximated residual lifetime distribution for Bayesian parametric method and proposed Bayesian semi-parametric method}.
    \scalebox{0.95}{\begin{tabular}{|c|c|c|c|c|}
\hline
\multirow{2}{*}{Unit} & \multicolumn{2}{c|}{$m=11$} & 
    \multicolumn{2}{c|}{$m=31$} \\
\cline{2-5}

 &  Parametric & Semi-parametric &  Parametric & Semi-parametric \\
\hline
1  & 0.553 & 0.543 & 0.632 & 0.604 \\
\hline
2  & 0.065 & 0.076 & 0.116 & 0.083 \\
\hline
3  & 0.267 & 0.276 & 0.188 & 0.193 \\
\hline
4  & 0.182 & 0.183 & 0.170 & 0.197 \\
\hline
5  & 0.388 & 0.393 & 0.177 & 0.198 \\
\hline
6  & 0.094 & 0.079 & 0.156 & 0.078 \\
\hline
7  & 0.268 & 0.260 & 0.388 & 0.413 \\
\hline
8  & 0.542 & 0.570 & 0.271 & 0.293 \\
\hline
9  & 0.161 & 0.164 & 0.177 & 0.189 \\
\hline
10 & 0.176 & 0.164 & 0.167 & 0.140 \\
\hline
\end{tabular}}
\end{table}

\begin{table}[hbt!]
    \centering
    \caption{Prediction accuracy for Parametric and Semi-Parametric method}.
    \scalebox{0.95}{\begin{tabular}{|c|p{1cm}|p{1cm}|p{1cm}|p{1cm}|p{1cm}|p{1cm}|p{1cm}|p{1cm}|}
\hline
\multirow{3}{*}{Scenario} & \multicolumn{4}{c|}{$M_1$} & \multicolumn{4}{c|}{$M_2$} \\ \cline{2-9}
 & \multicolumn{2}{c|}{Parametric} & \multicolumn{2}{c|}{Semi-parametric} & \multicolumn{2}{c|}{Parametric} & \multicolumn{2}{c|}{Semi-parametric} \\ \cline{2-9}
 
 &  RMSE & MAE & RMSE & MAE &  RMSE & MAE &  RMSE & MAE \\
\hline
$n = 10, m=11$  & 0.474 & 0.626 & 0.470 & 0.621 & 0.474 & 0.620 & 0.462 & 0.612 \\
\hline
$n= 10, m=31$  & 0.530 & 0.676 & 0.515 & 0.667 & 0.536 & 0.675 & 0.523 & 0.666 \\
\hline
$n= 30, m=11$  & 0.696 & 0.796 & 0.708 & 0.803 & 0.689 & 0.794 & 0.701 & 0.801 \\
\hline
$n= 30, m=31$ & 0.559 & 0.721 & 0.633 & 0.771 & 0.556 & 0.718 & 0.633 & 0.771 \\
\hline
$n =50, m=11$ & 0.561 & 0.692 & 0.824 & 0.871 & 0.559 & 0.688 & 0.820 & 0.869 \\
\hline
$n= 50, m=31$ & 0.500 & 0.678 & 0.726 & 0.830 & 0.499 & 0.675 & 0.726 & 0.829 \\
\hline
\end{tabular}}
\end{table}

Predictive intervals based on the 95\% highest probability density region of the generated samples for the individual units are presented in Table 16. The histograms of the generated residual lifetimes for units 1-3 along with the trace plot of and autocorrelation plot of generated samples for semi-parametric and parametric method for this case are provided in Figure 7-8 respectively in the Appendix A to this article. For most of the units, the produced predictive intervals for both parametric and semi-parametric methods contain the true value. Predictive interval produced by both parametric and semi-parametric method fails to contain true value for unit 1 and 9.

Kolmogorov-Smirnov (KS) statistic obtained for $m=11, 31$ and $n=10$ is given in Table 17. For most of the slowly degrading units for example unit 3, 8 value of KS statistic decreases for both the method as number of observations per unit increases. The same also happens for the fast degrading units. So, in this case performance of both the methods is quite similar, because the heterogeneity is comparatively less for this case. RMSE and MAE are calculated for each combination of $n$ and $m$. Similar to Case 1, It is evident from Table 13 that, as the number of observations per unit increases for $n= 30, 50$, both RMSE and MAE decreases under $M_1$ and $M_2$. Overall parametric method performs well in this case. The reason for the poor performance of the semi-parametric method is that the estimates are not accurate, for the parameters associated with the individual degrading which impacts the residual lifetime distributions. Though method $M_2$ produce smaller errors, but this methods violates the assumption of our model.

\subsection{Case 4}

For the case 4, we consider a 3-component mixture of normal distribution for random effect, where mean of first component is 2 and the second component is 4 and third component is 6, where variances of each individual mixture is taken to be small to make each sub-population more homogeneous. Figure 1(d) denotes the degradation observation generated for this case. The mean, standard deviation (s.d) and 95\% Credible interval of random effects for each of 10 units, fixed effect $\alpha$ and variance of error $\sigma_{\epsilon}^2$ with respect to the different prior combinations for both the parametric and semi-parametric model are given in Tables 19 and 20. The trace plot and autocorrelation plot for the generated mcmc samples of $\alpha$, $\beta_i$, $i=1,2, 3$ and $\sigma_{\epsilon}^2$ for semi-parametric and parametric method are provided in Figure 18 in Appendix B.

\renewcommand{\arraystretch}{1}
\begin{table}[ht!]
    \centering
    \caption{Posterior mean, standard deviation and 95\% Credible Interval of random effects ($\beta_i$, $i=1,\dots,10$ ), fixed effect ($\alpha$) and variance of measurement error ($\sigma_{\epsilon}^2$) based on Prior 2}.

    \scalebox{0.95}{\begin{tabular}{|c|c|c|c|c|c|c|}
\hline
\multirow{3}{*}{Parameter} & \multicolumn{3}{c|}{Parametric method} & 
    \multicolumn{3}{c|}{Semi-parametric method} \\
\cline{2-7}
 & mean & s.d & 95\% Credible interval & mean & s.d & 95\% Credible interval  \\
\hline
$\beta_1$ & 4.738 & 0.288 & (4.181, 5.303) & 4.745 & 0.295 & (4.164, 5.331) \\
\hline
$\beta_2$ & 4.209 & 0.293 & (3.649, 4.802) & 4.212 & 0.307 & (3.616, 4.822) \\
\hline
$\beta_3$ & 2.121 & 0.294 & (1.570, 2.715) & 2.131 & 0.301 & (1.566, 2.686) \\
\hline
$\beta_4$ & 3.411 & 0.300 & (2.808, 3.974) & 3.409 & 0.317 & (2.800, 4.002) \\
\hline
$\beta_5$ & 2.723 & 0.299 & (2.154, 3.303) & 2.710 & 0.315 & (2.157, 3.344) \\
\hline
$\beta_6$ & 5.906 & 0.304 & (5.354, 6.523) & 5.920 & 0.299 & (5.362, 6.519) \\
\hline
$\beta_7$ & 2.534 & 0.306 & (1.922, 3.138) & 2.521 & 0.309 & (1.954, 3.142) \\
\hline
$\beta_8$ & 2.615 & 0.305 & (2.017, 3.232) & 2.618 & 0.307 & (2.044, 3.212) \\
\hline
$\beta_9$ & 5.741 & 0.307 & (5.156, 6.360) & 5.742 & 0.313 & (5.152, 6.354) \\
\hline
$\beta_{10}$ & 5.863 & 0.298 & (5.298, 6.423) & 5.867 & 0.291 & (5.305, 6.436) \\
\hline
$\alpha$ & 2.043 & 0.095 & (1.865, 2.221) & 2.042 & 0.095 & (1.850, 2.222)\\
\hline
$\sigma_\epsilon^2$ & 0.808 & 0.091 & (0.648, 0.997) & 0.810 & 0.094 & (0.649, 1.022) \\
\hline
\end{tabular}}
\vspace{2mm}
\end{table}

\begin{table}[ht!]
    \centering
    \caption{Posterior mean, standard deviation and 95\% Credible Interval of random effects ($\beta_i$, $i=1,\dots,10$), fixed effect ($\alpha$) and variance of measurement error ($\sigma_{\epsilon}^2$) based on Prior 3 for parametric and prior 5 for semi-parametric method}. 

    \scalebox{0.95}{\begin{tabular}{|c|c|c|c|c|c|c|}
\hline
\multirow{3}{*}{Parameter} & \multicolumn{3}{c|}{Parametric method} & 
    \multicolumn{3}{c|}{Semi-parametric method} \\
\cline{2-7}
 & mean & s.d & 95\% Credible interval & mean & s.d & 95\% Credible interval  \\
\hline
$\beta_1$ & 4.777 & 0.301 & (4.165, 5.373) & 4.774 & 0.299 & (4.104, 5.311) \\
\hline
$\beta_2$ & 4.205 & 0.304 & (3.600, 4.768) & 4.218 & 0.314 & (3.613, 4.848) \\
\hline
$\beta_3$ & 2.094 & 0.316 & (1.522, 2.720) & 2.114 & 0.302 & (1.471, 2.603) \\
\hline
$\beta_4$ & 3.410 & 0.317 & (2.796, 4.042) & 3.404 & 0.320 & (2.843, 3.947) \\
\hline
$\beta_5$ & 2.715 & 0.307 & (2.132, 3.317) & 2.718 & 0.325 & (2.116, 3.264) \\
\hline
$\beta_6$ & 5.922 & 0.305 & (5.335, 6.535) & 5.924 & 0.280 & (5.432, 6.425) \\
\hline
$\beta_7$ & 2.528 & 0.300 & (1.943, 3.127) & 2.526 & 0.276 & (2.070, 3.144) \\
\hline
$\beta_8$ & 2.609 & 0.289 & (2.023, 3.182) & 2.609 & 0.311 & (1.913, 3.132) \\
\hline
$\beta_9$ & 5.755 & 0.313 & (5.119, 6.343) & 5.751 & 0.313 & (5.221, 6.270) \\
\hline
$\beta_{10}$ & 5.896 & 0.300 & (5.320, 6.489) & 5.885 & 0.304 & (5.344, 6.527) \\
\hline
$\alpha$ & 2.040 & 0.093 & (1.856, 2.222) & 2.039 & 0.092 & (1.867, 2.205) \\
\hline
$\sigma_\epsilon^2$ & 0.813 & 0.093 & (0.646, 1.005) & 0.804 & 0.089 & (0.638, 0.992) \\
\hline
\end{tabular}}
\vspace{2mm}
\end{table}


The 95\% credible intervals for both proposed semi-parametric and parametric method seems to overestimate true value of error variance, whereas the fixed effect is estimated accurately. The estimates of random effects for both model tends to be similar as number of observations are increased.

\begin{table}[hbt!]
    \centering
    \caption{Estimated median and 95\% predictive interval of residual lifetime using a Bayesian parametric method and proposed Bayesian semi-parametric method with $m=31$}
    \scalebox{0.95}{\begin{tabular}{|c|c|c|c|c|c|}
\hline
\multirow{2}{*}{Unit} & \multirow{2}{*}{True value} & \multicolumn{2}{c|}{Parametric} & 
    \multicolumn{2}{c|}{Semi-parametric} \\
\cline{3-6}
 & & median & 95\% Predictive interval  & median & 95\% Predictive interval  \\
\hline
1 & 0.189 & 0.475 & (0.073, 0.842) & 0.478 & (0.090, 0.846) \\
\hline
2 & 0.277 & 0.668 & (0.206, 1.177) & 0.667 & (0.117, 1.084) \\
\hline
3 & 1.653 & 2.272 & (1.187, 3.689) & 2.285 & (1.228, 3.692) \\
\hline
4 & 0.632 & 1.033 & (0.350, 1.662) & 1.033 & (0.390, 1.694) \\
\hline
5 & 0.905 & 1.584 & (0.797, 2.523) & 1.541 & (0.734, 2.466) \\
\hline
6 & 0.017 & 0.206 & (0.001, 0.469) & 0.205 & (0.001, 0.458) \\
\hline
7 & 0.928 & 1.761 & (0.893, 2.838) & 1.773 & (0.971, 2.894) \\
\hline
8 & 0.976 & 1.652 & (0.809, 2.676) & 1.652 & (0.772, 2.670) \\
\hline
9 & 0.027 & 0.239 & (0.001, 0.501) & 0.235 & (0.001, 0.499) \\
\hline
10 & 0.027 & 0.221 & (0.001, 0.461) & 0.222 & (0.008, 0.478) \\
\hline
\end{tabular}}
\end{table}

\begin{table}[hbt!]
    \centering
    \caption{KS distance between true and approximated residual lifetime distribution for Bayesian parametric method and proposed Bayesian semi-parametric method}.
    \scalebox{0.95}{\begin{tabular}{|c|c|c|c|c|}
\hline
\multirow{2}{*}{Unit} & \multicolumn{2}{c|}{$m=11$} & 
    \multicolumn{2}{c|}{$m=31$} \\
\cline{2-5}

 &  Parametric & Semi-parametric &  Parametric & Semi-parametric \\
\hline
1   & 0.151     & 0.171    & 0.222    & 0.220    \\
\hline
2   & 0.421     & 0.411    & 0.205    & 0.204    \\
\hline
3   & 0.571     & 0.552    & 0.241    & 0.235    \\
\hline
4   & 0.276     & 0.263    & 0.175    & 0.170    \\
\hline
5   & 0.198     & 0.173    & 0.190    & 0.182    \\
\hline
6   & 0.321     & 0.354    & 0.068    & 0.059    \\
\hline
7   & 0.454     & 0.458    & 0.125    & 0.136    \\
\hline
8   & 0.626     & 0.639    & 0.136    & 0.131    \\
\hline
9   & 0.213     & 0.213    & 0.143    & 0.139    \\
\hline
10  & 0.064     & 0.086    & 0.070    & 0.062    \\
\hline
\end{tabular}}
\end{table}

\begin{table}[hbt!]
    \centering
    \caption{Prediction accuracy for Parametric and Semi-Parametric method}
    \scalebox{0.95}{\begin{tabular}{|c|p{1cm}|p{1cm}|p{1cm}|p{1cm}|p{1cm}|p{1cm}|p{1cm}|p{1cm}|}
\hline
\multirow{3}{*}{Scenario} & \multicolumn{4}{c|}{$M_1$} & \multicolumn{4}{c|}{$M_2$} \\ \cline{2-9}
 & \multicolumn{2}{c|}{Parametric} & \multicolumn{2}{c|}{Semi-parametric} & \multicolumn{2}{c|}{Parametric} & \multicolumn{2}{c|}{Semi-parametric} \\ \cline{2-9}
 
 &  RMSE & MAE & RMSE & MAE &  RMSE & MAE &  RMSE & MAE \\
\hline
$n = 10, m=11$ & 0.637 & 0.688 & 0.632 & 0.685 & 0.605 & 0.674 & 0.611 & 0.674 \\
\hline
$n= 10, m=31$ & 0.501 & 0.669 & 0.499 & 0.668 & 0.486 & 0.657 & 0.492 & 0.660 \\
\hline
$n= 30, m=11$ & 0.792 & 0.780 & 0.829 & 0.791 & 0.782 & 0.774 & 0.822 & 0.785 \\
\hline
$n= 30, m=31$ & 0.720 & 0.742 & 0.725 & 0.742 & 0.726 & 0.740 & 0.738 & 0.743 \\
\hline
$n =50, m=11$ & 0.617 & 0.685 & 0.659 & 0.706 & 0.611 & 0.681 & 0.651 & 0.701 \\
\hline
$n= 50, m=31$ & 0.620 & 0.722 & 0.621 & 0.721 & 0.625 & 0.723 & 0.626 & 0.722 \\
\hline
\end{tabular}}
\end{table}

We present predictive intervals of residual lifetime for the individual units in Table 21. The estimates for residual lifetime are rounded off to three decimal places for each unit. The histograms of the generated residual lifetimes for units 1-3 along with the trace plot of and autocorrelation plot of generated samples for semi-parametric and parametric method are provided in Figures 9 and 10 respectively in the Appendix A to this article. For most of the units, the produced predictive intervals for both parametric and semi-parametric methods contain the true value.

Kolmogorov-Smirnov (KS) statistic obtained for $m=11, 31$ and $n=10$ for this case is given in Table 22. For most of the slowly degrading units for example unit 3, 8 value of KS statistic is less for semi-parametric method compared to parametric method for $m=11,31$. The semi-parametric model is able to accurately estimate the random effects associated with these units, due to the its multi-modality characteristics. For the moderate degrading units for example, unit 4 semi-parametric methods also produces lesser KS value. On the other side, for fastest degrading units the results are very similar. To asses the overall predictive performance of both the models,  RMSE and MAE are calculated for each combination of $n$ and $m$. It is evident from Table 23 that, as the number of observations per unit increases for $n= 10, 30, 50$, both RMSE and MAE decreases. It is found that the produced errors seems to similar with respect to both model under $M_1$ and $M_2$ as the number of units and observations per unit increases.

\subsection{Case 5}

Now, to asses the impact of proposed semi-parametric method, we consider a 5-component mixture of normal distribution for random effect. The degradation paths for the units generated for this case is given by Figure 1(e). Results obtained for estimates of posterior mean, standard deviation (s.d) and 95\% Credible interval of random effects for each of 10 units, fixed effect $\alpha$ and variance of error $\sigma_{\epsilon}^2$ with respect to the different prior combinations for both the parametric and semi-parametric model in Tables 24 and 25. The trace plot and autocorrelation plot for the generated mcmc samples of $\alpha$, $\beta_i$, $i=1,2, 3$ and $\sigma_{\epsilon}^2$ for semi-parametric and parametric method are provided in Figure 19 in Appendix B.

\renewcommand{\arraystretch}{1}
\begin{table}[ht!]
    \centering
    \caption{Posterior mean, standard deviation and 95\% Credible Interval of random effects ($\beta_i$, $i=1,\dots,10$ ), fixed effect ($\alpha$) and variance of measurement error ($\sigma_{\epsilon}^2$) based on Prior 2}. 

    \scalebox{0.95}{\begin{tabular}{|c|c|c|c|c|c|c|}
\hline
\multirow{3}{*}{Parameter} & \multicolumn{3}{c|}{Parametric method} & 
    \multicolumn{3}{c|}{Semi-parametric method} \\
\cline{2-7}
 & mean & s.d & 95\% Credible interval & mean & s.d & 95\% Credible interval  \\
\hline
$\beta_1$ & 1.329 & 0.345 & (0.639, 1.991) & 1.343 & 0.384 & (0.544, 2.079) \\
\hline
$\beta_2$ & 4.626 & 0.350 & (3.908, 5.284) & 4.596 & 0.377 & (3.812, 5.311) \\
\hline
$\beta_3$ & 1.844 & 0.344 & (1.104, 2.503) & 1.837 & 0.349 & (1.158, 2.513) \\
\hline
$\beta_4$ & 3.958 & 0.339 & (3.286, 4.638) & 3.938 & 0.376 & (3.201, 4.655) \\
\hline
$\beta_5$ & 2.622 & 0.346 & (1.940, 3.320) & 2.608 & 0.370 & (1.883, 3.305) \\
\hline
$\beta_6$ & 9.643 & 0.352 & (8.909, 10.326) & 9.625 & 0.422 & (8.598, 10.330) \\
\hline
$\beta_7$ & 2.147 & 0.334 & (1.476, 2.825) & 2.133 & 0.373 & (1.356, 2.882) \\
\hline
$\beta_8$ & 5.694 & 0.352 & (5.069, 6.394) & 5.686 & 0.386 & (4.863, 6.398) \\
\hline
$\beta_9$ & 2.140 & 0.347 & (1.456, 2.866) & 2.108 & 0.368 & (1.353, 2.795) \\
\hline
$\beta_{10}$ & 8.207 & 0.338 & (7.550, 8.889) & 8.185 & 0.407 & (7.308, 8.920)\\
\hline
$\alpha$ & 2.052 & 0.110 & (1.846, 2.264) & 2.057 & 0.136 & (1.827, 2.346) \\
\hline
$\sigma_\epsilon^2$ & 0.994 & 0.122 & (0.784, 1.264) & 0.990 & 0.135 & (0.714, 1.253) \\
\hline
\end{tabular}}
\vspace{2mm}
\end{table}

\begin{table}[ht!]
    \centering
    \caption{Posterior mean, standard deviation and 95\% Credible Interval of random effects ($\beta_i$, $i=1,\dots,10$), fixed effect ($\alpha$) and variance of measurement error ($\sigma_{\epsilon}^2$) based on Prior 5}. 

    \scalebox{0.95}{\begin{tabular}{|c|c|c|c|c|c|c|}
\hline
\multirow{3}{*}{Parameter} & \multicolumn{3}{c|}{Parametric method} & 
    \multicolumn{3}{c|}{Semi-parametric method} \\
\cline{2-7}
 & mean & s.d & 95\% Credible interval & mean & s.d & 95\% Credible interval  \\
\hline
$\beta_1$ & 1.309 & 0.352 & (0.643, 2.029) & 1.327 & 0.342 & (0.666, 1.995) \\
\hline
$\beta_2$ & 4.620 & 0.365 & (3.928, 5.338) & 4.629 & 0.353 & (3.994, 5.359) \\
\hline
$\beta_3$ & 1.793 & 0.346 & (1.105, 2.522) & 1.829 & 0.366 & (1.166, 2.567) \\
\hline
$\beta_4$ & 3.950 & 0.354 & (3.303, 4.649) & 3.963 & 0.344 & (3.326, 4.633) \\
\hline
$\beta_5$ & 2.583 & 0.367 & (1.879, 3.311) & 2.609 & 0.358 & (1.916, 3.360) \\
\hline
$\beta_6$ & 9.725 & 0.355 & (9.032, 10.425) & 9.690 & 0.347 & (8.961, 10.353)\\
\hline
$\beta_7$ & 2.116 & 0.347 & (1.461, 2.777) & 2.133 & 0.358 & (1.475, 2.858) \\
\hline
$\beta_8$ & 5.734 & 0.369 & (5.012, 6.431) & 5.722 & 0.344 & (5.019, 6.361) \\
\hline
$\beta_9$ & 2.082 & 0.362 & (1.379, 2.839) & 2.111 & 0.350 & (1.446, 2.865) \\
\hline
$\beta_{10}$ & 8.281 & 0.353 & (7.548, 8.943) & 8.279 & 0.347 & (7.630, 8.983) \\
\hline
$\alpha$ & 2.052 & 0.115 & (1.808, 2.276) & 2.048 & 0.113 & (1.826, 2.267) \\
\hline
$\sigma_\epsilon^2$ & 1.016 & 0.126 & (0.804, 1.301) & 1.011 & 0.123 & (0.788, 1.267) \\
\hline
\end{tabular}}
\vspace{2mm}
\end{table}


It is evident from Tables 24 and 25 that for the proposed semi-parametric and parametric method seems to overestimate the true error variance. On the other side, 95\% credible interval contains true fixed value for both the methods. The posterior mean for semi-parametric model is slightly higher than parametric method under prior 5. Similar to the previous cases, we find that the estimates produced by both the methods are not significantly different under different prior combination as the number of observations increases. We estimate residual lifetime distribution with respect to prior 2.

\begin{table}[hbt!]
    \centering
    \caption{Estimated median and 95\% predictive interval of residual lifetime using a Bayesian parametric method and proposed Bayesian semi-parametric method with $m=31$}
    \scalebox{0.95}{\begin{tabular}{|c|c|c|c|c|c|}
\hline
\multirow{2}{*}{Unit} & \multirow{2}{*}{True value} & \multicolumn{2}{c|}{Parametric} & 
    \multicolumn{2}{c|}{Semi-parametric} \\
\cline{3-6}
 & & median & 95\% Predictive interval  & median & 95\% Predictive interval  \\
\hline
1    & 3.284 & 7.356 & (3.482, 14.451) & 7.342 & (3.209, 15.710) \\
\hline
2    & 1.035 & 1.359 & (0.832, 1.873)  & 1.376 & (0.881, 1.972)  \\
\hline
3    & 3.842 & 4.853 & (2.912, 8.097)  & 4.956 & (2.955, 8.178)  \\
\hline
4    & 1.196 & 1.757 & (1.111, 2.413)  & 1.779 & (1.107, 2.422)  \\
\hline
5    & 2.702 & 3.206 & (2.054, 4.544)  & 3.199 & (1.974, 4.711)  \\
\hline
6    & 0.014 & 0.145 & (0.001, 0.313)  & 0.146 & (0.002, 0.333)  \\
\hline
7    & 3.703 & 4.165 & (2.528, 6.506)  & 4.146 & (2.361, 6.701)  \\
\hline
8    & 0.701 & 0.925 & (0.491, 1.268)  & 0.920 & (0.525, 1.348)  \\
\hline
9    & 2.756 & 4.126 & (2.438, 6.355)  & 4.237 & (2.675, 6.704)  \\
\hline
10   & 0.043 & 0.335 & (0.116, 0.602)  & 0.341 & (0.105, 0.612)  \\
\hline
\end{tabular}}
\end{table}

\begin{table}[hbt!]
    \centering
    \caption{KS distance between true and approximated residual lifetime distribution for Bayesian parametric method and proposed Bayesian semi-parametric method}.
    \scalebox{0.95}{\begin{tabular}{|c|c|c|c|c|}
\hline
\multirow{2}{*}{Unit} & \multicolumn{2}{c|}{$m=11$} & 
    \multicolumn{2}{c|}{$m=31$} \\
\cline{2-5}

 &  Parametric & Semi-parametric &  Parametric & Semi-parametric \\
\hline
1 & 0.719 & 0.707  & 0.877  & 0.867  \\
\hline
2 & 0.376 & 0.412  & 0.259  & 0.247  \\
\hline
3 & 0.469 & 0.471  & 0.302  & 0.279  \\
\hline
4 & 0.173 & 0.199  & 0.273  & 0.302  \\
\hline
5 & 0.595 & 0.595  & 0.305  & 0.321  \\
\hline
6 & 0.092 & 0.105  & 0.117  & 0.128  \\
\hline
7 & 0.403 & 0.464  & 0.592  & 0.571  \\
\hline
8 & 0.412 & 0.391  & 0.191  & 0.192  \\
\hline
9 & 0.319 & 0.312  & 0.345 & 0.375  \\
\hline
10 & 0.121 & 0.121  & 0.108  & 0.131  \\
\hline
\end{tabular}}
\end{table}

\begin{table}[hbt!]
    \centering
    \caption{Prediction accuracy for Parametric and Semi-Parametric method}.
    \scalebox{0.95}{\begin{tabular}{|c|p{1cm}|p{1cm}|p{1cm}|p{1cm}|p{1cm}|p{1cm}|p{1cm}|p{1cm}|}
\hline
\multirow{3}{*}{Scenario} & \multicolumn{4}{c|}{$M_1$} & \multicolumn{4}{c|}{$M_2$} \\ \cline{2-9}
 & \multicolumn{2}{c|}{Parametric} & \multicolumn{2}{c|}{Semi-parametric} & \multicolumn{2}{c|}{Parametric} & \multicolumn{2}{c|}{Semi-parametric} \\ \cline{2-9}
 &  RMSE & MAE & RMSE & MAE &  RMSE & MAE &  RMSE & MAE \\
\hline
$n = 10, m=11$ & 1.176 & 0.832 & 1.181 & 0.824 & 1.013 & 0.797 & 0.950 & 0.767 \\
\hline
$n= 10, m=31$ & 1.433 & 0.946 & 1.447 & 0.957 & 1.411 & 0.949 & 1.425 & 0.960 \\
\hline
$n= 30, m=11$ & 1.536 & 1.019 & 1.488 & 1.003 & 1.501 & 1.012 & 1.488 & 1.003 \\
\hline
$n= 30, m=31$ & 0.916 & 0.842 & 0.923 & 0.842 & 0.904 & 0.838 & 0.902 & 0.834 \\
\hline
$n =50, m=11$ & 0.967 & 0.811 & 0.975 & 0.810 & 0.916 & 0.794 & 0.991 & 0.811 \\
\hline
$n= 50, m=31$ & 0.716 & 0.751 & 0.801 & 0.794 & 0.710 & 0.747 & 0.807 & 0.795 \\
\hline
\end{tabular}}
\end{table}

The predictive intervals of residual lifetime for the individual units is given in Table 26. For most of the units, the produced predictive intervals for both parametric and semi-parametric methods contain the true value. Although for unit 1, predictive interval for parametric method fails to contain true residual lifetime whereas semi-parametric method does. For unit 10 both the method fails to contain the true value. It is also evident that both parametric and semi-parametric method produces predictive interval of very large length for the most slowly degrading units, which indicates to the fact that both the model is unable to predict occurrences of failures which will happen after a significantly long time. The histograms of the generated residual lifetimes for units 1-3 along with the trace plot of and autocorrelation plot of generated samples for semi-parametric and parametric method are provided in Figure 11 and 12 respectively in the Appendix A of this article.

Kolmogorov-Smirnov (KS) distance obtained for $m=11, 31$ and $n=10$ is given in Table 27. For most of the slowly degrading units for example unit 1, 3 value of KS statistic is less for semi-parametric method compared to parametric method for $m=31$. For most of the moderate degrading units for example unit 8, both parametric and semi-parametric methods produces almost similar KS value as $m$ increases. On the other side, for fastest degrading units value of KS distance is less for parametric method. It is evident from Table 28 that, as the number of observations per unit increases for $n= 30, 50$, both RMSE and MAE decreases. It is found that the produced errors seems to similar with respect to both model under $M_1$ and $M_2$ as $n$ and $m$ both increases. In each scenarios for case 3,4 and 5, the estimated median of residual lifetime is higher than true value which is also observed for case 1 and 2. The reason is similar the distribution functions for the individual residual lifetime contains terms which are average of many distribution functions and longer upper tail.

\section{Data Analysis}

Now we demonstrate analysis of fatigue crack size dataset given in Lu \& Meeker\cite{Lu}. The measurements are taken for 21 units. The crack lengths are measured at interval of $0.01$ million of cycles. It is considered that failure occurs when crack length exceeded the threshold value $D= 1.6$ inches (see Lawless and Crowder \cite{Lawless}, Lu and Meeker\cite{Lu}, Giorgio et al. \cite{Giorgio}). For each unit the initial crack size was 0.9 inch and data collection was terminated at the first inspection after a unit’s crack size reached 1.6 inches or censored after 0.12 million cycles, whichever came first. Degradation paths of the units are presented in Figure 2. It may be noted from the plot that some of the units are degrading faster than other units, so it can be considered that the units are coming from heterogeneous population. For semi-parametric method we take truncation number $N$ as 21. 

\vspace{5mm}

\begin{figure}[hbt!]
    \centering
    \includegraphics[width=0.7\textwidth]{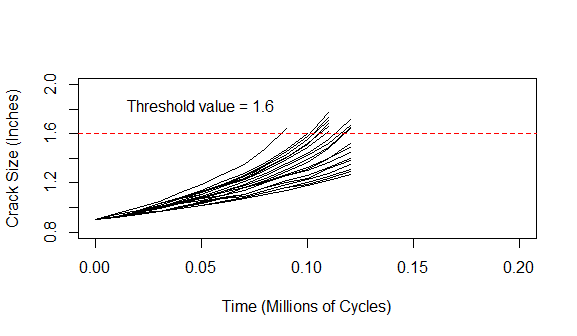}
    \caption{Development of crack sizes over time \cite{Lu}}
    \label{Fig:Figure 5}
\end{figure}

\renewcommand{\arraystretch}{1}
\begin{table}[ht!]
    \centering
    \caption{Posterior mean, standard deviation and 95\% Credible Interval of random effects ($\beta_i$, $i=1,\dots,21$ ), fixed effect ($\alpha$) and variance of measurement error ($\sigma_{\epsilon}^2$) based on Prior 2.} 

    \scalebox{0.95}{\begin{tabular}{|c|c|c|c|c|c|c|}
\hline
\multirow{3}{*}{Parameter} & \multicolumn{3}{c|}{Parametric method} & 
    \multicolumn{3}{c|}{Semi-parametric method} \\
\cline{2-7}
 & mean & s.d & 95\% Credible interval & mean & s.d & 95\% Credible interval  \\
\hline
$\beta_1$ & 5.665 & 0.879 & (4.010, 7.456) & 5.690 & 0.790 & (4.166, 7.368) \\
\hline
$\beta_2$ & 5.617 & 0.812 & (3.944, 7.206) & 5.617 & 0.794 & (4.073, 7.120) \\
\hline
$\beta_3$ & 5.749 & 0.775 & (4.280, 7.348) & 5.759 & 0.741 & (4.429, 7.362) \\
\hline
$\beta_4$ & 5.653 & 0.794 & (4.193, 7.398) & 5.670 & 0.755 & (4.171, 7.181) \\
\hline
$\beta_5$ & 5.679 & 0.788 & (4.271, 7.409) & 5.633 & 0.763 & (4.140, 7.247)\\
\hline
$\beta_6$ & 5.626 & 0.757 & (4.162, 7.171) & 5.644 & 0.719 & (4.212, 7.037) \\
\hline
$\beta_7$ & 5.614 & 0.784 & (4.038, 7.094) & 5.539 & 0.756 & (4.048, 7.110) \\
\hline
$\beta_8$ & 5.585 & 0.792 & (3.916, 7.270) & 5.530 & 0.714 & (4.230, 7.023) \\
\hline
$\beta_9$ & 5.573 & 0.766 & (4.048, 6.988) & 5.563 & 0.703 & (4.153, 6.980) \\
\hline
$\beta_{10}$ & 5.455 & 0.757 & (4.064, 6.904) & 5.487 & 0.707 & (4.131, 6.881) \\
\hline
$\beta_{11}$ & 5.444 & 0.777 & (3.964, 6.926) & 5.412 & 0.720 & (3.947, 6.837) \\
\hline
$\beta_{12}$ & 5.415 & 0.770 & (3.986, 6.902) & 5.380 & 0.734 & (3.928, 6.971) \\
\hline
$\beta_{13}$ & 5.235 & 0.741 & (3.762, 6.755) & 5.296 & 0.762 & (3.859, 6.900) \\
\hline
$\beta_{14}$ & 5.146 & 0.757 & (3.514, 6.549) & 5.185 & 0.711 & (3.739, 6.579) \\
\hline
$\beta_{15}$ & 5.278 & 0.775 & (3.764, 6.762) & 5.218 & 0.718 & (3.800, 6.683) \\
\hline
$\beta_{16}$ & 5.037 & 0.802 & (3.368, 6.638) & 5.063 & 0.752 & (3.588, 6.545) \\
\hline
$\beta_{17}$ & 5.052 & 0.762 & (3.549, 6.603) & 5.027 & 0.729 & (3.505, 6.435) \\
\hline
$\beta_{18}$ & 4.934 & 0.807 & (3.249, 6.428) & 5.014 & 0.766 & (3.373, 6.416) \\
\hline
$\beta_{19}$ & 4.895 & 0.798 & (3.192, 6.411) & 4.943 & 0.733 & (3.379, 6.352) \\
\hline
$\beta_{20}$ & 4.886 & 0.787 & (3.292, 6.357) & 4.880 & 0.740 & (3.328, 6.271) \\
\hline
$\beta_{21}$ & 4.837 & 0.812 & (3.094, 6.359) & 4.835 & 0.739 & (3.316, 6.164) \\
\hline
$\alpha$ & 0.845 & 0.031 & (0.783, 0.905) & 0.844 & 0.028 & (0.787, 0.896) \\
\hline
$\sigma_\epsilon^2$ & 0.114 & 0.019 & (0.077, 0.152) & 0.113 & 0.017 & (0.082, 0.152) \\
\hline
\end{tabular}}
\vspace{2mm}
\end{table}

The mean, standard deviation (s.d) and 95\% credible interval (C.I) of posterior samples of random effects for each of 21 units, fixed effect $\alpha$ and variance of measurement error $\sigma_{\epsilon}^2$ obtained for both parametric and semi-parametric method under prior 2, are presented in Table 29. Observe that for some units, posterior mean of random effect is much higher than some other units. For example for unit 1, the estimated posterior mean for random effects is much higher than that of unit 20, which can motivate us to analyze this data using the proposed semi-parametric model. The trace plot and autocorrelation plot for the generated mcmc samples of $\alpha$, $\beta_i$, $i=1,2$ and 3 and $\sigma_{\epsilon}^2$ for semi-parametric and parametric method are provided in Figure 20 in Appendix B to this article.

\begin{table}[hbt!]
    \centering
    \caption{Estimated median and 95\% predictive interval of residual lifetime using a Bayesian parametric method and proposed Bayesian semi-parametric method with $m=31$}
    \scalebox{0.95}{\begin{tabular}{|c|c|c|c|c|c|}
\hline
\multirow{2}{*}{Unit} & \multirow{2}{*}{True value} & \multicolumn{2}{c|}{Parametric} & 
    \multicolumn{2}{c|}{Semi-parametric} \\
\cline{3-6}
 & & median & 95\% Predictive interval  & median & 95\% Predictive interval  \\
\hline
1 & 0.008  & 0.07 & (0.0003,0.191) & 0.068 & (0.0003, 0.175) \\
\hline
2 & 0.010  & 0.063 & (0.0001, 0.174) & 0.061 & (0.0007, 0.164) \\
\hline
3 & 0.001  & 0.056 & (0.0001, 0.152) & 0.059 & (0.0002, 0.153) \\
\hline
4 & 0.003  & 0.056 & (0.0002, 0.148) & 0.056 & (0.0002, 0.158) \\
\hline
5 & 0.003  & 0.057 & (0.0008, 0.162) & 0.061 & (0.0002, 0.160) \\
\hline
6 & 0.006  & 0.058 & (0.0001, 0.156) & 0.055 & (0.0001, 0.153) \\
\hline
7 & 0.006  & 0.057 & (0.0001, 0.163) & 0.055 & (0.0001, 0.150) \\
\hline
8 & 0.009  & 0.057 & (0.0001, 0.167) & 0.058 & (0.0001, 0.151) \\
\hline
9 & 0.003  & 0.057 & (0.0002, 0.152) & 0.055 & (0.0001, 0.157) \\
\hline
10 & 0.005 & 0.051 & (0.0001, 0.153) & 0.059 & (0.0002, 0.161) \\
\hline
11 & 0.008 & 0.059 & (0.0003, 0.157) & 0.056 & (0.0001, 0.159) \\
\hline
12 & 0.008 & 0.056 & (0.0001, 0.151) & 0.057 & (0.0001, 0.164) \\
\hline
13 & 0.009 & 0.063 & (0.0001, 0.161) & 0.055 & (0.0001, 0.159) \\
\hline
14 & 0.013 & 0.056 & (0.0001, 0.167) & 0.055 & (0.0006, 0.161) \\
\hline
15 & 0.018 & 0.057 & (0.0004, 0.158) & 0.054 & (0.0003, 0.163) \\
\hline
16 & 0.024 & 0.062 & (0.0002, 0.178) & 0.053 & (0.0001, 0.166) \\
\hline
17 & 0.026 & 0.060 & (0.0002, 0.165) & 0.063 & (0.0007, 0.174) \\
\hline
18 & 0.031 & 0.062 & (0.0001, 0.185) & 0.062 & (0.0001, 0.178) \\
\hline
19 & 0.040 & 0.068 & (0.0002, 0.189) & 0.063 & (0.0002, 0.172) \\
\hline
20 & 0.047 & 0.068 & (0.0002, 0.192) & 0.067 & (0.0001, 0.183) \\
\hline
21 & 0.050 & 0.069 & (0.0003, 0.191) & 0.066 & (0.0001, 0.186) \\
\hline
\end{tabular}}
\end{table}

\vspace{2mm}

We derive residual lifetime distribution of individual units corresponding to the last time point at which degrading unit has not crossed the threshold value with respect each prior combination. Table 30 presents residual lifetime estimates where the lower limit of the predictive interval is rounded off to four decimal places and rest of the values are rounded off to three decimal places along with the true values for each of 21 units. Histograms of the generated residual lifetimes of units 1-3 along with the trace plot and autocorrelation plot for the generated samples for semi-parametric and parametric method are provided in Figures 13 and 14 respectively in the Appendix A to this article. It is evident from Table 31 and 32 that both proposed semi-parametric and parametric method perform equally well in almost all the cases. Also, the choice of prior is not impacting the prediction accuracy for both models. It is also evident that for most of the cases the produced predictive intervals for both the methods contain true value for the residual lifetime.

\begin{table}[hbt!]
    \centering
    \caption{Prediction accuracy for Parametric method}
    \scalebox{0.95}{\begin{tabular}{|c|p{1cm}|p{1cm}|p{1cm}|p{1cm}|}
\hline
\multirow{2}{*}{Scenario} & \multicolumn{2}{c|}{$M_1$} & \multicolumn{2}{c|}{$M_2$} \\ \cline{2-5}
 
 &  RMSE & MAE & RMSE & MAE  \\
\hline
Prior 1 & 0.047 & 0.209 & 0.048 & 0.211 \\
\hline
Prior 2 & 0.046 & 0.212 & 0.047 & 0.211 \\
\hline
Prior 3 & 0.045 & 0.201 & 0.046 & 0.202 \\
\hline
\end{tabular}}
\end{table}

\begin{table}[hbt!]
    \centering
    \caption{Prediction accuracy for Semi-parametric method}
    \scalebox{0.95}{\begin{tabular}{|c|p{1cm}|p{1cm}|p{1cm}|p{1cm}|}
\hline
\multirow{2}{*}{Scenario} & \multicolumn{2}{c|}{$M_1$} & \multicolumn{2}{c|}{$M_2$} \\ \cline{2-5}
 
 &  RMSE & MAE & RMSE & MAE  \\
\hline
Prior 1 & 0.047 & 0.206 & 0.045 & 0.201 \\
\hline
Prior 2 & 0.046 & 0.209 & 0.048 & 0.214 \\
\hline
Prior 3 & 0.045 & 0.207 & 0.045 & 0.207 \\
\hline
Prior 4 & 0.045 & 0.204 & 0.044 & 0.200 \\
\hline
Prior 5 & 0.046 & 0.206 & 0.047 & 0.207 \\
\hline
\end{tabular}}
\end{table}

\newpage
 
\section{Conclusion}

The objective of this article is to predict residual lifetime of units in a heterogeneous situation. To deal with heterogeneity, we proposed a Bayesian semi-parametric degradation model. In the first part we assumed a general path model where the random effects are modeled using the Dirichlet process mixture of normal distributions. Model hierarchy is represented according to Ishwaran \& Zarrerpur \cite{Ishwaran2002}. Gibbs sampling is used to draw samples from the posterior distribution of the model parameters. In the second part of the problem we developed the residual lifetime distribution of each unit which depends on the samples drawn from posterior distribution. Finally we simulate samples from this distribution using MCMC technique and considered the sample median as the predicted residual lifetime of each unit. For implementation of the proposed method we use the R programming language. The proposed Bayesian semi-parametric method is compared with a Bayesian parametric method with respect to the error criteria discussed earlier in this paper. It is found that in both parametric and semi-parametric model, the estimated median of residual lifetime is higher than true value for most of the cases. The reason for this is the distribution functions derived to simulate observations involves terms which are average of many distribution functions and as a result this induces a much longer upper tail. In simulation study it is found that, in most of the cases the proposed method is performing well compared to the parametric method for the the units which degrade slowly compared to others, whereas the parametric method performs better than the proposed method for the fastest degrading units in most of scenarios. Finally, we applied the model to Fatigue Crack-Size dataset to evaluate the performance of our model. In this case, it is found that the prediction is almost similar for both the proposed semi-parametric and parametric method, but for most of the units, Bayesian semi-parametric method produces more accurate estimates for residual lifetime. The proposed Bayesian semi-parametric degradation model may be used for similar problems like in medical diagnosis contexts as well.

\newpage

\section*{APPENDIX A : MCMC convergence for residual lifetime}

\subsection*{Case 1}

\begin{figure}[ht!]
\begin{subfigure}[t]{0.3\textwidth}
    \includegraphics[width=\linewidth]{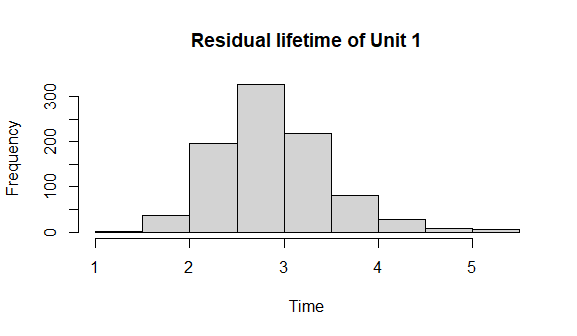}
\end{subfigure}\hfill
\begin{subfigure}[t]{0.3\textwidth}
  \includegraphics[width=\linewidth]{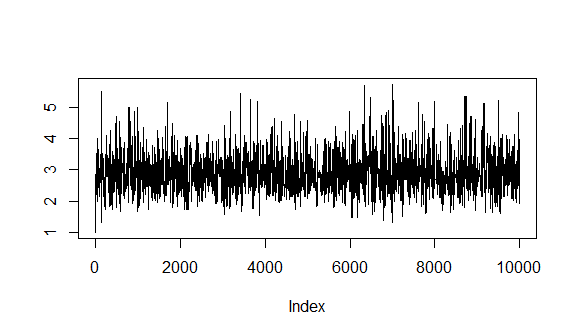}
\end{subfigure}\hfill
\begin{subfigure}[t]{0.3\textwidth}
    \includegraphics[width=\linewidth]{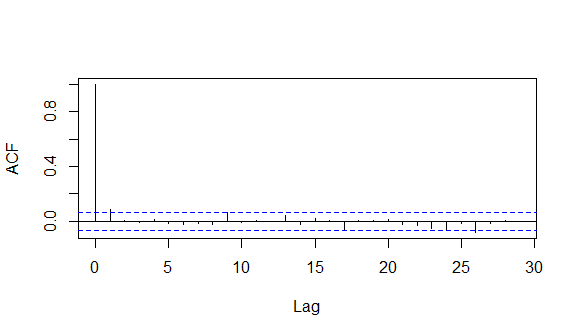}
\end{subfigure}

\begin{subfigure}[t]{0.3\textwidth}
    \includegraphics[width=\linewidth]{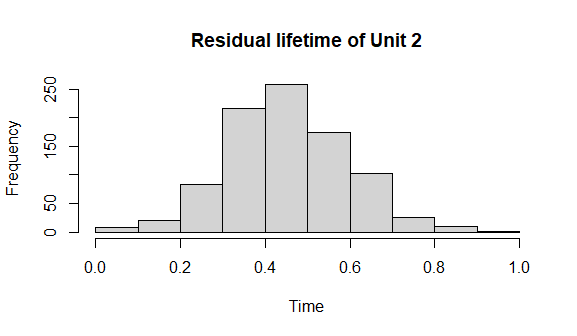}
\end{subfigure}\hfill
\begin{subfigure}[t]{0.3\textwidth}
    \includegraphics[width=\linewidth]{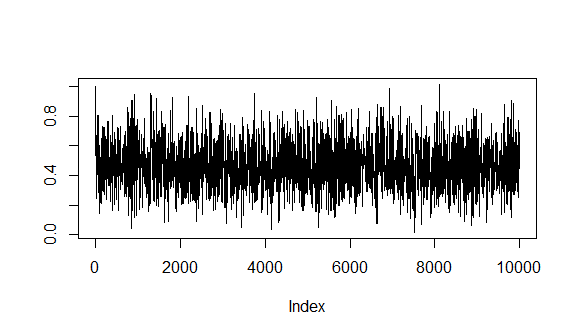}
\end{subfigure}\hfill
\begin{subfigure}[t]{0.3\textwidth}
    \includegraphics[width=\textwidth]{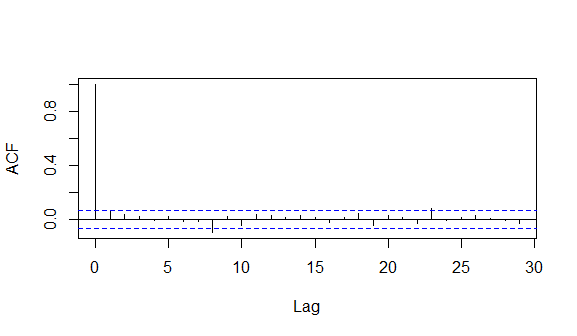}
\end{subfigure}

\begin{subfigure}[t]{0.3\textwidth}
    \includegraphics[width=\linewidth]{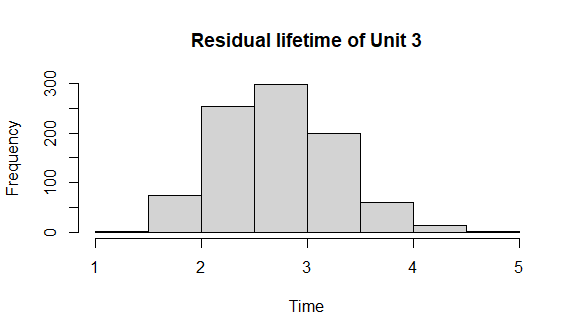}
\end{subfigure}\hfill
\begin{subfigure}[t]{0.3\textwidth}
    \includegraphics[width=\linewidth]{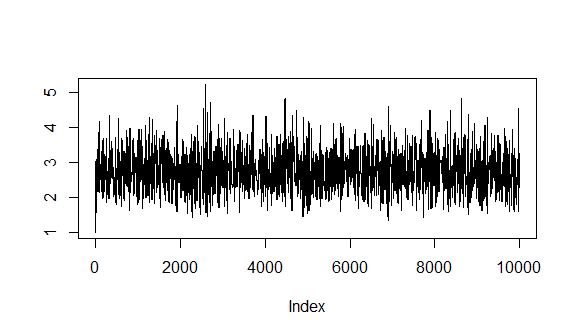}
\end{subfigure}\hfill
\begin{subfigure}[t]{0.3\textwidth}
    \includegraphics[width=\textwidth]{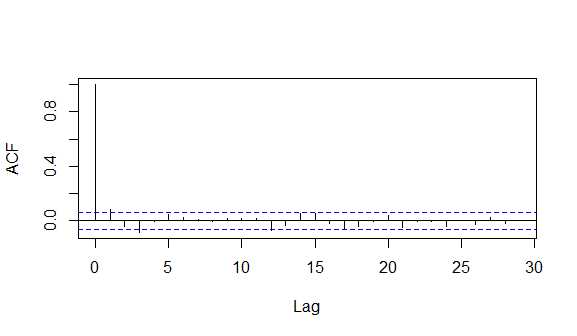}
\end{subfigure}

\caption{For each row, the left side plot is for histogram for Residual lifetimes, middle one is for trace plot and right side denotes autocorrelation plot of the samples produced for Unit 1-3 and $n=10$, $m=31$ by semi-parametric method.}
\end{figure}


\begin{figure}[ht!]
\begin{subfigure}[t]{0.3\textwidth}
    \includegraphics[width=\linewidth]{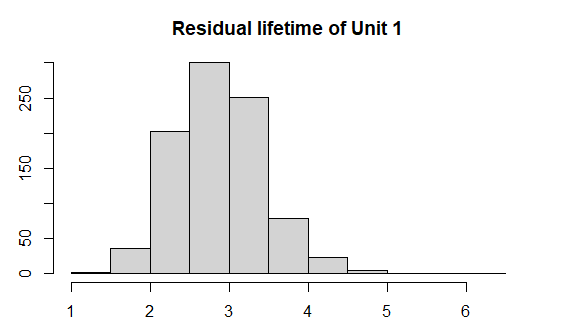}
\end{subfigure}\hfill
\begin{subfigure}[t]{0.3\textwidth}
  \includegraphics[width=\linewidth]{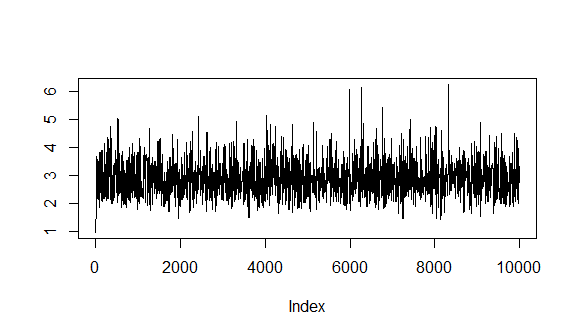}
\end{subfigure}\hfill
\begin{subfigure}[t]{0.3\textwidth}
    \includegraphics[width=\linewidth]{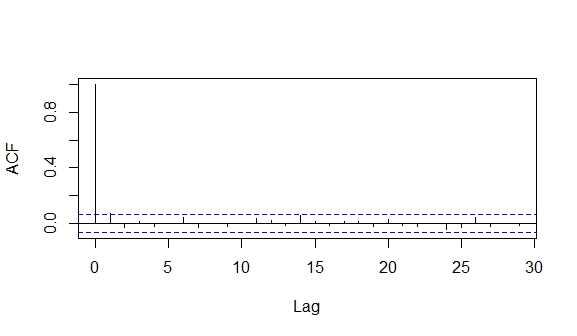}
\end{subfigure}

\begin{subfigure}[t]{0.3\textwidth}
    \includegraphics[width=\linewidth]{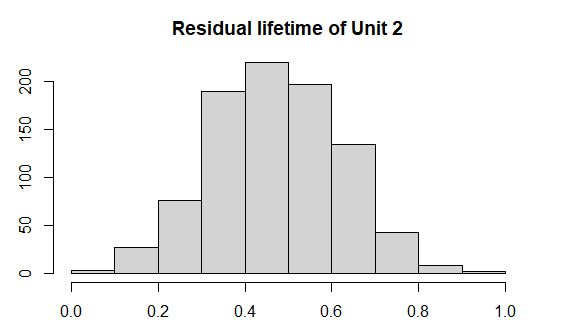}
\end{subfigure}\hfill
\begin{subfigure}[t]{0.3\textwidth}
    \includegraphics[width=\linewidth]{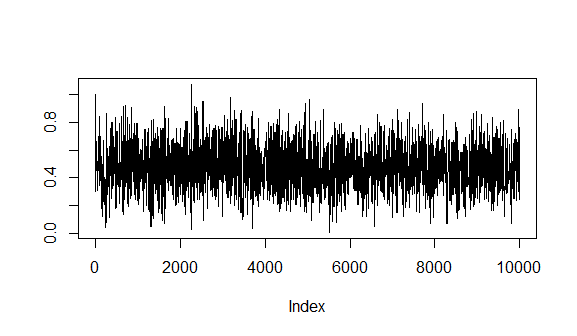}
\end{subfigure}\hfill
\begin{subfigure}[t]{0.3\textwidth}
    \includegraphics[width=\textwidth]{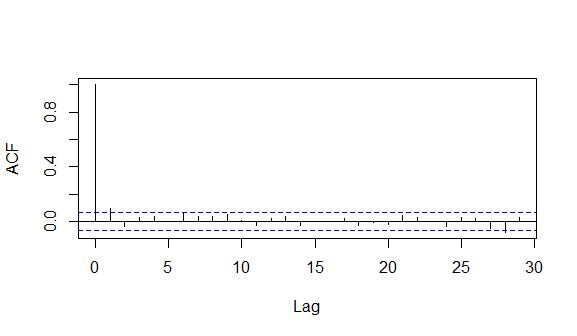}
\end{subfigure}

\begin{subfigure}[t]{0.3\textwidth}
    \includegraphics[width=\linewidth]{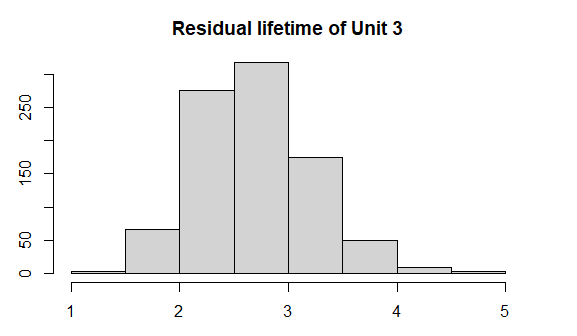}
\end{subfigure}\hfill
\begin{subfigure}[t]{0.3\textwidth}
    \includegraphics[width=\linewidth]{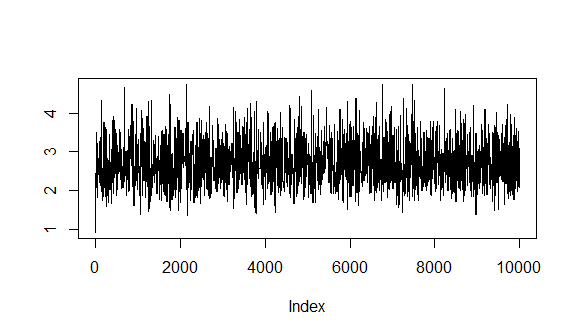}
\end{subfigure}\hfill
\begin{subfigure}[t]{0.3\textwidth}
    \includegraphics[width=\textwidth]{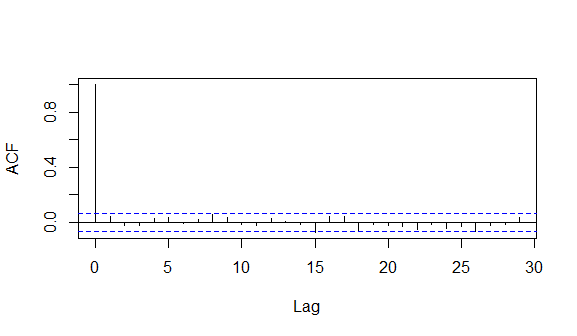}
\end{subfigure}

\caption{For each row, the left side plot is for histogram for Residual lifetimes, middle one is for trace plot and right side denotes autocorrelation plot of the samples produced for Unit 1-3 and $n=10$, $m=31$ by parametric method.}

\end{figure}

\subsection*{Case 2}

\begin{figure}[ht!]
\begin{subfigure}[t]{0.3\textwidth}
    \includegraphics[width=\linewidth]{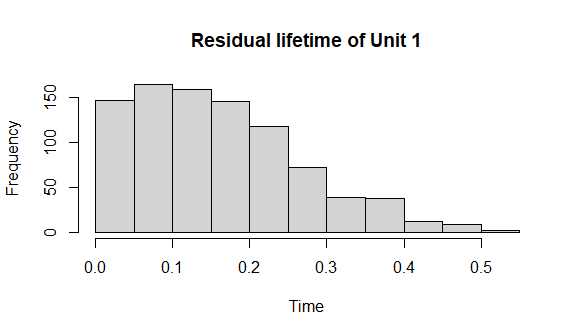}
\end{subfigure}\hfill
\begin{subfigure}[t]{0.3\textwidth}
  \includegraphics[width=\linewidth]{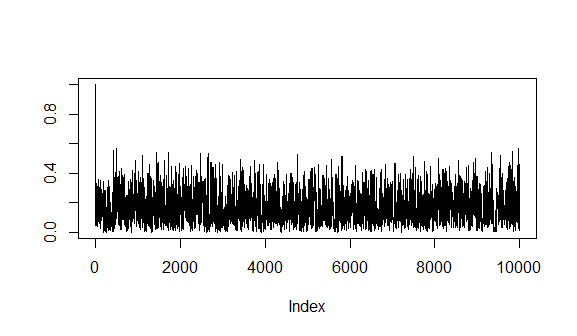}
\end{subfigure}\hfill
\begin{subfigure}[t]{0.3\textwidth}
    \includegraphics[width=\linewidth]{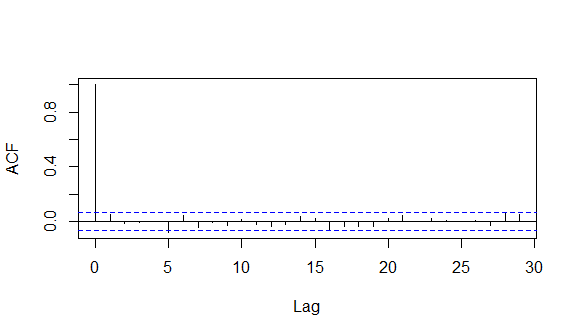}
\end{subfigure}

\begin{subfigure}[t]{0.3\textwidth}
    \includegraphics[width=\linewidth]{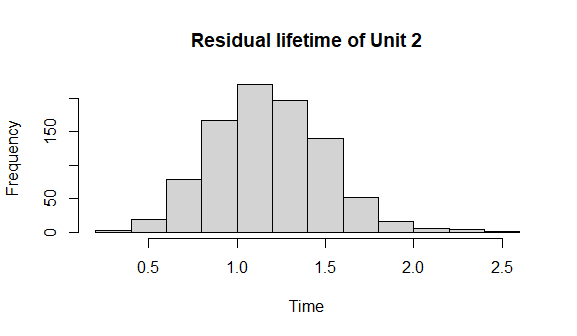}
\end{subfigure}\hfill
\begin{subfigure}[t]{0.3\textwidth}
    \includegraphics[width=\linewidth]{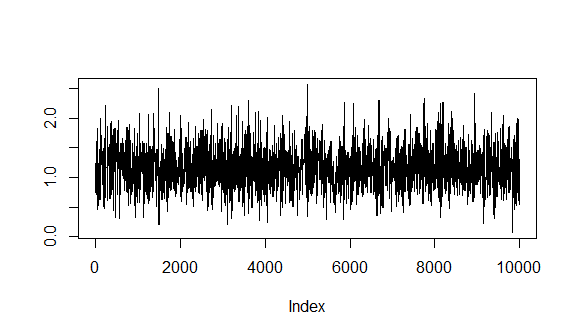}
\end{subfigure}\hfill
\begin{subfigure}[t]{0.3\textwidth}
    \includegraphics[width=\textwidth]{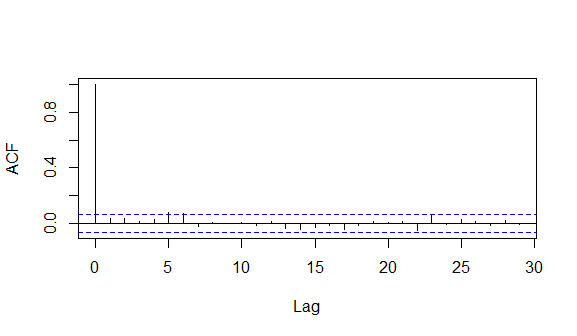}
\end{subfigure}

\begin{subfigure}[t]{0.3\textwidth}
    \includegraphics[width=\linewidth]{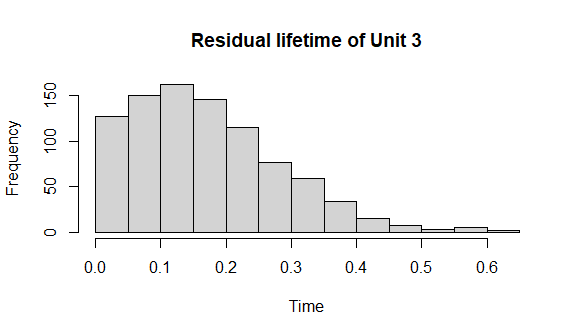}
\end{subfigure}\hfill
\begin{subfigure}[t]{0.3\textwidth}
    \includegraphics[width=\linewidth]{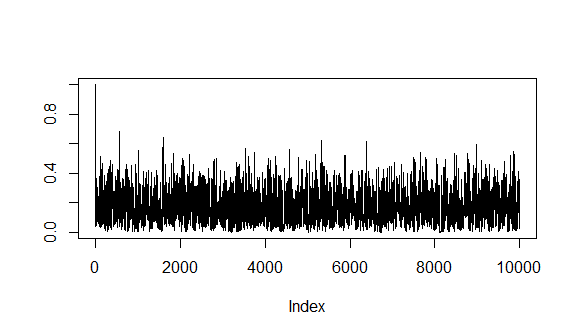}
\end{subfigure}\hfill
\begin{subfigure}[t]{0.3\textwidth}
    \includegraphics[width=\textwidth]{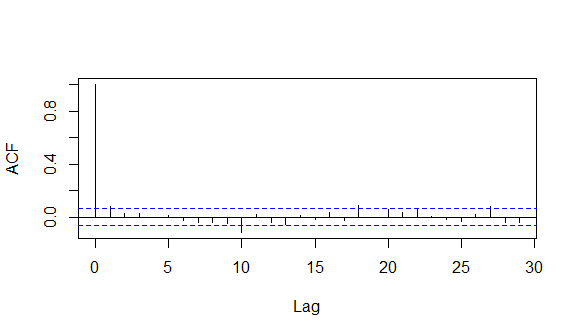}
\end{subfigure}

\caption{For each row, the left side plot is for histogram for Residual lifetimes, middle one is for trace plot and right side denotes autocorrelation plot of the samples produced for Unit 1-3 and $n=10$, $m=31$ by semi-parametric method.}
\end{figure}


\begin{figure}[ht!]
\begin{subfigure}[t]{0.3\textwidth}
    \includegraphics[width=\linewidth]{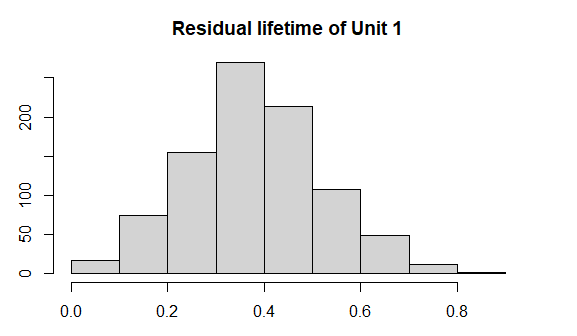}
\end{subfigure}\hfill
\begin{subfigure}[t]{0.3\textwidth}
  \includegraphics[width=\linewidth]{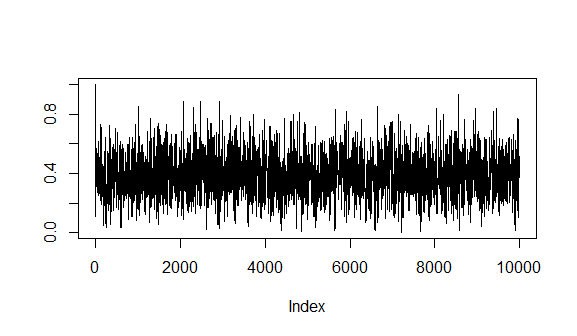}
\end{subfigure}\hfill
\begin{subfigure}[t]{0.3\textwidth}
    \includegraphics[width=\linewidth]{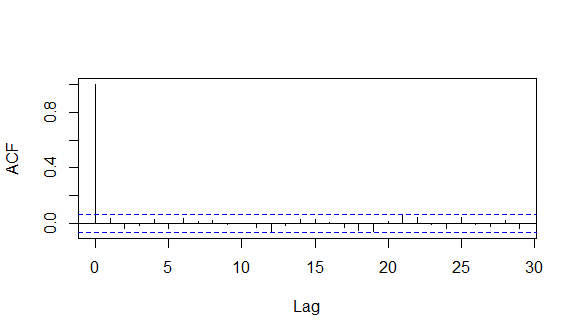}
\end{subfigure}

\begin{subfigure}[t]{0.3\textwidth}
    \includegraphics[width=\linewidth]{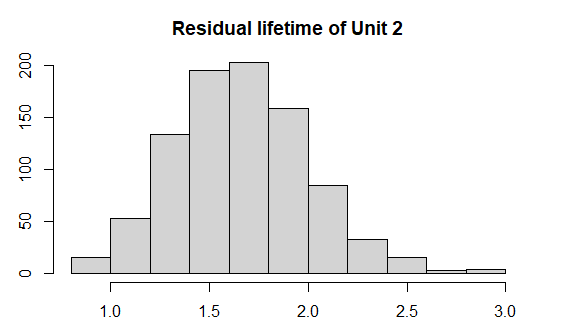}
\end{subfigure}\hfill
\begin{subfigure}[t]{0.3\textwidth}
    \includegraphics[width=\linewidth]{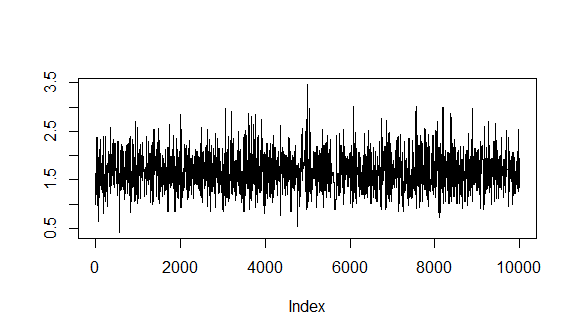}
\end{subfigure}\hfill
\begin{subfigure}[t]{0.3\textwidth}
    \includegraphics[width=\textwidth]{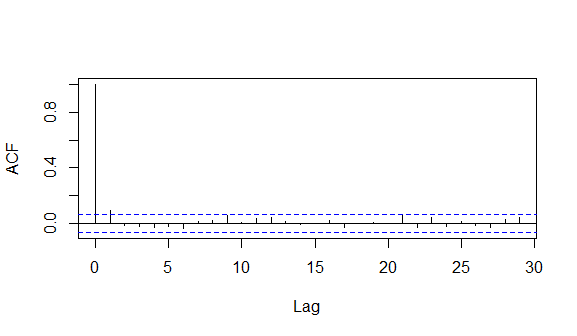}
\end{subfigure}

\begin{subfigure}[t]{0.3\textwidth}
    \includegraphics[width=\linewidth]{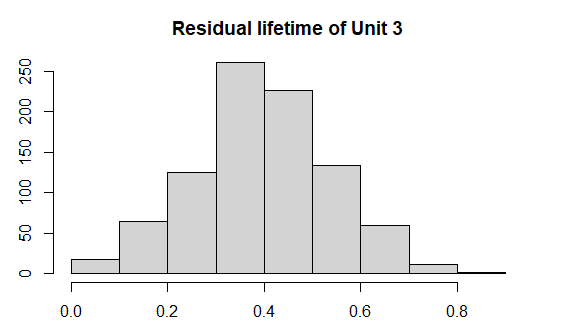}
\end{subfigure}\hfill
\begin{subfigure}[t]{0.3\textwidth}
    \includegraphics[width=\linewidth]{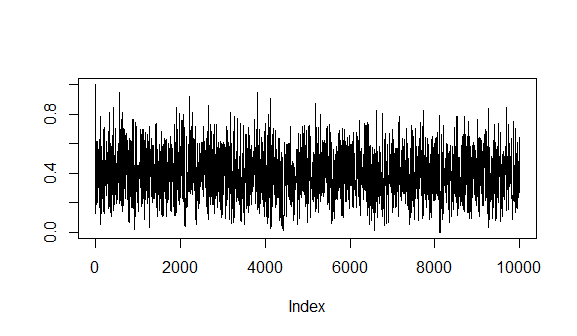}
\end{subfigure}\hfill
\begin{subfigure}[t]{0.3\textwidth}
    \includegraphics[width=\textwidth]{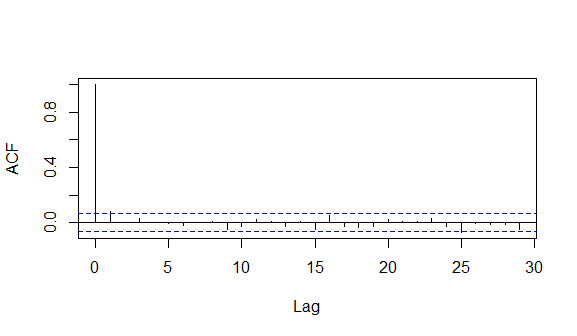}
\end{subfigure}

\caption{For each row, the left side plot is for histogram for Residual lifetimes, middle one is for trace plot and right side denotes autocorrelation plot of the samples produced for Unit 1-3 and $n=10$, $m=31$ by parametric method.}

\end{figure}

\newpage 

\subsection*{Case 3}

\begin{figure}[ht!]
\begin{subfigure}[t]{0.3\textwidth}
    \includegraphics[width=\linewidth]{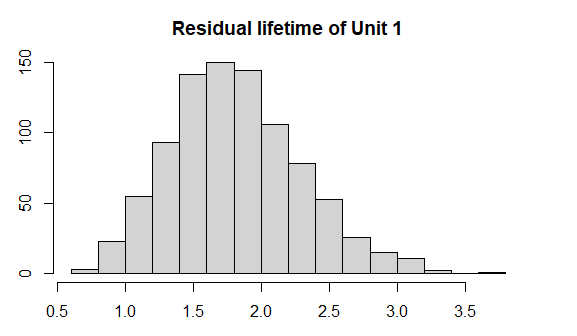}
\end{subfigure}\hfill
\begin{subfigure}[t]{0.3\textwidth}
  \includegraphics[width=\linewidth]{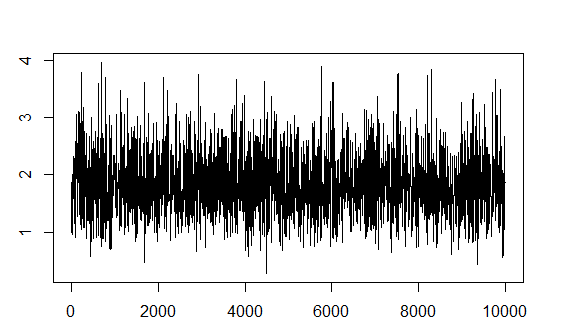}
\end{subfigure}\hfill
\begin{subfigure}[t]{0.3\textwidth}
    \includegraphics[width=\linewidth]{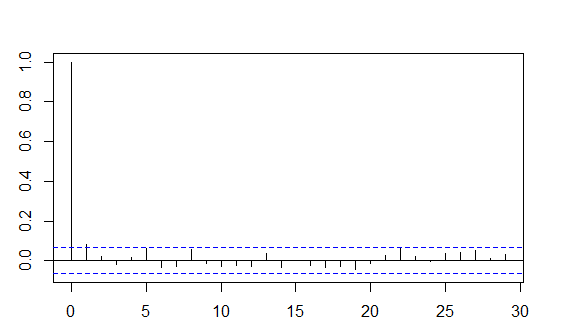}
\end{subfigure}

\begin{subfigure}[t]{0.3\textwidth}
    \includegraphics[width=\linewidth]{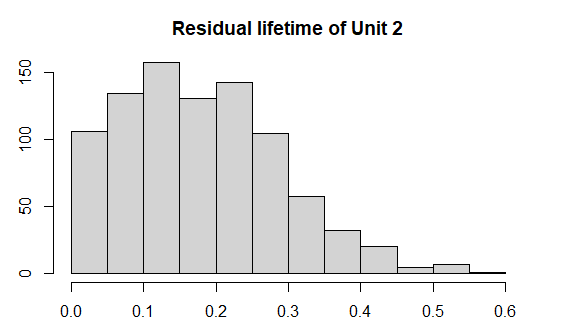}
\end{subfigure}\hfill
\begin{subfigure}[t]{0.3\textwidth}
    \includegraphics[width=\linewidth]{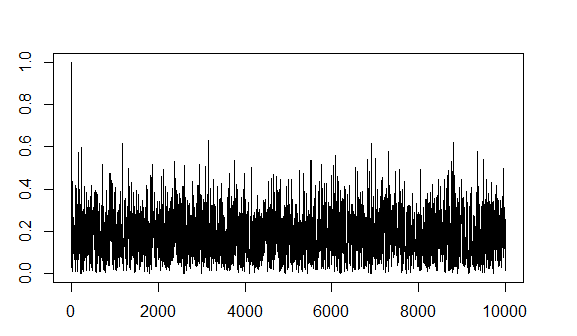}
\end{subfigure}\hfill
\begin{subfigure}[t]{0.3\textwidth}
    \includegraphics[width=\textwidth]{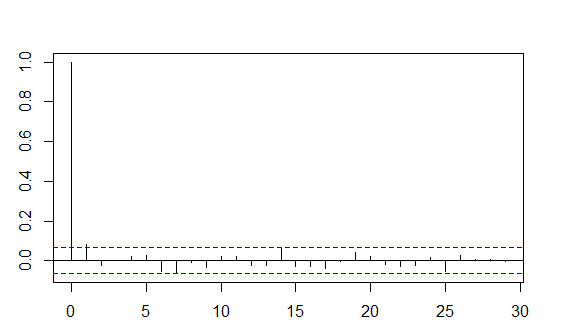}
\end{subfigure}

\begin{subfigure}[t]{0.3\textwidth}
    \includegraphics[width=\linewidth]{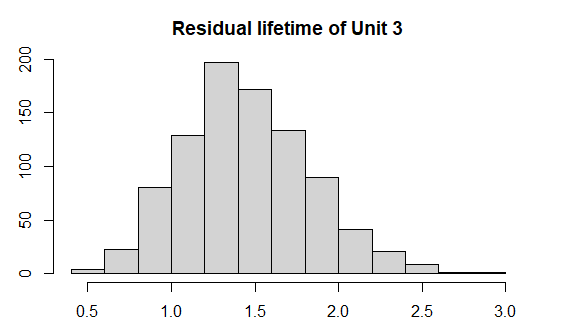}
\end{subfigure}\hfill
\begin{subfigure}[t]{0.3\textwidth}
    \includegraphics[width=\linewidth]{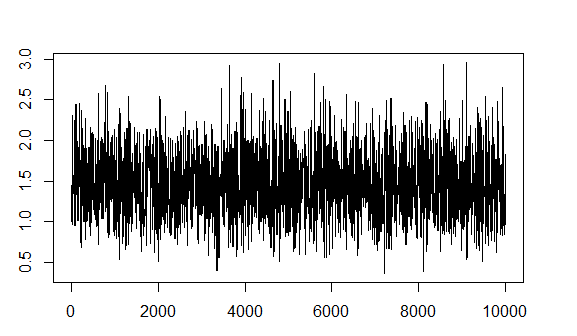}
\end{subfigure}\hfill
\begin{subfigure}[t]{0.3\textwidth}
    \includegraphics[width=\textwidth]{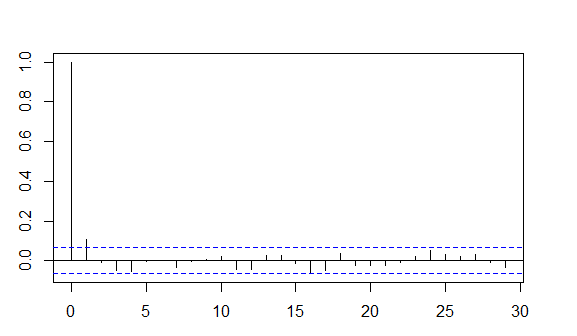}
\end{subfigure}

\caption{For each row, the left side plot is for histogram for Residual lifetimes, middle one is for trace plot and right side denotes autocorrelation plot of the samples produced for Unit 1-3 and $n=10$, $m=31$ by semi-parametric method.}
\end{figure}


\begin{figure}[ht!]
\begin{subfigure}[t]{0.3\textwidth}
    \includegraphics[width=\linewidth]{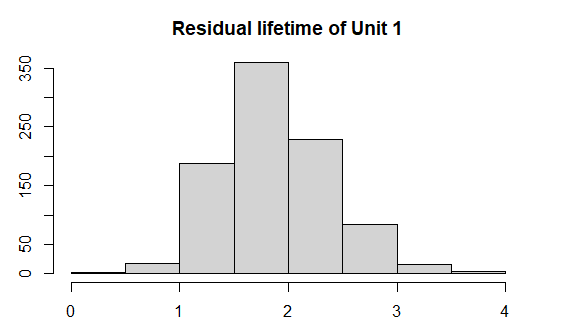}
\end{subfigure}\hfill
\begin{subfigure}[t]{0.3\textwidth}
  \includegraphics[width=\linewidth]{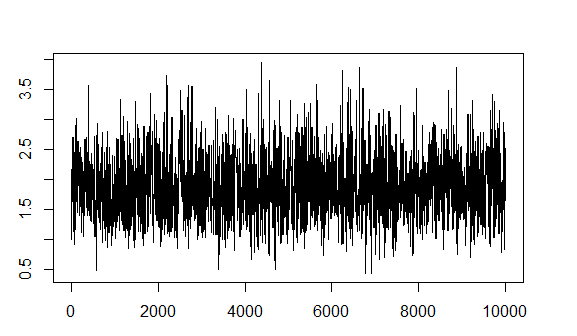}
\end{subfigure}\hfill
\begin{subfigure}[t]{0.3\textwidth}
    \includegraphics[width=\linewidth]{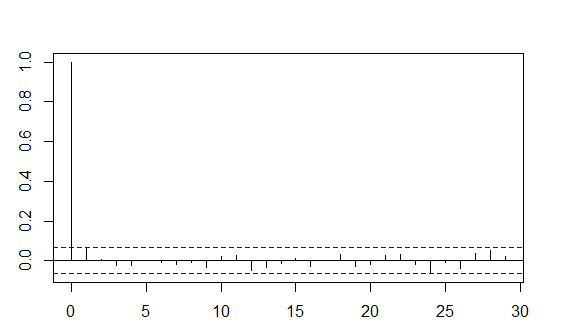}
\end{subfigure}

\begin{subfigure}[t]{0.3\textwidth}
    \includegraphics[width=\linewidth]{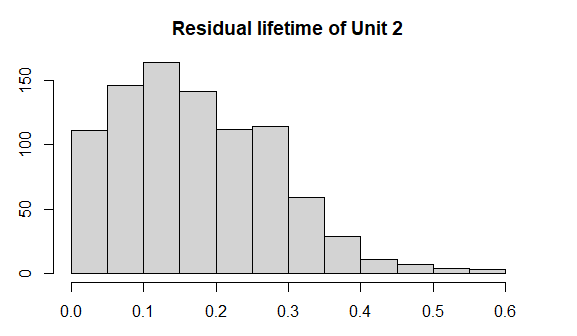}
\end{subfigure}\hfill
\begin{subfigure}[t]{0.3\textwidth}
    \includegraphics[width=\linewidth]{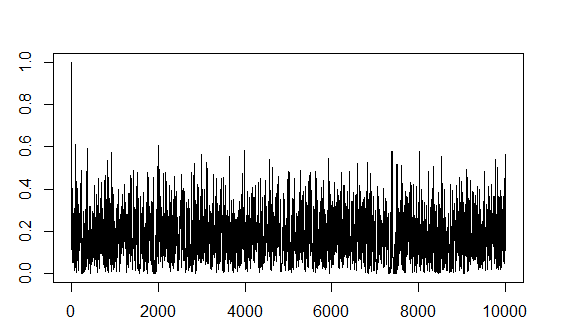}
\end{subfigure}\hfill
\begin{subfigure}[t]{0.3\textwidth}
    \includegraphics[width=\textwidth]{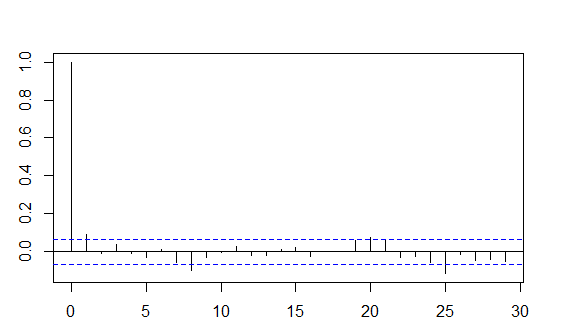}
\end{subfigure}

\begin{subfigure}[t]{0.3\textwidth}
    \includegraphics[width=\linewidth]{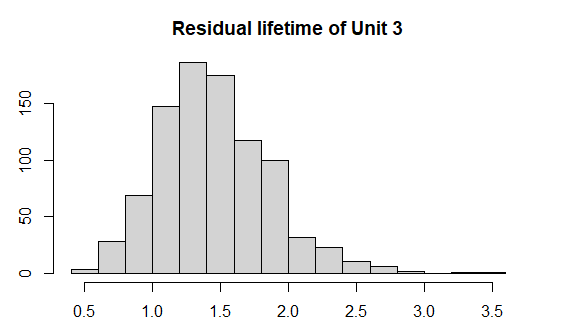}
\end{subfigure}\hfill
\begin{subfigure}[t]{0.3\textwidth}
    \includegraphics[width=\linewidth]{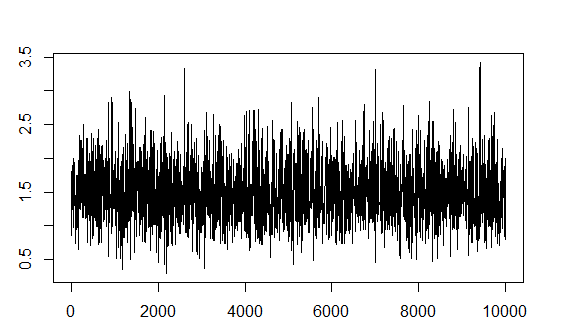}
\end{subfigure}\hfill
\begin{subfigure}[t]{0.3\textwidth}
    \includegraphics[width=\textwidth]{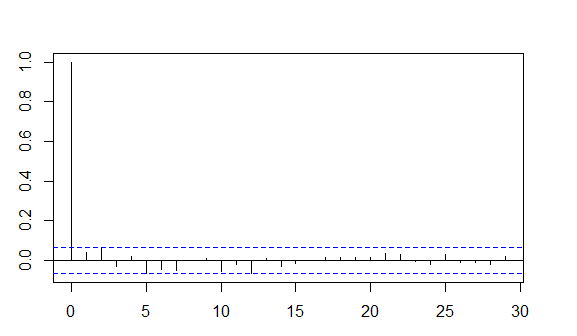}
\end{subfigure}

\caption{For each row, the left side plot is for histogram for Residual lifetimes, middle one is for trace plot and right side denotes autocorrelation plot of the samples produced for Unit 1-3 and $n=10$, $m=31$ by parametric method.}

\end{figure}

\newpage 

\subsection*{Case 4}

\begin{figure}[ht!]
\begin{subfigure}[t]{0.3\textwidth}
    \includegraphics[width=\linewidth]{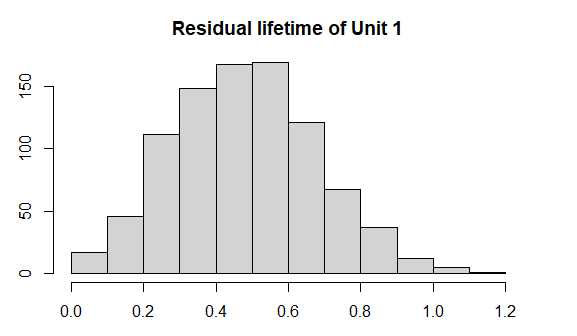}
\end{subfigure}\hfill
\begin{subfigure}[t]{0.3\textwidth}
  \includegraphics[width=\linewidth]{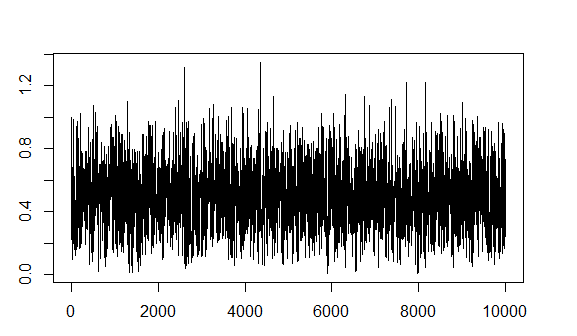}
\end{subfigure}\hfill
\begin{subfigure}[t]{0.3\textwidth}
    \includegraphics[width=\linewidth]{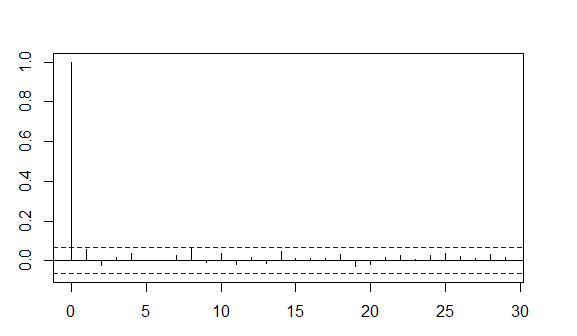}
\end{subfigure}

\begin{subfigure}[t]{0.3\textwidth}
    \includegraphics[width=\linewidth]{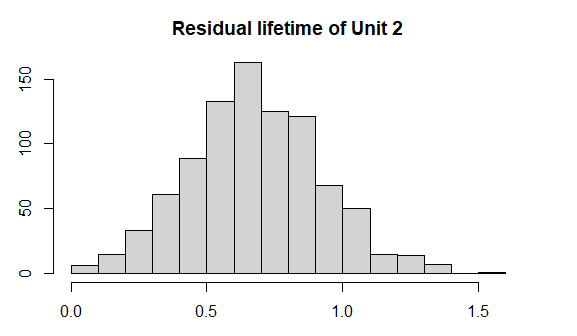}
\end{subfigure}\hfill
\begin{subfigure}[t]{0.3\textwidth}
    \includegraphics[width=\linewidth]{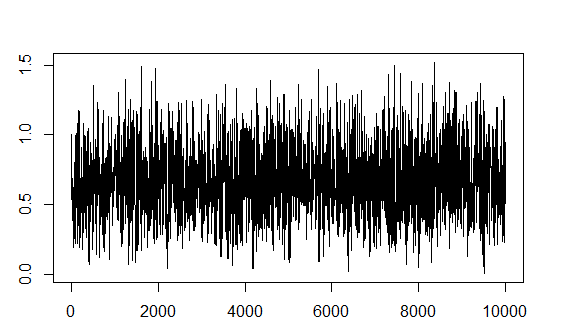}
\end{subfigure}\hfill
\begin{subfigure}[t]{0.3\textwidth}
    \includegraphics[width=\textwidth]{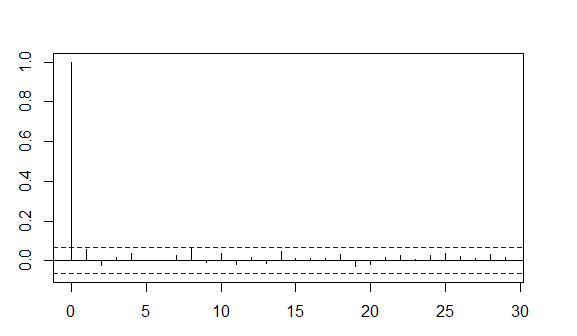}
\end{subfigure}

\begin{subfigure}[t]{0.3\textwidth}
    \includegraphics[width=\linewidth]{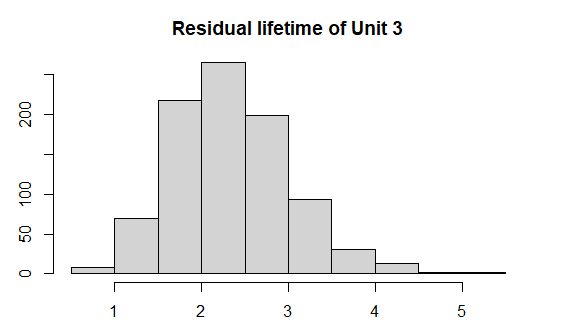}
\end{subfigure}\hfill
\begin{subfigure}[t]{0.3\textwidth}
    \includegraphics[width=\linewidth]{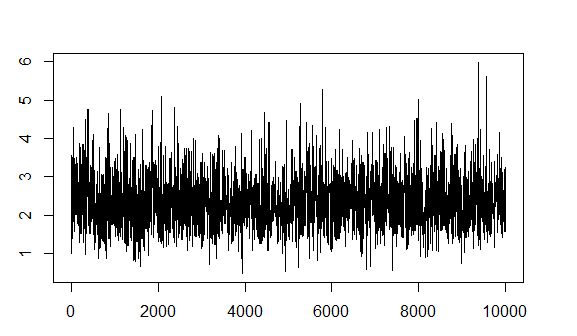}
\end{subfigure}\hfill
\begin{subfigure}[t]{0.3\textwidth}
    \includegraphics[width=\textwidth]{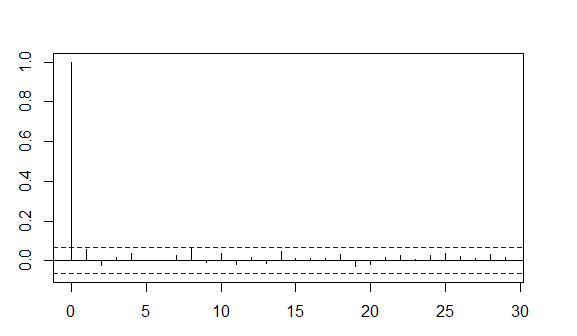}
\end{subfigure}

\caption{For each row, the left side plot is for histogram for Residual lifetimes, middle one is for trace plot and right side denotes autocorrelation plot of the samples produced for Unit 1-3 and $n=10$, $m=31$ by semi-parametric method.}
\end{figure}


\begin{figure}[ht!]
\begin{subfigure}[t]{0.3\textwidth}
    \includegraphics[width=\linewidth]{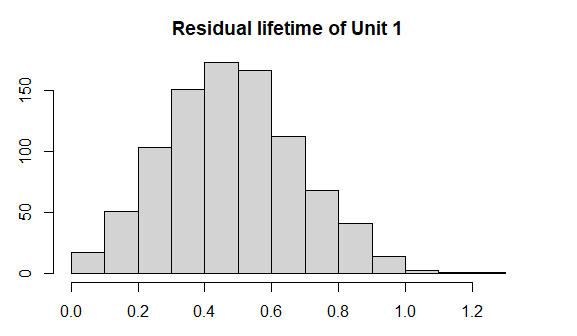}
\end{subfigure}\hfill
\begin{subfigure}[t]{0.3\textwidth}
  \includegraphics[width=\linewidth]{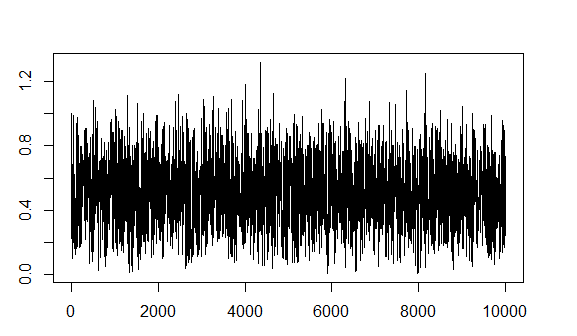}
\end{subfigure}\hfill
\begin{subfigure}[t]{0.3\textwidth}
    \includegraphics[width=\linewidth]{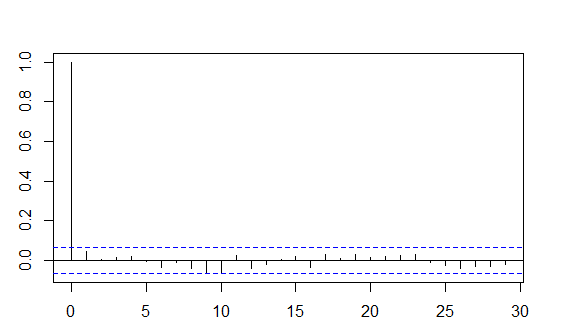}
\end{subfigure}

\begin{subfigure}[t]{0.3\textwidth}
    \includegraphics[width=\linewidth]{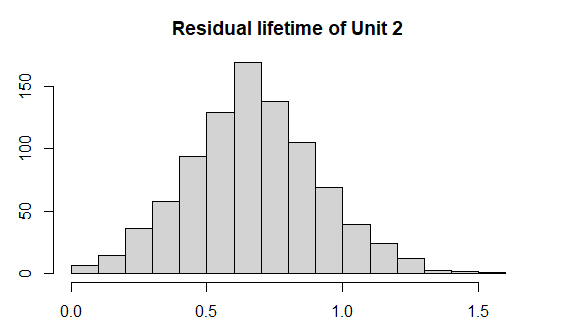}
\end{subfigure}\hfill
\begin{subfigure}[t]{0.3\textwidth}
    \includegraphics[width=\linewidth]{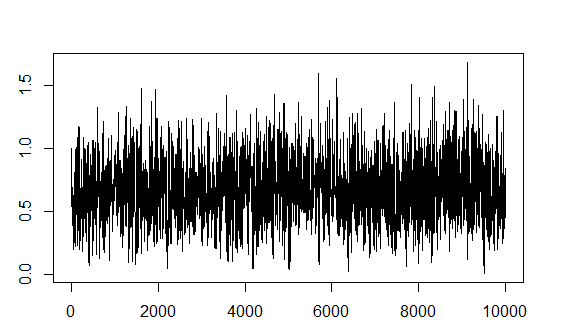}
\end{subfigure}\hfill
\begin{subfigure}[t]{0.3\textwidth}
    \includegraphics[width=\textwidth]{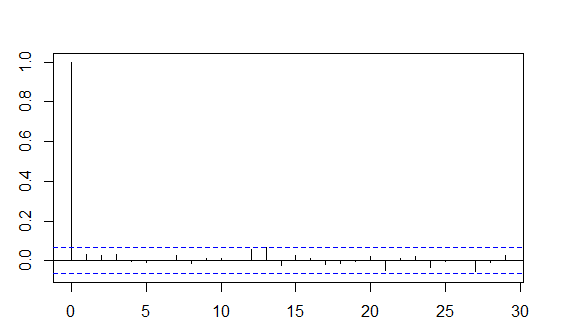}
\end{subfigure}

\begin{subfigure}[t]{0.3\textwidth}
    \includegraphics[width=\linewidth]{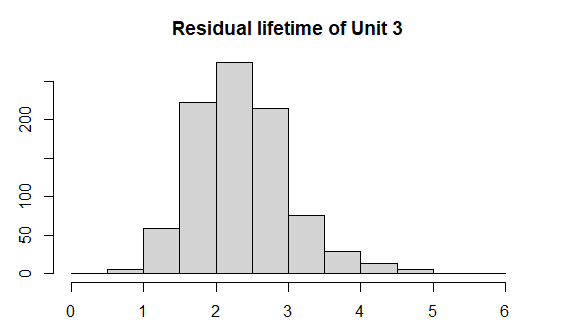}
\end{subfigure}\hfill
\begin{subfigure}[t]{0.3\textwidth}
    \includegraphics[width=\linewidth]{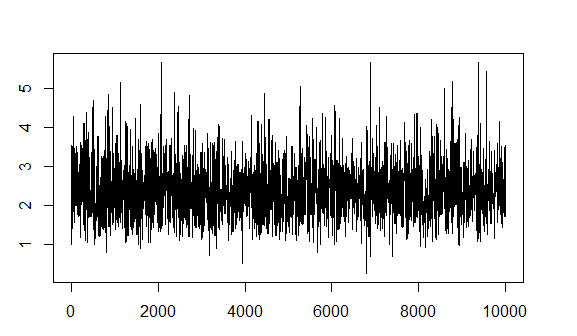}
\end{subfigure}\hfill
\begin{subfigure}[t]{0.3\textwidth}
    \includegraphics[width=\textwidth]{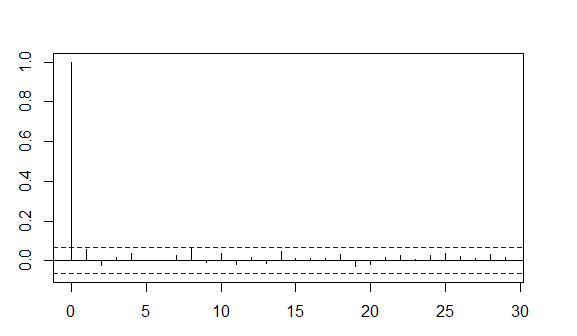}
\end{subfigure}

\caption{For each row, the left side plot is for histogram for Residual lifetimes, middle one is for trace plot and right side denotes autocorrelation plot of the samples produced for Unit 1-3 and $n=10$, $m=31$ by parametric method.}

\end{figure}

\newpage 

\subsection*{Case 5}

\begin{figure}[ht!]
\begin{subfigure}[t]{0.3\textwidth}
    \includegraphics[width=\linewidth]{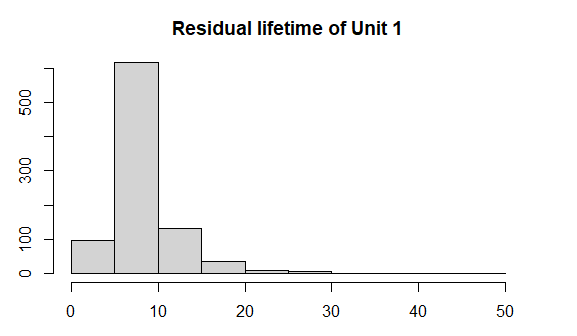}
\end{subfigure}\hfill
\begin{subfigure}[t]{0.3\textwidth}
  \includegraphics[width=\linewidth]{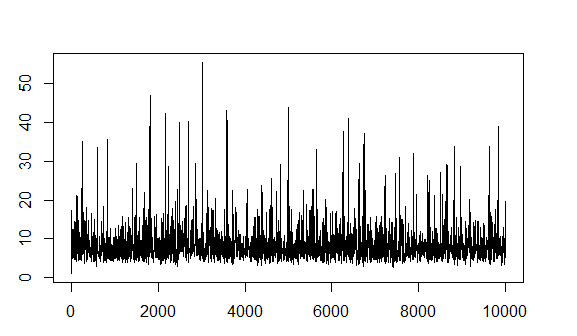}
\end{subfigure}\hfill
\begin{subfigure}[t]{0.3\textwidth}
    \includegraphics[width=\linewidth]{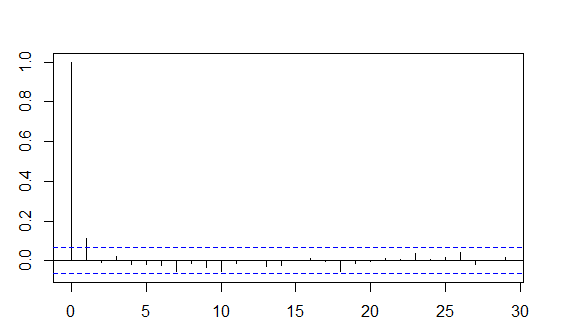}
\end{subfigure}

\begin{subfigure}[t]{0.3\textwidth}
    \includegraphics[width=\linewidth]{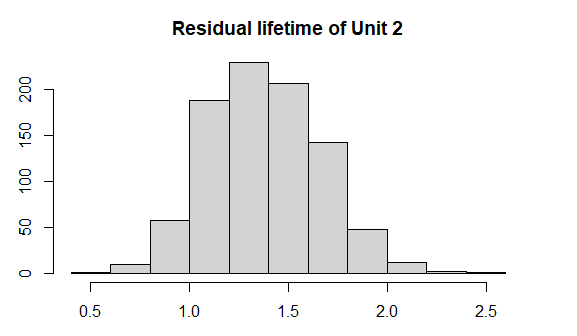}
\end{subfigure}\hfill
\begin{subfigure}[t]{0.3\textwidth}
    \includegraphics[width=\linewidth]{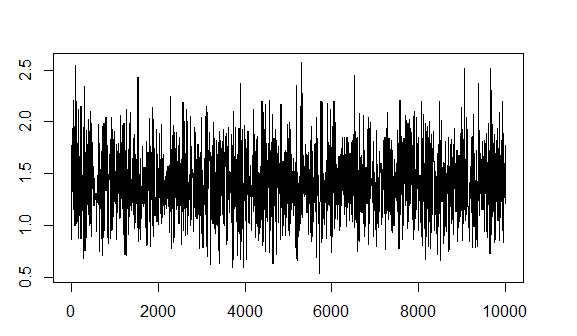}
\end{subfigure}\hfill
\begin{subfigure}[t]{0.3\textwidth}
    \includegraphics[width=\textwidth]{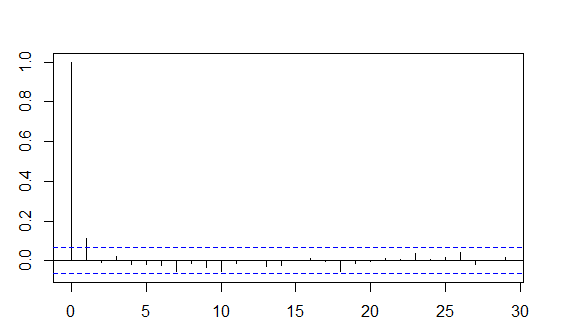}
\end{subfigure}

\begin{subfigure}[t]{0.3\textwidth}
    \includegraphics[width=\linewidth]{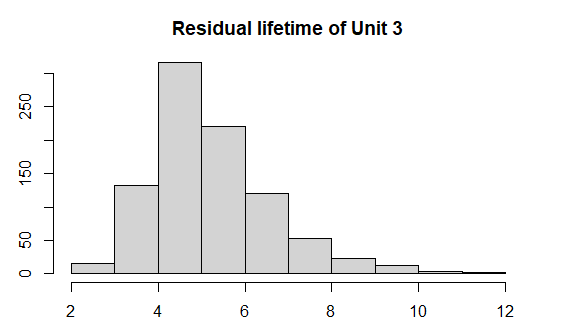}
\end{subfigure}\hfill
\begin{subfigure}[t]{0.3\textwidth}
    \includegraphics[width=\linewidth]{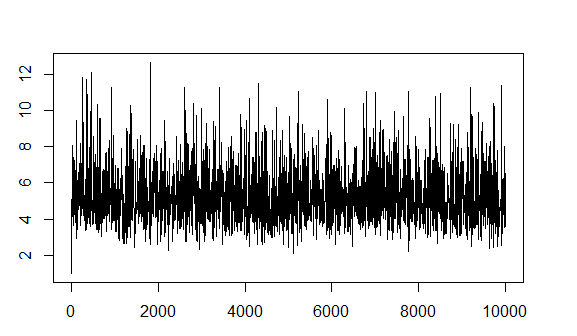}
\end{subfigure}\hfill
\begin{subfigure}[t]{0.3\textwidth}
    \includegraphics[width=\textwidth]{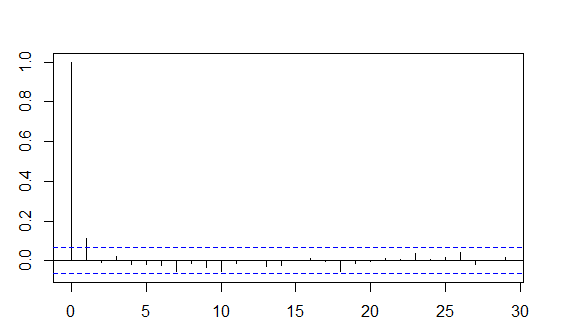}
\end{subfigure}

\caption{For each row, the left side plot is for histogram for Residual lifetimes, middle one is for trace plot and right side denotes autocorrelation plot of the samples produced for Unit 1-3 and $n=10$, $m=31$ by semi-parametric method.}
\end{figure}


\begin{figure}[ht!]
\begin{subfigure}[t]{0.3\textwidth}
    \includegraphics[width=\linewidth]{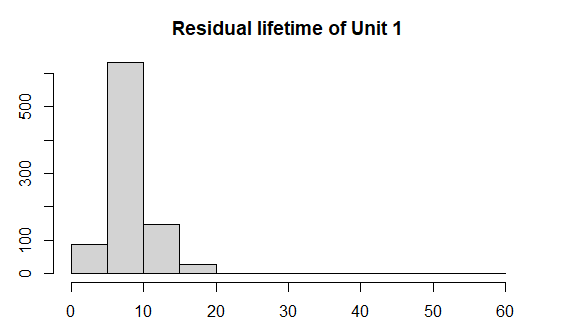}
\end{subfigure}\hfill
\begin{subfigure}[t]{0.3\textwidth}
  \includegraphics[width=\linewidth]{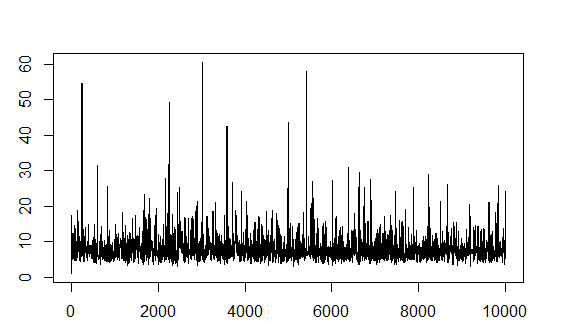}
\end{subfigure}\hfill
\begin{subfigure}[t]{0.3\textwidth}
    \includegraphics[width=\linewidth]{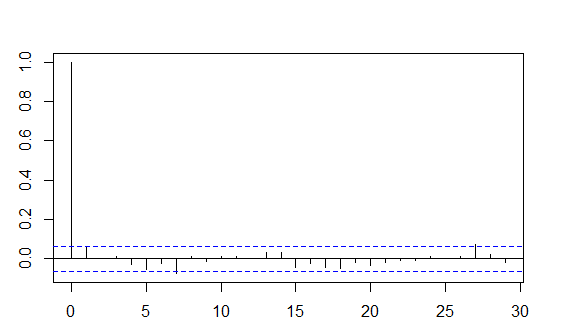}
\end{subfigure}

\begin{subfigure}[t]{0.3\textwidth}
    \includegraphics[width=\linewidth]{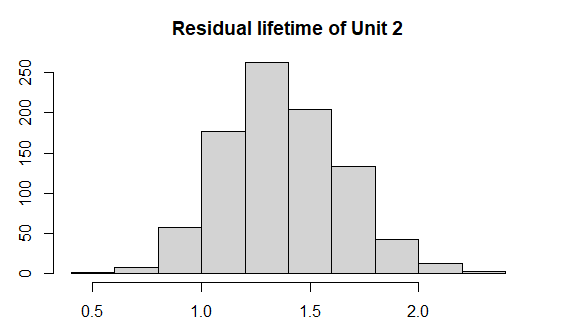}
\end{subfigure}\hfill
\begin{subfigure}[t]{0.3\textwidth}
    \includegraphics[width=\linewidth]{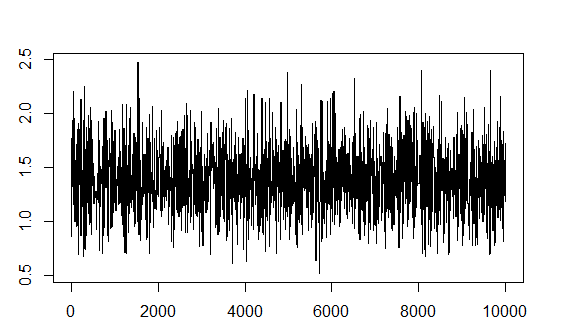}
\end{subfigure}\hfill
\begin{subfigure}[t]{0.3\textwidth}
    \includegraphics[width=\textwidth]{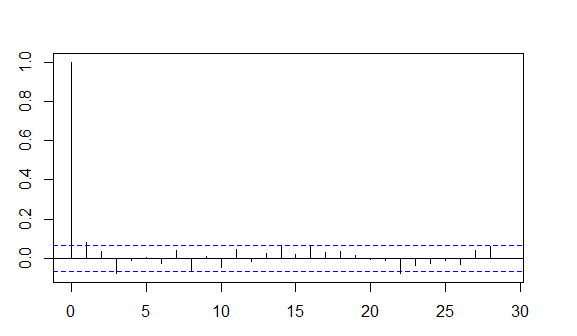}
\end{subfigure}

\begin{subfigure}[t]{0.3\textwidth}
    \includegraphics[width=\linewidth]{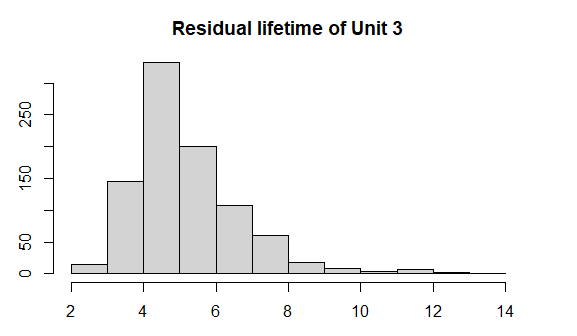}
\end{subfigure}\hfill
\begin{subfigure}[t]{0.3\textwidth}
    \includegraphics[width=\linewidth]{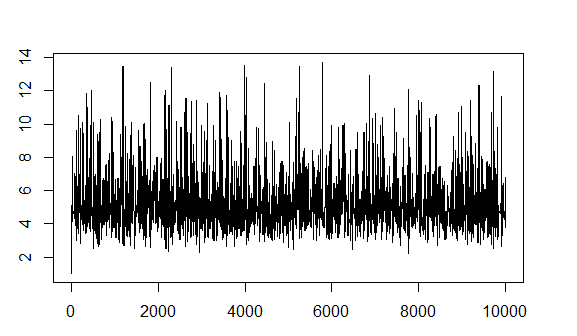}
\end{subfigure}\hfill
\begin{subfigure}[t]{0.3\textwidth}
    \includegraphics[width=\textwidth]{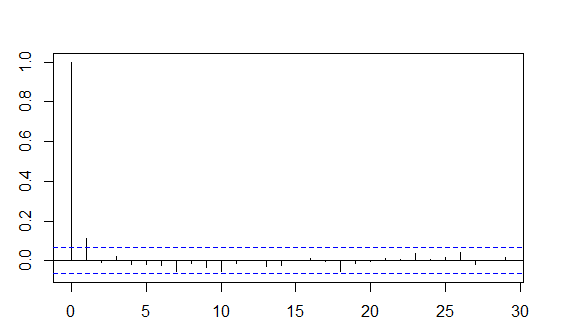}
\end{subfigure}

\caption{For each row, the left side plot is for histogram for Residual lifetimes, middle one is for trace plot and right side denotes autocorrelation plot of the samples produced for Unit 1-3 and $n=10$, $m=31$ by parametric method.}

\end{figure}

\newpage 

\subsection*{Fatigue-Crack Size dataset}

\begin{figure}[ht!]
\begin{subfigure}[t]{0.3\textwidth}
    \includegraphics[width=\linewidth]{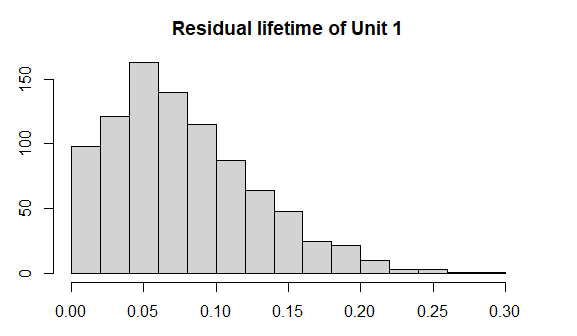}
\end{subfigure}\hfill
\begin{subfigure}[t]{0.3\textwidth}
  \includegraphics[width=\linewidth]{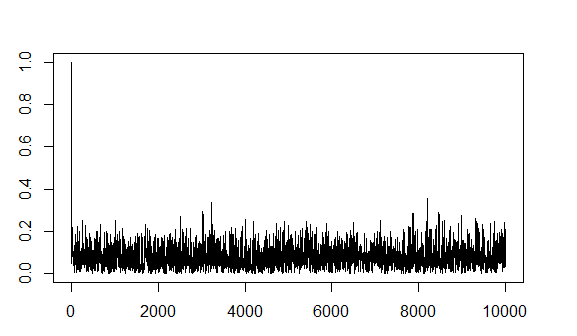}
\end{subfigure}\hfill
\begin{subfigure}[t]{0.3\textwidth}
    \includegraphics[width=\linewidth]{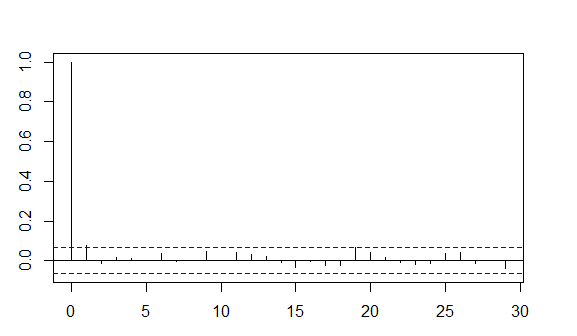}
\end{subfigure}

\begin{subfigure}[t]{0.3\textwidth}
    \includegraphics[width=\linewidth]{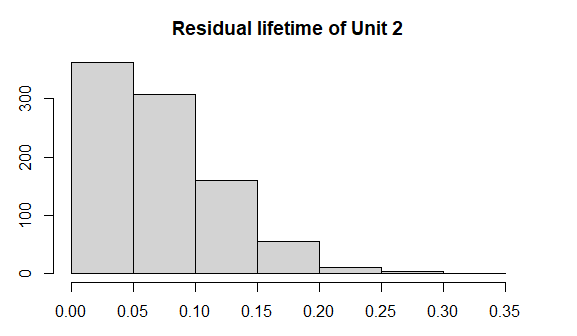}
\end{subfigure}\hfill
\begin{subfigure}[t]{0.3\textwidth}
    \includegraphics[width=\linewidth]{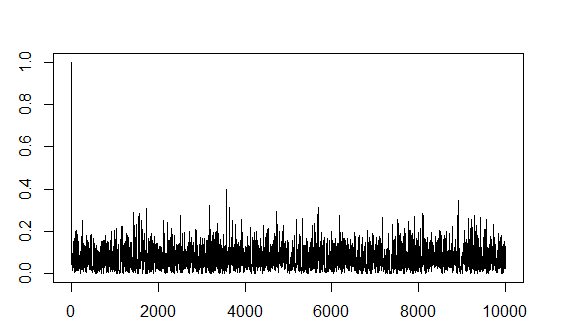}
\end{subfigure}\hfill
\begin{subfigure}[t]{0.3\textwidth}
    \includegraphics[width=\textwidth]{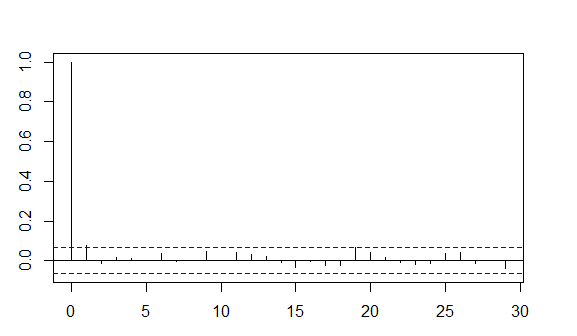}
\end{subfigure}

\begin{subfigure}[t]{0.3\textwidth}
    \includegraphics[width=\linewidth]{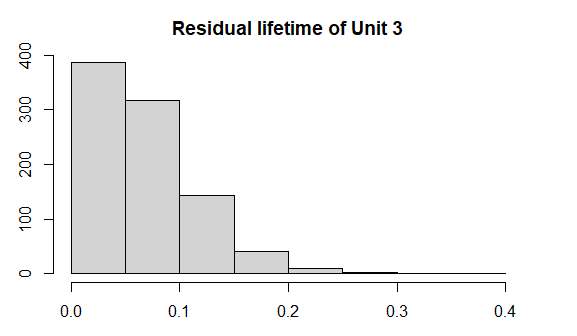}
\end{subfigure}\hfill
\begin{subfigure}[t]{0.3\textwidth}
    \includegraphics[width=\linewidth]{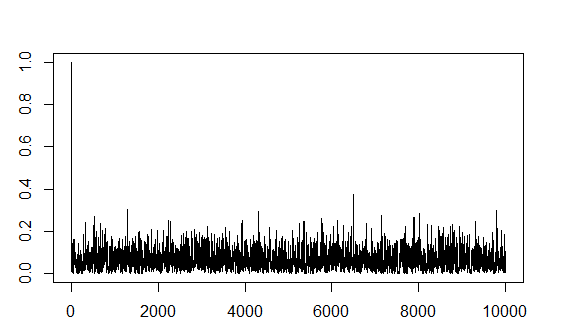}
\end{subfigure}\hfill
\begin{subfigure}[t]{0.3\textwidth}
    \includegraphics[width=\textwidth]{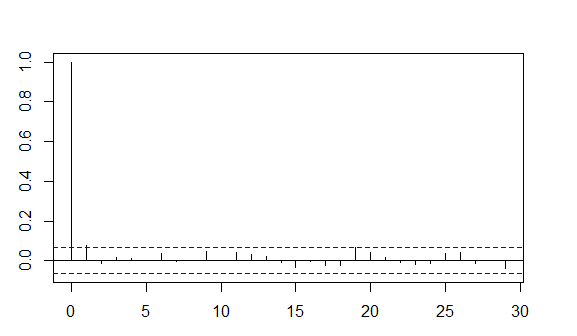}
\end{subfigure}

\caption{For each row, the left side plot is for histogram for Residual lifetimes, middle one is for trace plot and right side denotes autocorrelation plot of the samples produced for Unit 1-3 and $n=10$, $m=31$ by semi-parametric method.}
\end{figure}


\begin{figure}[ht!]
\begin{subfigure}[t]{0.3\textwidth}
    \includegraphics[width=\linewidth]{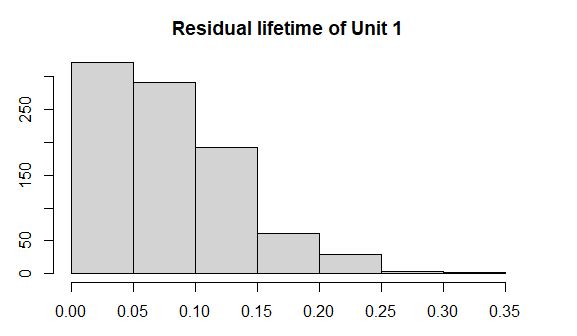}
\end{subfigure}\hfill
\begin{subfigure}[t]{0.3\textwidth}
  \includegraphics[width=\linewidth]{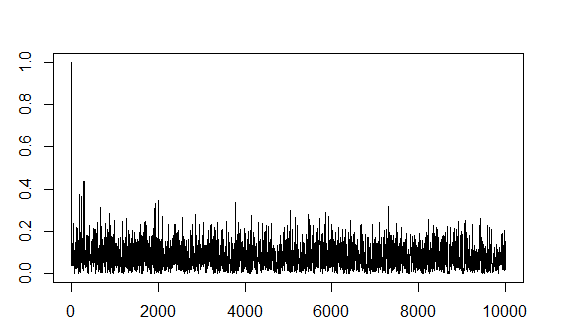}
\end{subfigure}\hfill
\begin{subfigure}[t]{0.3\textwidth}
    \includegraphics[width=\linewidth]{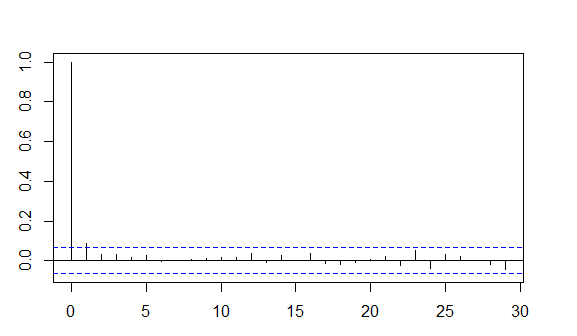}
\end{subfigure}

\begin{subfigure}[t]{0.3\textwidth}
    \includegraphics[width=\linewidth]{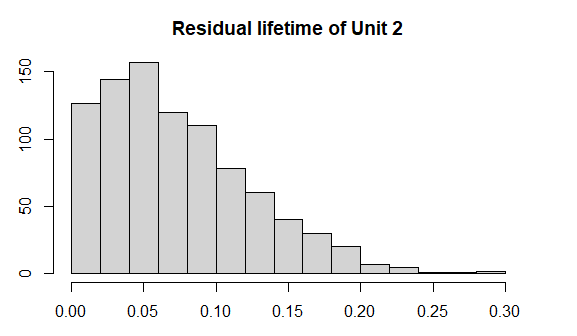}
\end{subfigure}\hfill
\begin{subfigure}[t]{0.3\textwidth}
    \includegraphics[width=\linewidth]{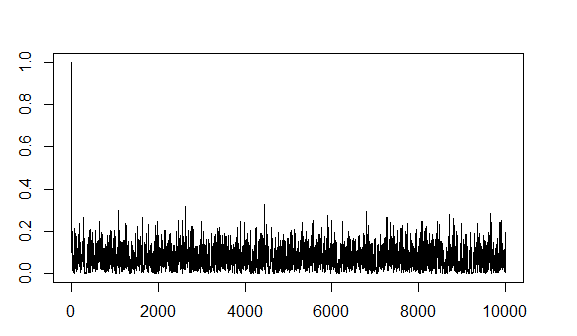}
\end{subfigure}\hfill
\begin{subfigure}[t]{0.3\textwidth}
    \includegraphics[width=\textwidth]{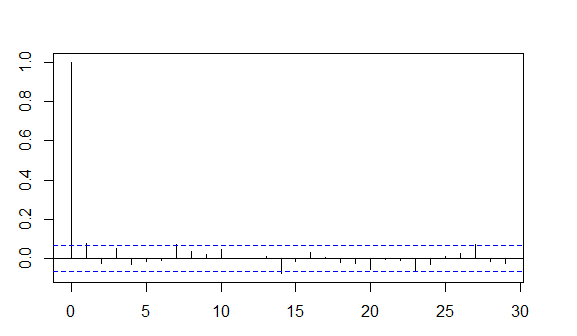}
\end{subfigure}

\begin{subfigure}[t]{0.3\textwidth}
    \includegraphics[width=\linewidth]{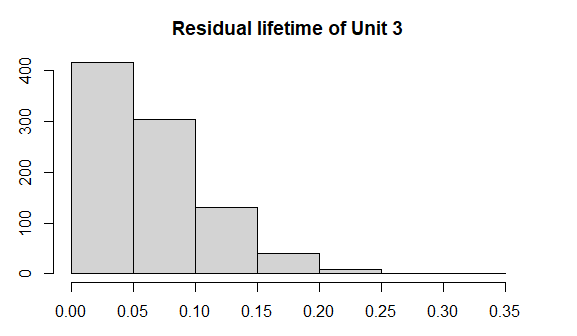}
\end{subfigure}\hfill
\begin{subfigure}[t]{0.3\textwidth}
    \includegraphics[width=\linewidth]{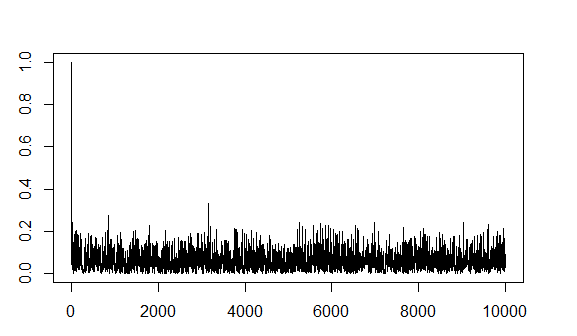}
\end{subfigure}\hfill
\begin{subfigure}[t]{0.3\textwidth}
    \includegraphics[width=\textwidth]{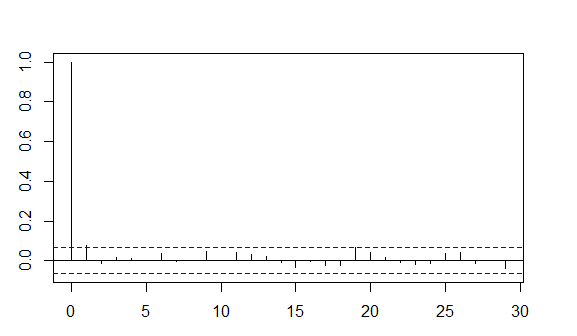}
\end{subfigure}

\caption{For each row, the left side plot is for histogram for Residual lifetimes, middle one is for trace plot and right side denotes autocorrelation plot of the samples produced for Unit 1-3 and $n=10$, $m=31$ by parametric method.}

\end{figure}

\section*{APPENDIX B : MCMC convergence for model parameters}
\subsection*{Case 1}

\begin{figure}[htbp]
\begin{subfigure}[t]{0.22\textwidth}
    \includegraphics[width=\linewidth]{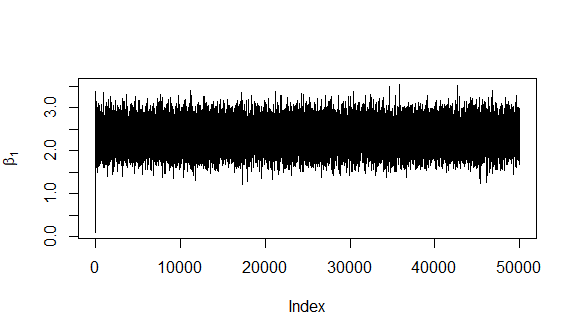}
    \end{subfigure}\hfill
\begin{subfigure}[t]{0.22\textwidth}
    \includegraphics[width=\linewidth]{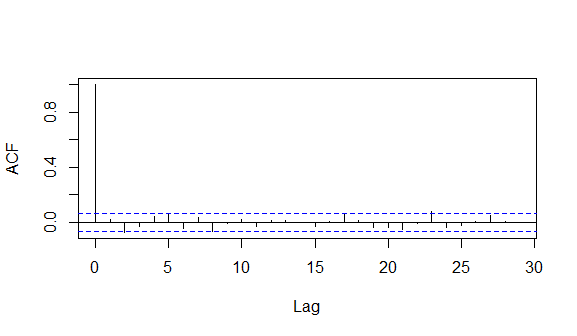}
\end{subfigure}\hfill
\begin{subfigure}[t]{0.22\textwidth}
    \includegraphics[width=\linewidth]{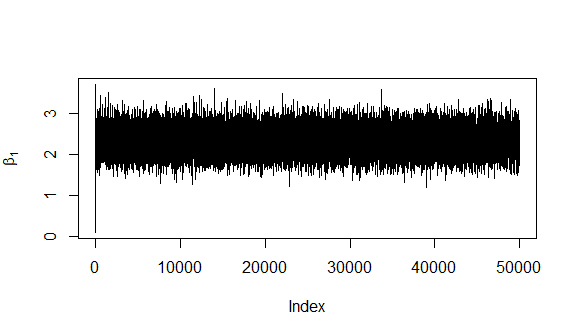}
    \end{subfigure}\hfill
\begin{subfigure}[t]{0.22\textwidth}
    \includegraphics[width=\linewidth]{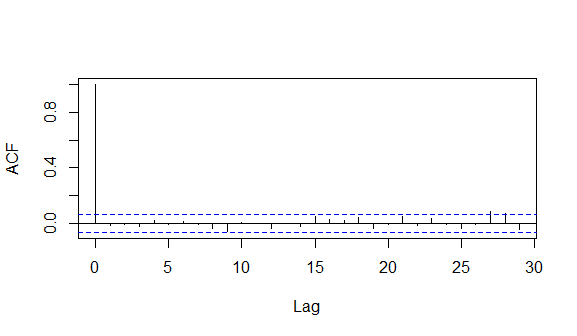}
\end{subfigure}\hfill
\begin{subfigure}[t]{0.22\textwidth}
    \includegraphics[width=\linewidth]{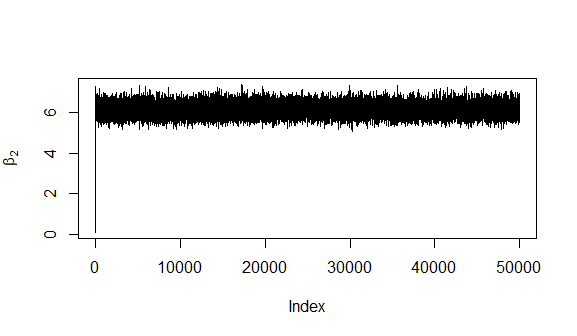}
    \end{subfigure}\hfill
\begin{subfigure}[t]{0.22\textwidth}
    \includegraphics[width=\linewidth]{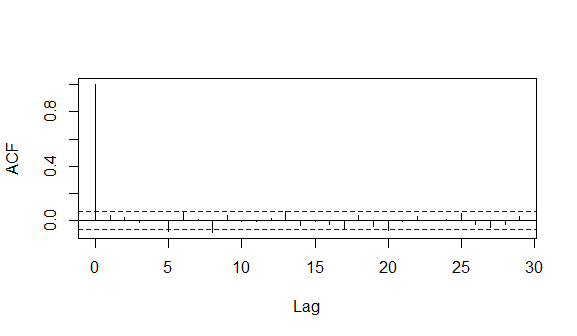}
\end{subfigure}\hfill
\begin{subfigure}[t]{0.22\textwidth}
    \includegraphics[width=\linewidth]{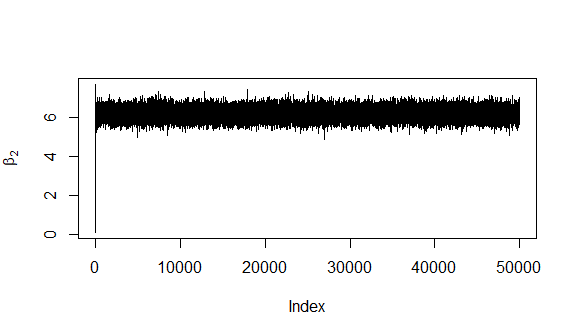}
    \end{subfigure}\hfill
\begin{subfigure}[t]{0.22\textwidth}
    \includegraphics[width=\linewidth]{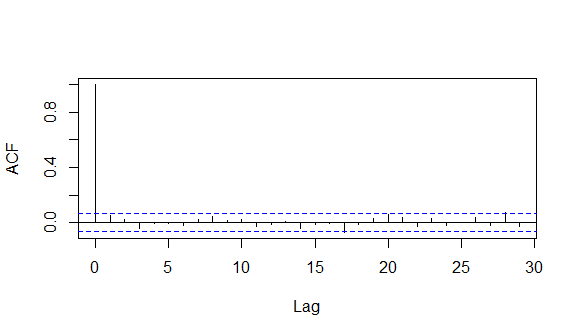}
\end{subfigure}\hfill
\begin{subfigure}[t]{0.22\textwidth}
    \includegraphics[width=\linewidth]{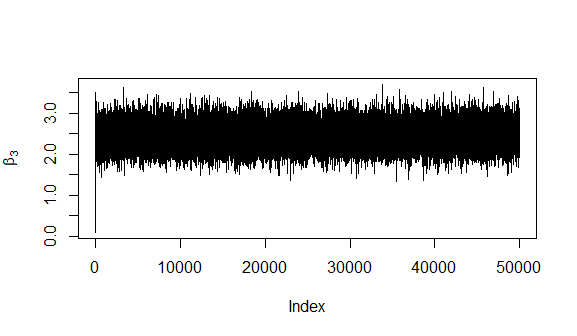}
    \end{subfigure}\hfill
\begin{subfigure}[t]{0.22\textwidth}
    \includegraphics[width=\linewidth]{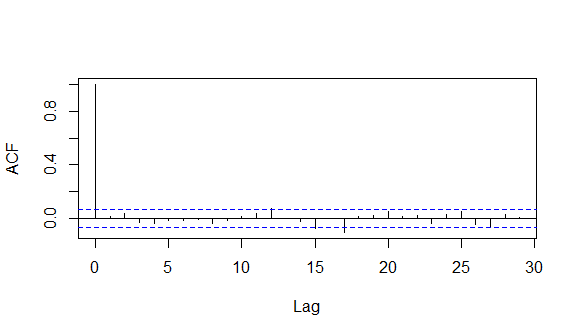}
\end{subfigure}\hfill
\begin{subfigure}[t]{0.22\textwidth}
    \includegraphics[width=\linewidth]{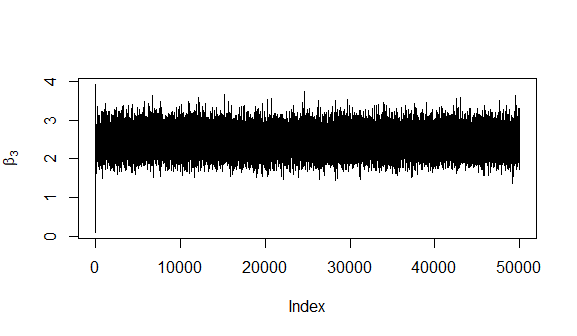}
    \end{subfigure}\hfill
\begin{subfigure}[t]{0.22\textwidth}
    \includegraphics[width=\linewidth]{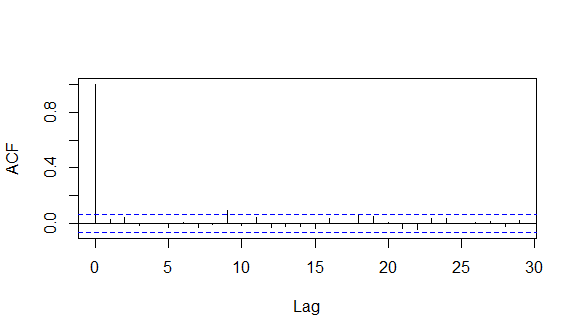}
\end{subfigure}\hfill
\begin{subfigure}[t]{0.22\textwidth}
    \includegraphics[width=\linewidth]{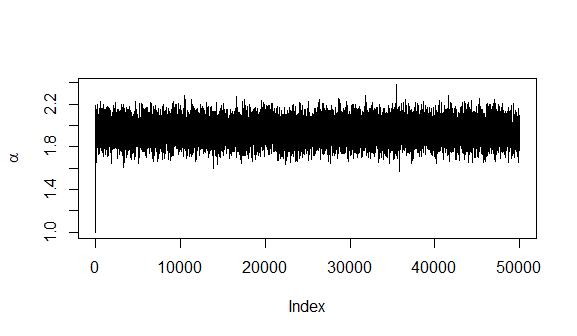}
    \end{subfigure}\hfill
\begin{subfigure}[t]{0.22\textwidth}
    \includegraphics[width=\linewidth]{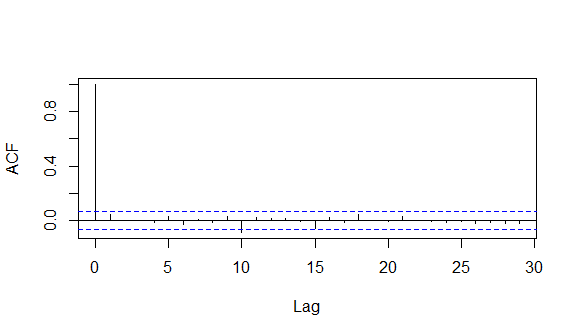}
\end{subfigure}\hfill
\begin{subfigure}[t]{0.22\textwidth}
    \includegraphics[width=\linewidth]{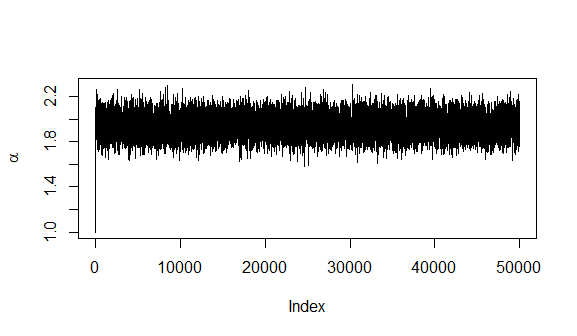}
    \end{subfigure}\hfill
\begin{subfigure}[t]{0.22\textwidth}
    \includegraphics[width=\linewidth]{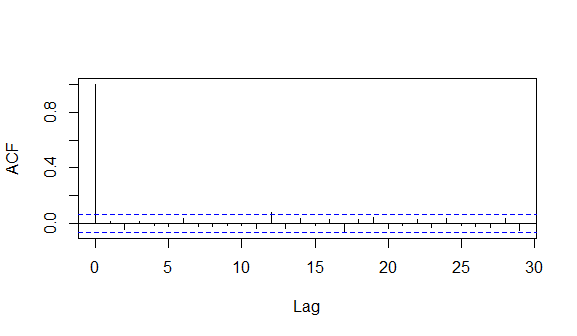}
\end{subfigure}\hfill
\begin{subfigure}[t]{0.22\textwidth}
    \includegraphics[width=\linewidth]{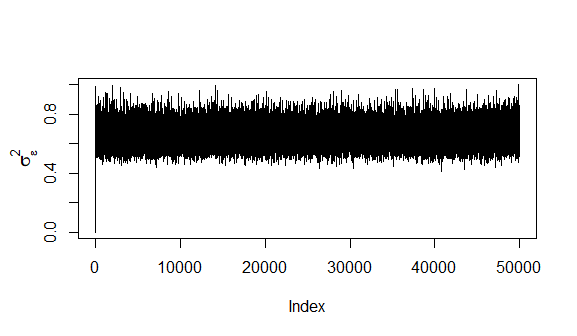}
    \end{subfigure}\hfill
\begin{subfigure}[t]{0.22\textwidth}
    \includegraphics[width=\linewidth]{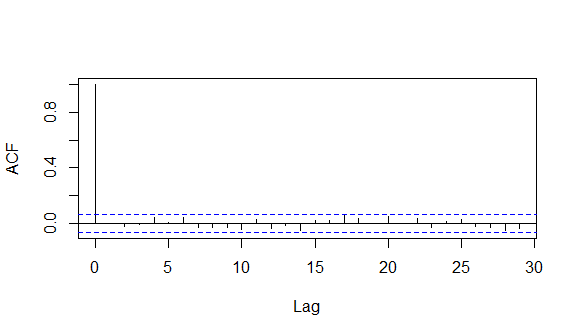}
\end{subfigure}\hfill
\begin{subfigure}[t]{0.22\textwidth}
    \includegraphics[width=\linewidth]{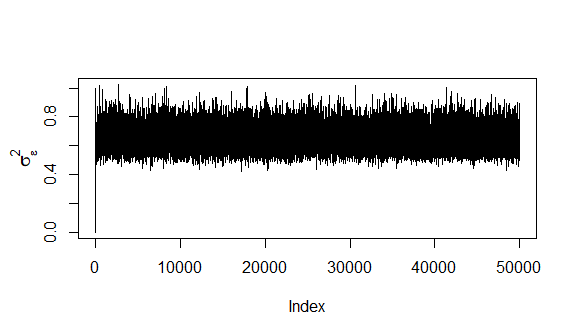}
    \end{subfigure}\hfill
\begin{subfigure}[t]{0.22\textwidth}
    \includegraphics[width=\linewidth]{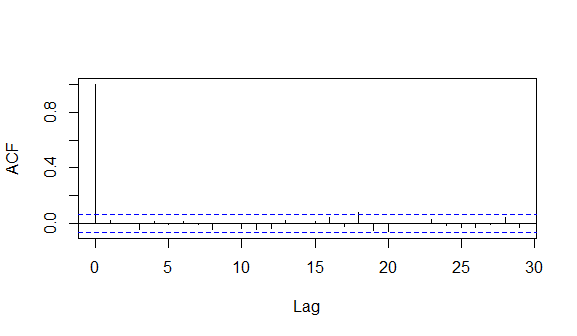}
\end{subfigure}
\caption{For each row, first two plots from the left side is trace plot and autocorrelation plot produced by semi-parametric method and last two plots denotes trace plot and autocorrelation plot produced by parametric method based on the generated samples for $\beta_i$, $i=1,2,3$, $\alpha$ and $\sigma_\epsilon^2$ respectively. Autocorrelation plots are constructed based on the samples at lag size 50.}
\end{figure}

\newpage 

\subsection*{Case 2}

\begin{figure}[htbp]
\begin{subfigure}[t]{0.22\textwidth}
    \includegraphics[width=\linewidth]{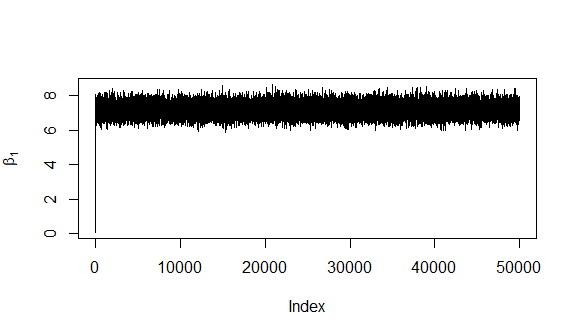}
    \end{subfigure}\hfill
\begin{subfigure}[t]{0.22\textwidth}
    \includegraphics[width=\linewidth]{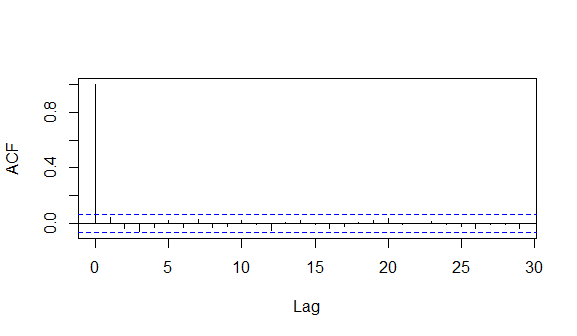}
\end{subfigure}\hfill
\begin{subfigure}[t]{0.22\textwidth}
    \includegraphics[width=\linewidth]{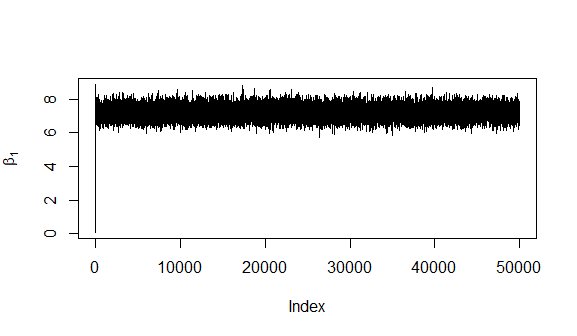}
    \end{subfigure}\hfill
\begin{subfigure}[t]{0.22\textwidth}
    \includegraphics[width=\linewidth]{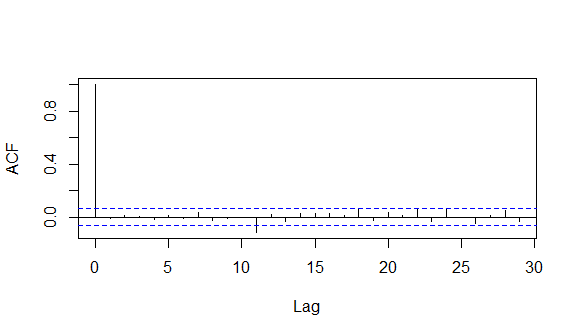}
\end{subfigure}\hfill
\begin{subfigure}[t]{0.22\textwidth}
    \includegraphics[width=\linewidth]{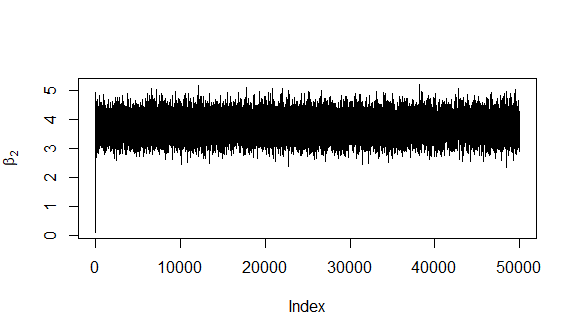}
    \end{subfigure}\hfill
\begin{subfigure}[t]{0.22\textwidth}
    \includegraphics[width=\linewidth]{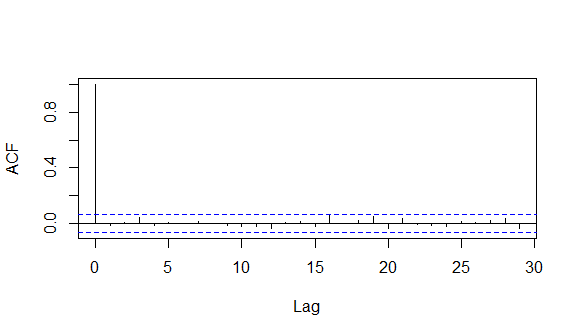}
\end{subfigure}\hfill
\begin{subfigure}[t]{0.22\textwidth}
    \includegraphics[width=\linewidth]{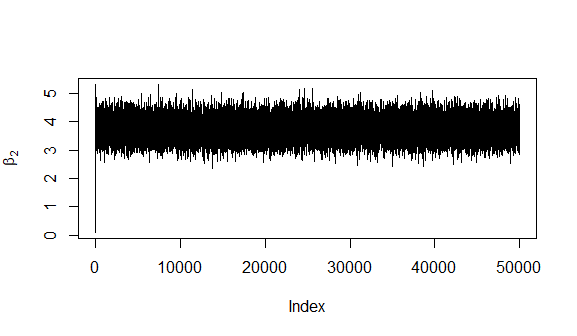}
    \end{subfigure}\hfill
\begin{subfigure}[t]{0.22\textwidth}
    \includegraphics[width=\linewidth]{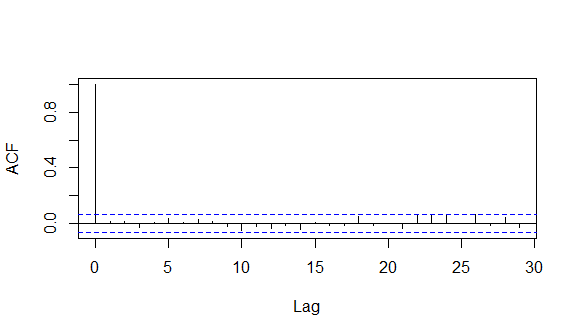}
\end{subfigure}\hfill
\begin{subfigure}[t]{0.22\textwidth}
    \includegraphics[width=\linewidth]{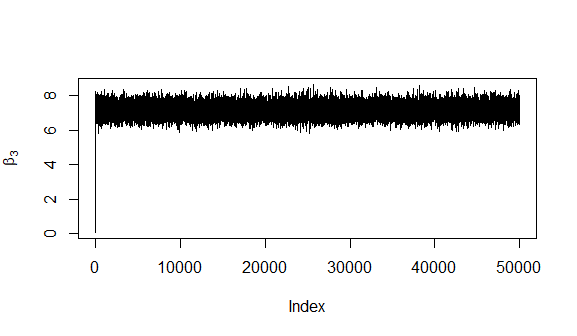}
    \end{subfigure}\hfill
\begin{subfigure}[t]{0.22\textwidth}
    \includegraphics[width=\linewidth]{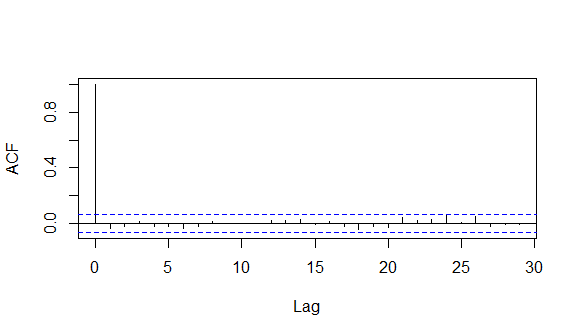}
\end{subfigure}\hfill
\begin{subfigure}[t]{0.22\textwidth}
    \includegraphics[width=\linewidth]{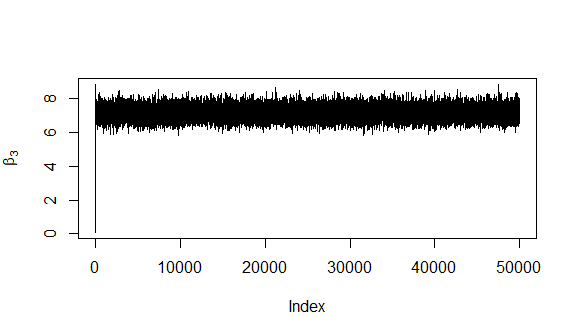}
    \end{subfigure}\hfill
\begin{subfigure}[t]{0.22\textwidth}
    \includegraphics[width=\linewidth]{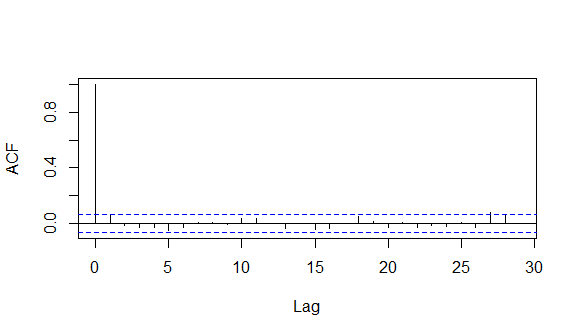}
\end{subfigure}\hfill
\begin{subfigure}[t]{0.22\textwidth}
    \includegraphics[width=\linewidth]{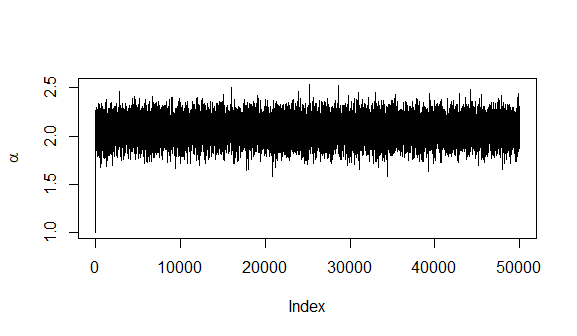}
    \end{subfigure}\hfill
\begin{subfigure}[t]{0.22\textwidth}
    \includegraphics[width=\linewidth]{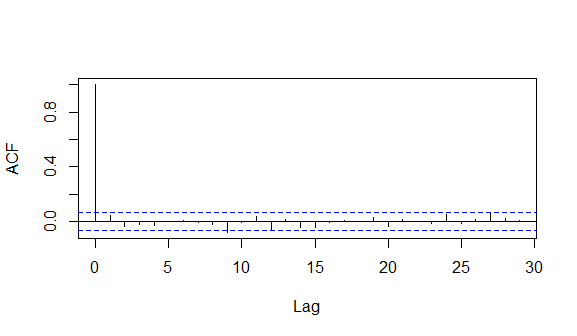}
\end{subfigure}\hfill
\begin{subfigure}[t]{0.22\textwidth}
    \includegraphics[width=\linewidth]{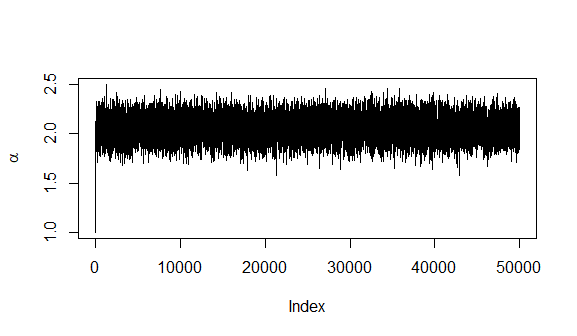}
    \end{subfigure}\hfill
\begin{subfigure}[t]{0.22\textwidth}
    \includegraphics[width=\linewidth]{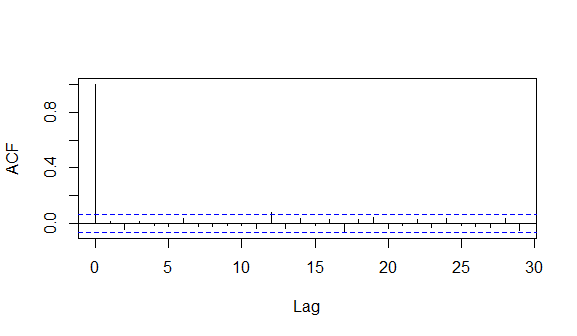}
\end{subfigure}\hfill
\begin{subfigure}[t]{0.22\textwidth}
    \includegraphics[width=\linewidth]{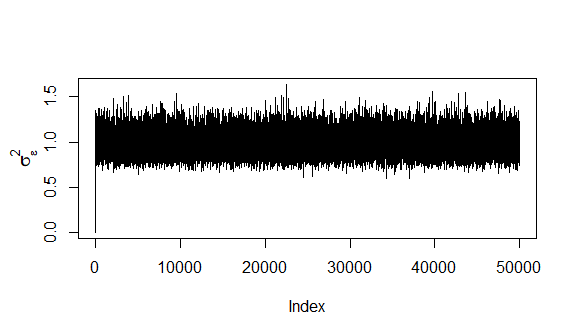}
    \end{subfigure}\hfill
\begin{subfigure}[t]{0.22\textwidth}
    \includegraphics[width=\linewidth]{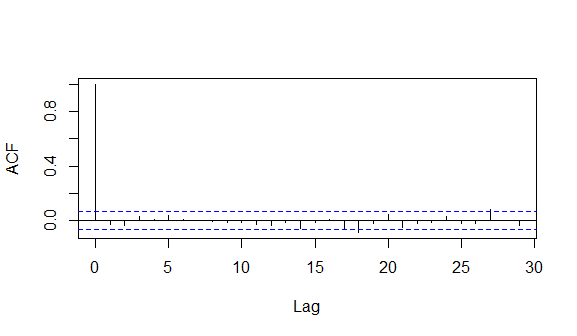}
\end{subfigure}\hfill
\begin{subfigure}[t]{0.22\textwidth}
    \includegraphics[width=\linewidth]{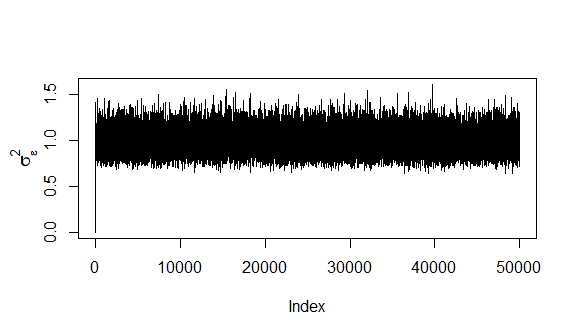}
    \end{subfigure}\hfill
\begin{subfigure}[t]{0.22\textwidth}
    \includegraphics[width=\linewidth]{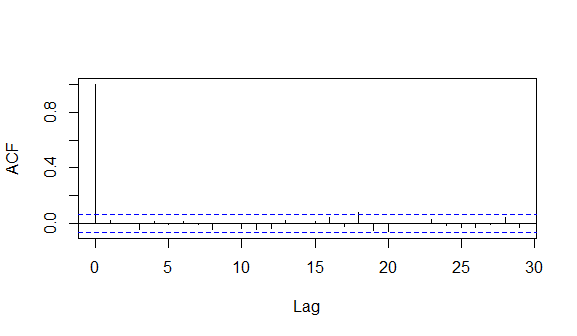}
\end{subfigure}
\caption{For each row, first two plots from the left side is trace plot and autocorrelation plot produced by semi-parametric method and last two plots denotes trace plot and autocorrelation plot produced by parametric method based on the generated samples for $\beta_i$, $i=1,2,3$, $\alpha$ and $\sigma_\epsilon^2$ respectively. Autocorrelation plots are constructed based on the samples at lag size 50.}
\end{figure}

\newpage

\subsection*{Case 3}

\begin{figure}[htbp]
\begin{subfigure}[t]{0.22\textwidth}
    \includegraphics[width=\linewidth]{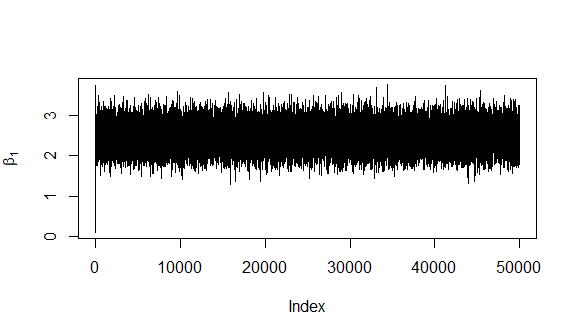}
    \end{subfigure}\hfill
\begin{subfigure}[t]{0.22\textwidth}
    \includegraphics[width=\linewidth]{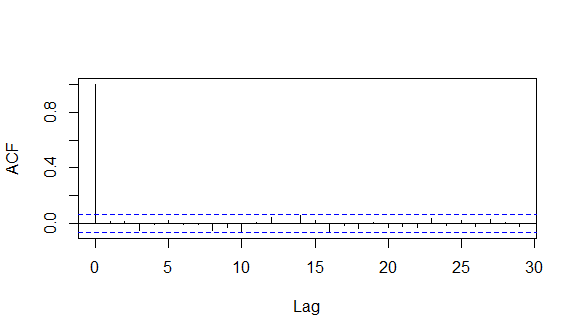}
\end{subfigure}\hfill
\begin{subfigure}[t]{0.22\textwidth}
    \includegraphics[width=\linewidth]{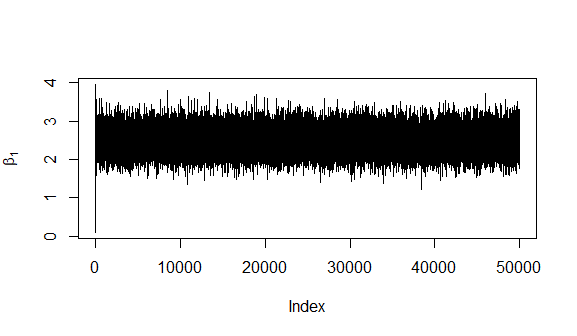}
    \end{subfigure}\hfill
\begin{subfigure}[t]{0.22\textwidth}
    \includegraphics[width=\linewidth]{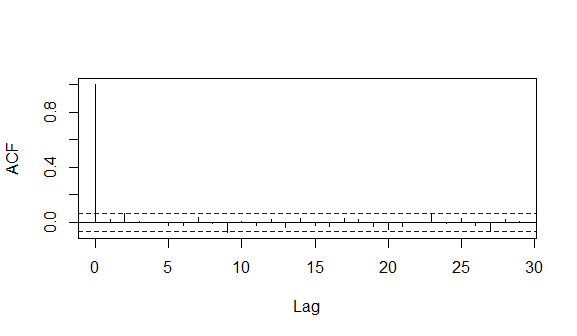}
\end{subfigure}\hfill
\begin{subfigure}[t]{0.22\textwidth}
    \includegraphics[width=\linewidth]{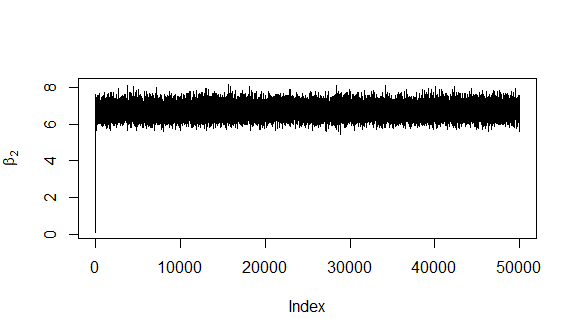}
    \end{subfigure}\hfill
\begin{subfigure}[t]{0.22\textwidth}
    \includegraphics[width=\linewidth]{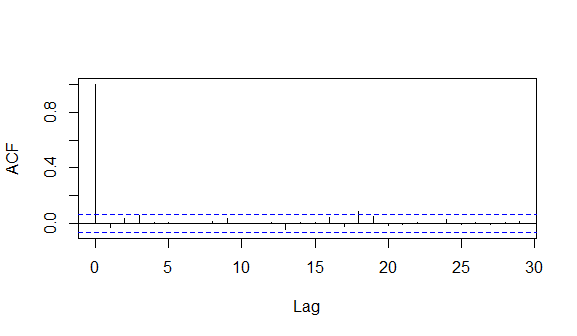}
\end{subfigure}\hfill
\begin{subfigure}[t]{0.22\textwidth}
    \includegraphics[width=\linewidth]{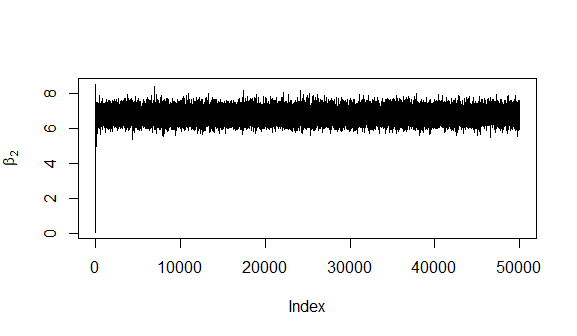}
    \end{subfigure}\hfill
\begin{subfigure}[t]{0.22\textwidth}
    \includegraphics[width=\linewidth]{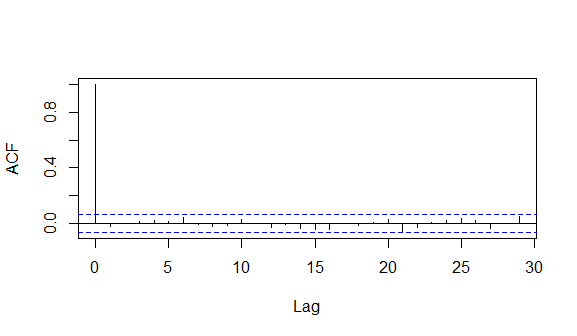}
\end{subfigure}\hfill
\begin{subfigure}[t]{0.22\textwidth}
    \includegraphics[width=\linewidth]{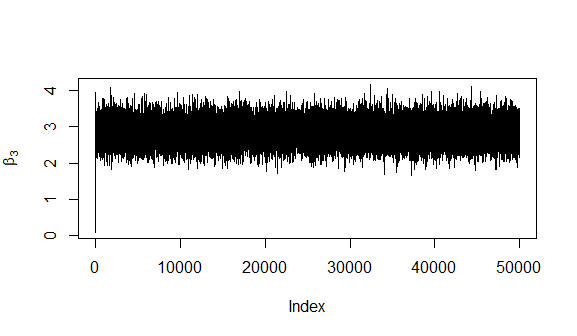}
    \end{subfigure}\hfill
\begin{subfigure}[t]{0.22\textwidth}
    \includegraphics[width=\linewidth]{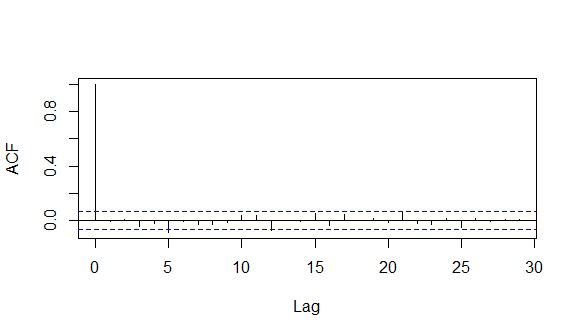}
\end{subfigure}\hfill
\begin{subfigure}[t]{0.22\textwidth}
    \includegraphics[width=\linewidth]{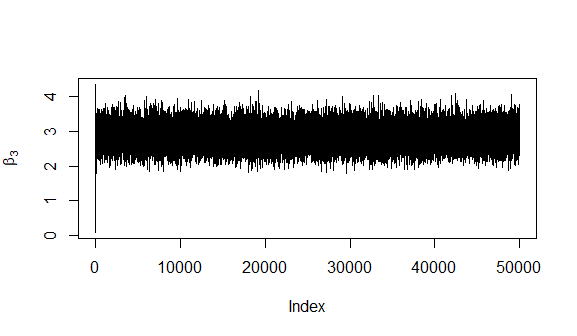}
    \end{subfigure}\hfill
\begin{subfigure}[t]{0.22\textwidth}
    \includegraphics[width=\linewidth]{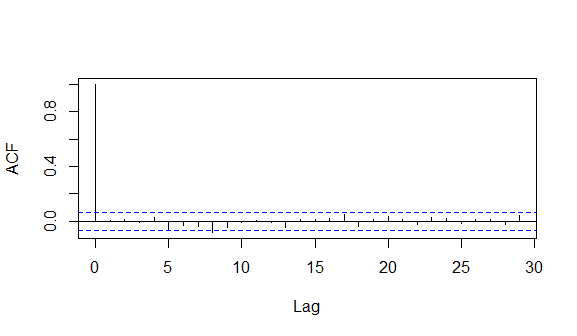}
\end{subfigure}\hfill
\begin{subfigure}[t]{0.22\textwidth}
    \includegraphics[width=\linewidth]{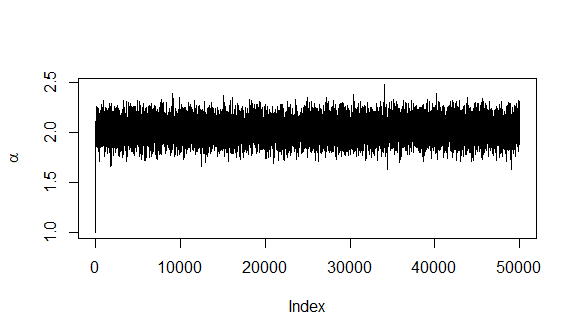}
    \end{subfigure}\hfill
\begin{subfigure}[t]{0.22\textwidth}
    \includegraphics[width=\linewidth]{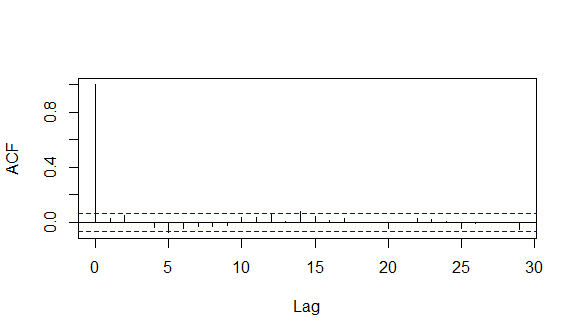}
\end{subfigure}\hfill
\begin{subfigure}[t]{0.22\textwidth}
    \includegraphics[width=\linewidth]{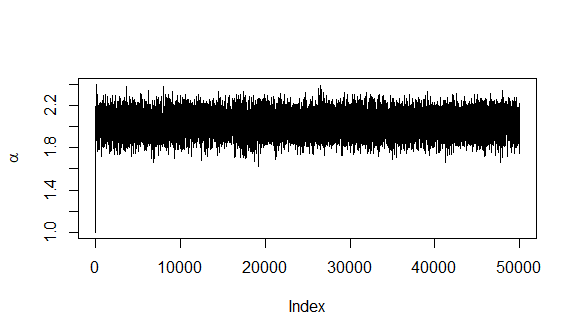}
    \end{subfigure}\hfill
\begin{subfigure}[t]{0.22\textwidth}
    \includegraphics[width=\linewidth]{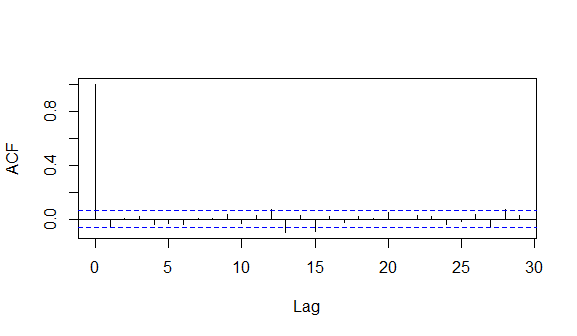}
\end{subfigure}\hfill
\begin{subfigure}[t]{0.22\textwidth}
    \includegraphics[width=\linewidth]{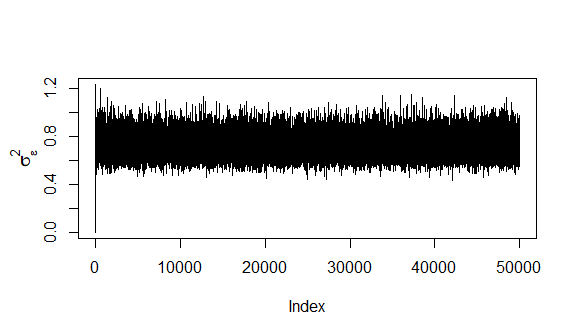}
    \end{subfigure}\hfill
\begin{subfigure}[t]{0.22\textwidth}
    \includegraphics[width=\linewidth]{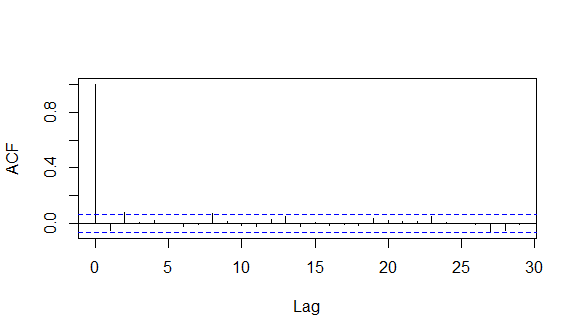}
\end{subfigure}\hfill
\begin{subfigure}[t]{0.22\textwidth}
    \includegraphics[width=\linewidth]{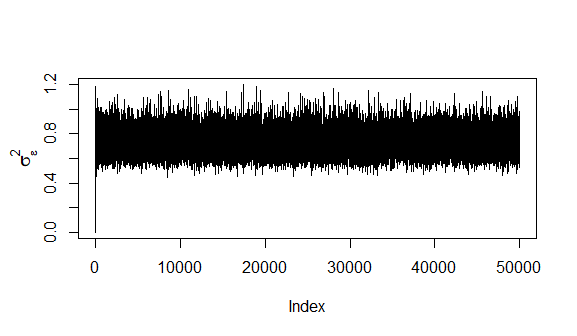}
    \end{subfigure}\hfill
\begin{subfigure}[t]{0.22\textwidth}
    \includegraphics[width=\linewidth]{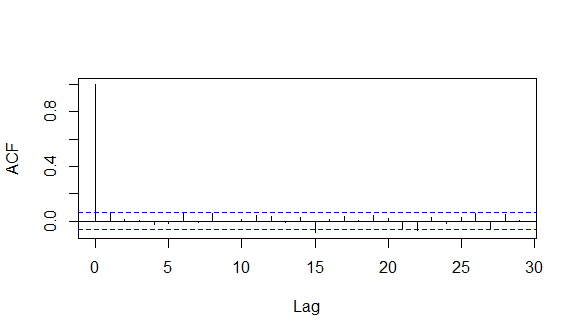}
\end{subfigure}
\caption{For each row, first two plots from the left side is trace plot and autocorrelation plot produced by semi-parametric method and last two plots denotes trace plot and autocorrelation plot produced by parametric method based on the generated samples for $\beta_i$, $i=1,2,3$, $\alpha$ and $\sigma_\epsilon^2$ respectively. Autocorrelation plots are constructed based on the samples at lag size 50.}
\end{figure}

\newpage

\subsection*{Case 4}

\begin{figure}[htbp]
\begin{subfigure}[t]{0.22\textwidth}
    \includegraphics[width=\linewidth]{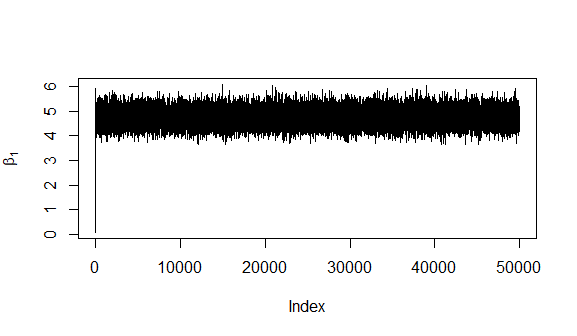}
    \end{subfigure}\hfill
\begin{subfigure}[t]{0.22\textwidth}
    \includegraphics[width=\linewidth]{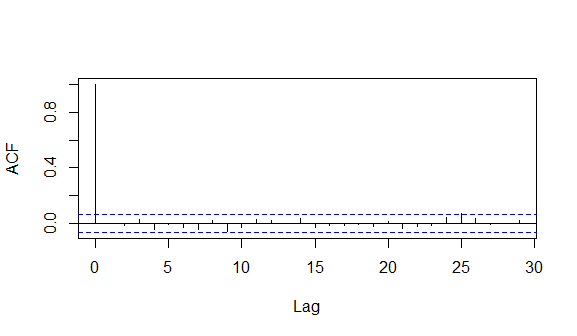}
\end{subfigure}\hfill
\begin{subfigure}[t]{0.22\textwidth}
    \includegraphics[width=\linewidth]{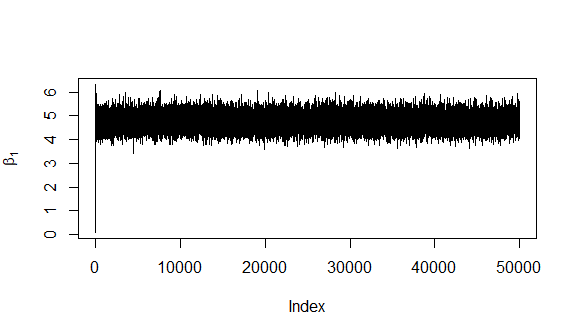}
    \end{subfigure}\hfill
\begin{subfigure}[t]{0.22\textwidth}
    \includegraphics[width=\linewidth]{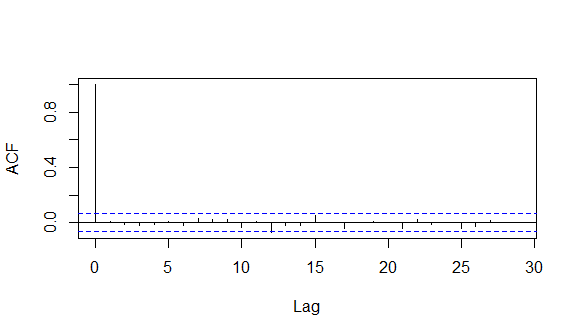}
\end{subfigure}\hfill
\begin{subfigure}[t]{0.22\textwidth}
    \includegraphics[width=\linewidth]{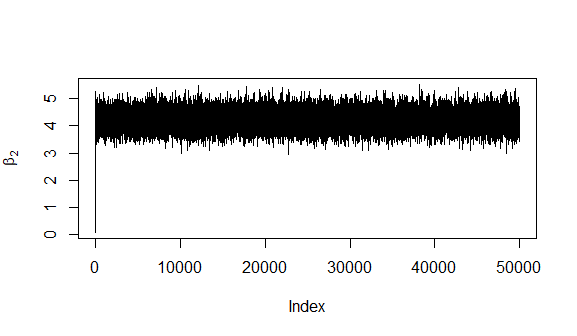}
    \end{subfigure}\hfill
\begin{subfigure}[t]{0.22\textwidth}
    \includegraphics[width=\linewidth]{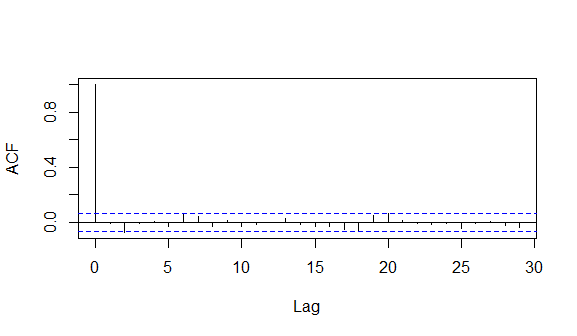}
\end{subfigure}\hfill
\begin{subfigure}[t]{0.22\textwidth}
    \includegraphics[width=\linewidth]{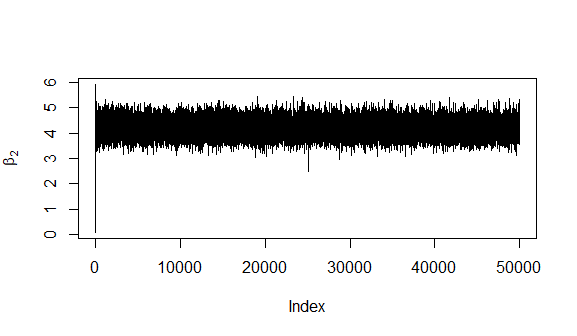}
    \end{subfigure}\hfill
\begin{subfigure}[t]{0.22\textwidth}
    \includegraphics[width=\linewidth]{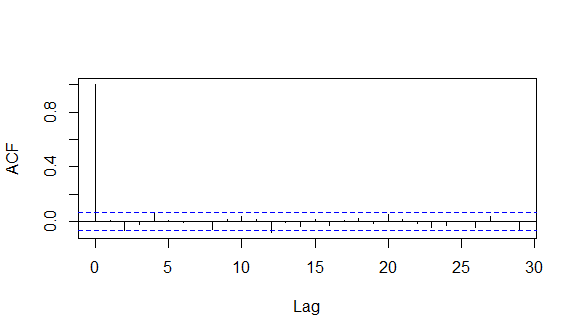}
\end{subfigure}\hfill
\begin{subfigure}[t]{0.22\textwidth}
    \includegraphics[width=\linewidth]{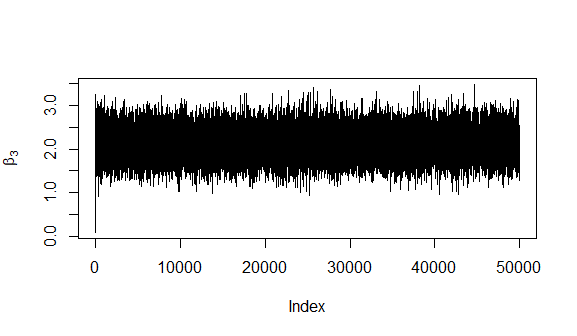}
    \end{subfigure}\hfill
\begin{subfigure}[t]{0.22\textwidth}
    \includegraphics[width=\linewidth]{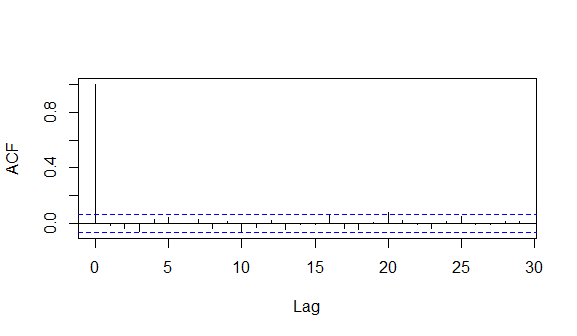}
\end{subfigure}\hfill
\begin{subfigure}[t]{0.22\textwidth}
    \includegraphics[width=\linewidth]{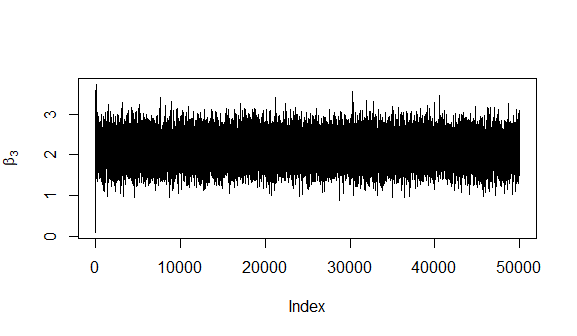}
    \end{subfigure}\hfill
\begin{subfigure}[t]{0.22\textwidth}
    \includegraphics[width=\linewidth]{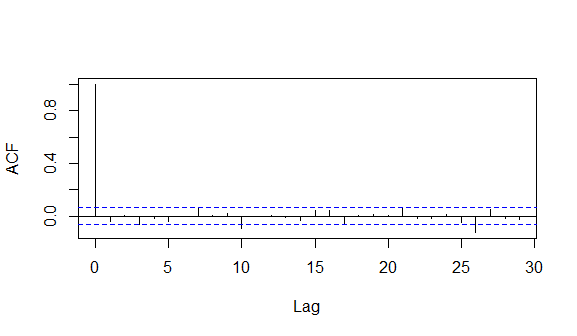}
\end{subfigure}\hfill
\begin{subfigure}[t]{0.22\textwidth}
    \includegraphics[width=\linewidth]{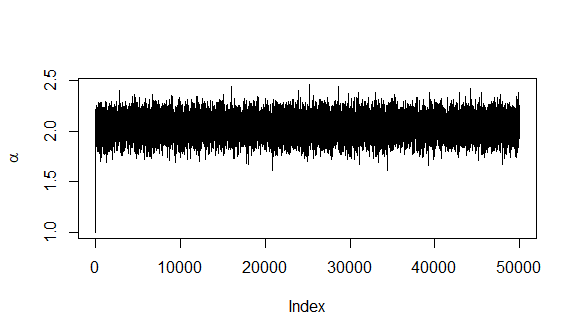}
    \end{subfigure}\hfill
\begin{subfigure}[t]{0.22\textwidth}
    \includegraphics[width=\linewidth]{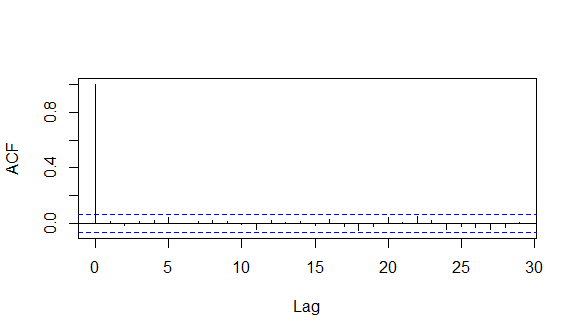}
\end{subfigure}\hfill
\begin{subfigure}[t]{0.22\textwidth}
    \includegraphics[width=\linewidth]{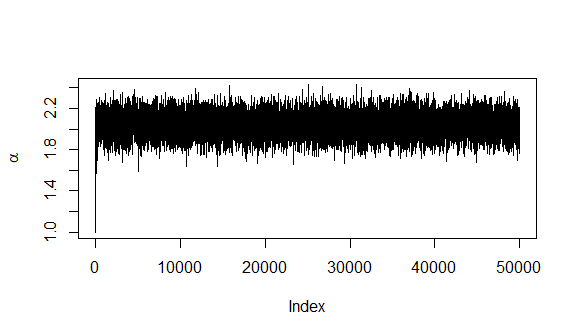}
    \end{subfigure}\hfill
\begin{subfigure}[t]{0.22\textwidth}
    \includegraphics[width=\linewidth]{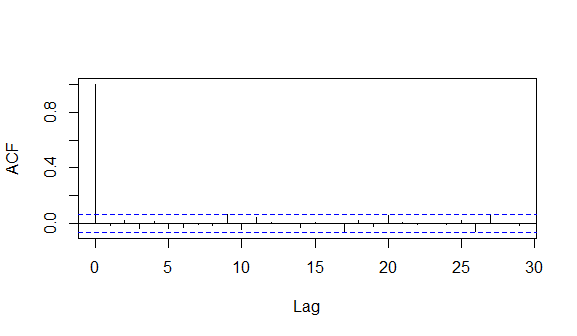}
\end{subfigure}\hfill
\begin{subfigure}[t]{0.22\textwidth}
    \includegraphics[width=\linewidth]{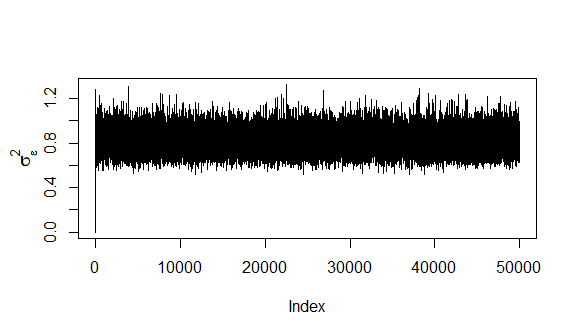}
    \end{subfigure}\hfill
\begin{subfigure}[t]{0.22\textwidth}
    \includegraphics[width=\linewidth]{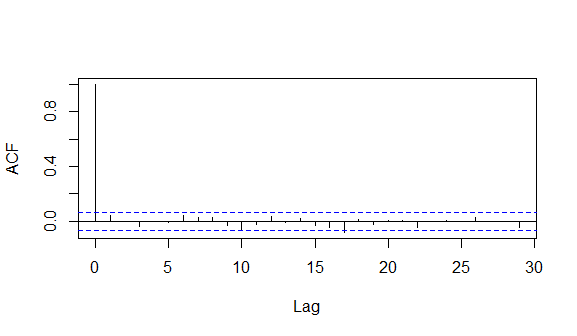}
\end{subfigure}\hfill
\begin{subfigure}[t]{0.22\textwidth}
    \includegraphics[width=\linewidth]{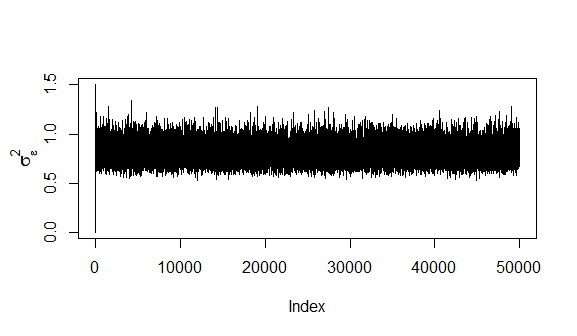}
    \end{subfigure}\hfill
\begin{subfigure}[t]{0.22\textwidth}
    \includegraphics[width=\linewidth]{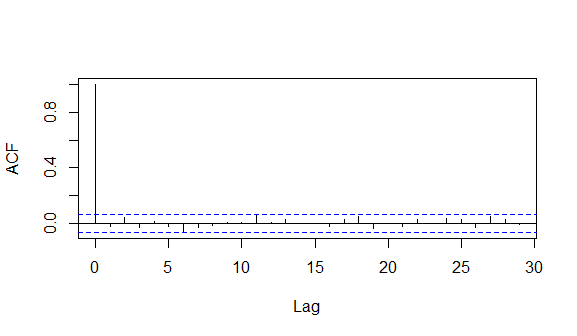}
\end{subfigure}
\caption{For each row, first two plots from the left side is trace plot and autocorrelation plot produced by semi-parametric method and last two plots denotes trace plot and autocorrelation plot produced by parametric method based on the generated samples for $\beta_i$, $i=1,2,3$, $\alpha$ and $\sigma_\epsilon^2$ respectively. Autocorrelation plots are constructed based on the samples at lag size 50.}
\end{figure}

\newpage

\subsection*{Case 5}

\begin{figure}[htbp]
\begin{subfigure}[t]{0.22\textwidth}
    \includegraphics[width=\linewidth]{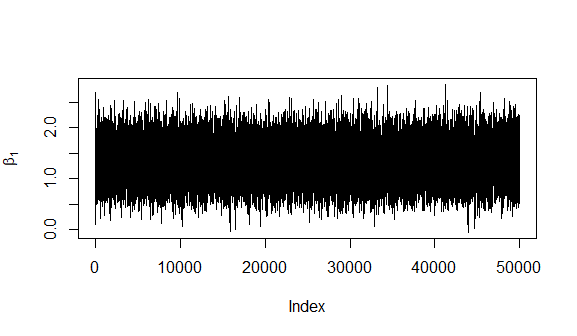}
    \end{subfigure}\hfill
\begin{subfigure}[t]{0.22\textwidth}
    \includegraphics[width=\linewidth]{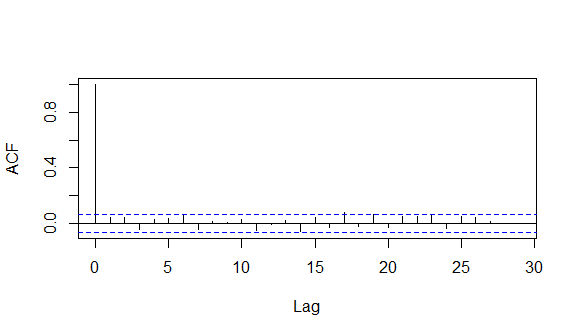}
\end{subfigure}\hfill
\begin{subfigure}[t]{0.22\textwidth}
    \includegraphics[width=\linewidth]{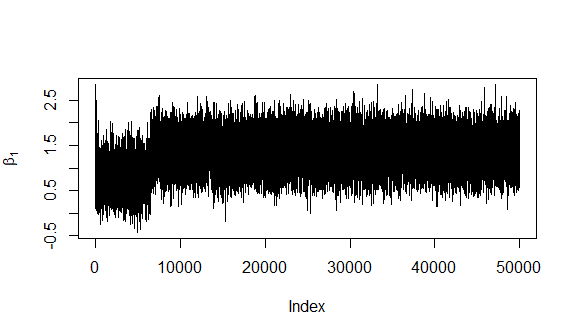}
    \end{subfigure}\hfill
\begin{subfigure}[t]{0.22\textwidth}
    \includegraphics[width=\linewidth]{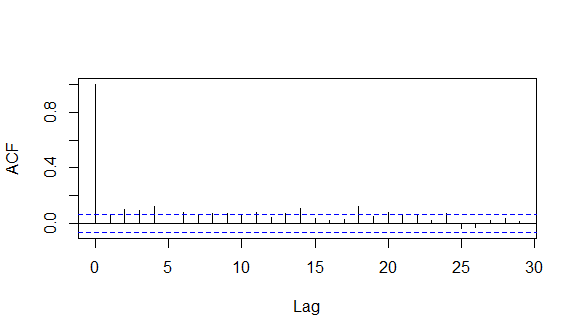}
\end{subfigure}\hfill
\begin{subfigure}[t]{0.22\textwidth}
    \includegraphics[width=\linewidth]{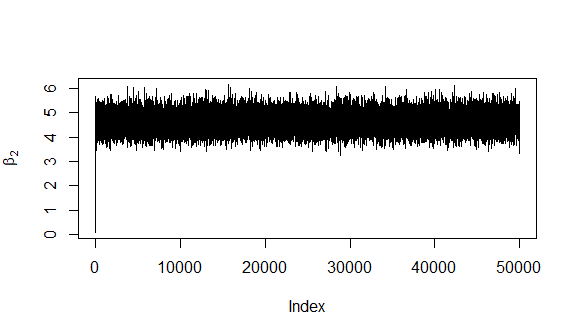}
    \end{subfigure}\hfill
\begin{subfigure}[t]{0.22\textwidth}
    \includegraphics[width=\linewidth]{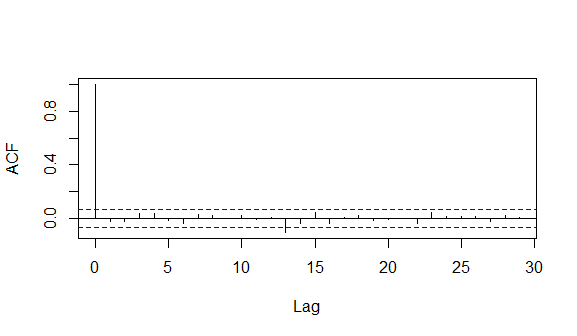}
\end{subfigure}\hfill
\begin{subfigure}[t]{0.22\textwidth}
    \includegraphics[width=\linewidth]{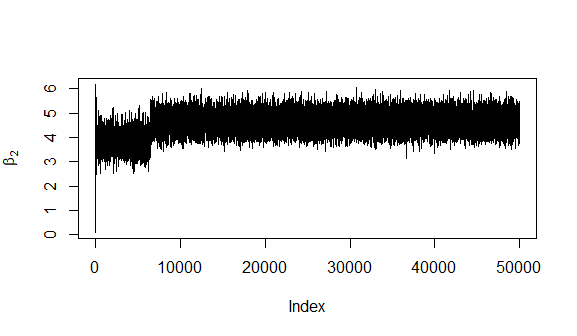}
    \end{subfigure}\hfill
\begin{subfigure}[t]{0.22\textwidth}
    \includegraphics[width=\linewidth]{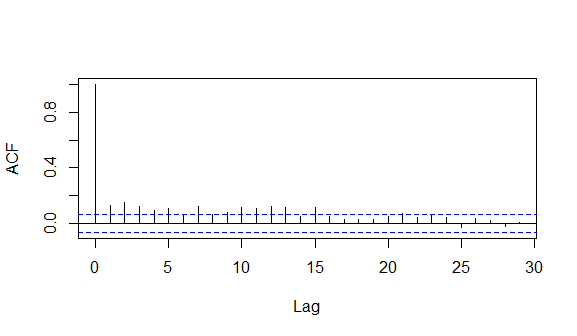}
\end{subfigure}\hfill
\begin{subfigure}[t]{0.22\textwidth}
    \includegraphics[width=\linewidth]{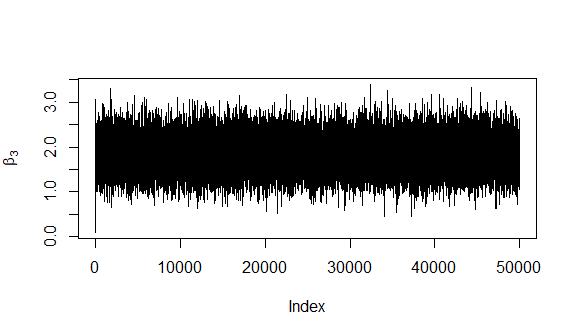}
    \end{subfigure}\hfill
\begin{subfigure}[t]{0.22\textwidth}
    \includegraphics[width=\linewidth]{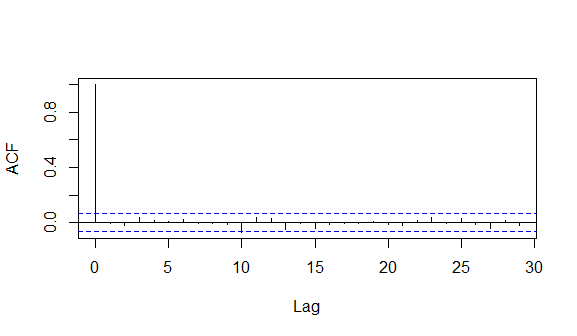}
\end{subfigure}\hfill
\begin{subfigure}[t]{0.22\textwidth}
    \includegraphics[width=\linewidth]{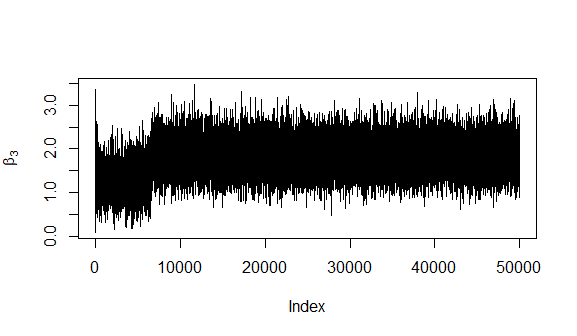}
    \end{subfigure}\hfill
\begin{subfigure}[t]{0.22\textwidth}
    \includegraphics[width=\linewidth]{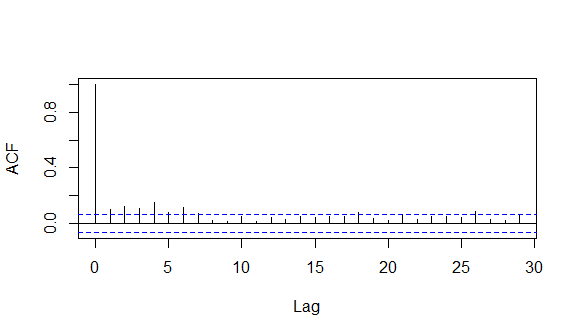}
\end{subfigure}\hfill
\begin{subfigure}[t]{0.22\textwidth}
    \includegraphics[width=\linewidth]{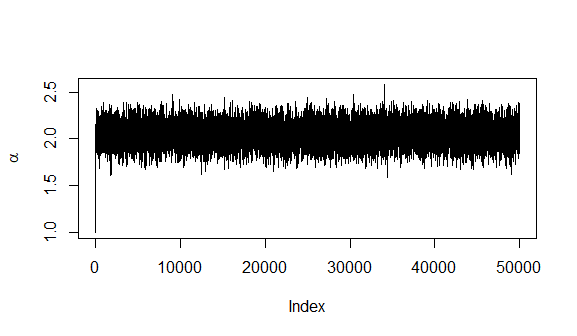}
    \end{subfigure}\hfill
\begin{subfigure}[t]{0.22\textwidth}
    \includegraphics[width=\linewidth]{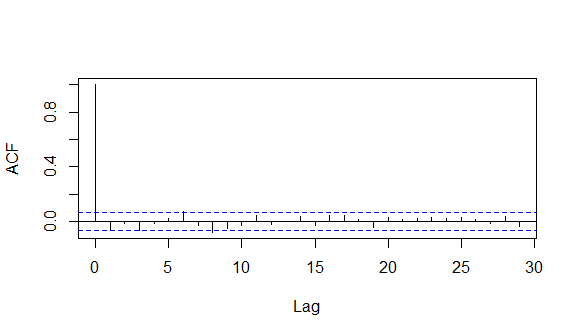}
\end{subfigure}\hfill
\begin{subfigure}[t]{0.22\textwidth}
    \includegraphics[width=\linewidth]{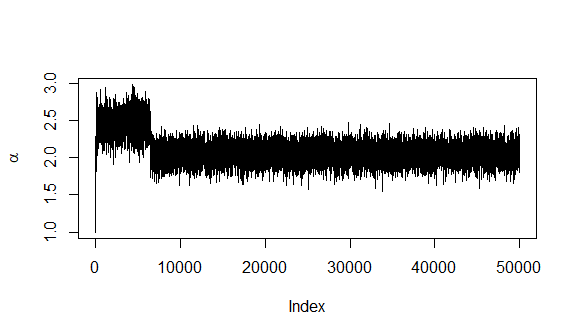}
    \end{subfigure}\hfill
\begin{subfigure}[t]{0.22\textwidth}
    \includegraphics[width=\linewidth]{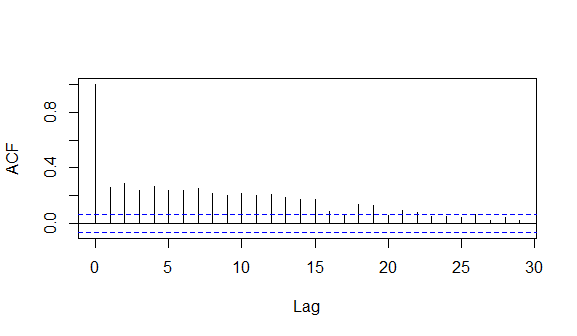}
\end{subfigure}\hfill
\begin{subfigure}[t]{0.22\textwidth}
    \includegraphics[width=\linewidth]{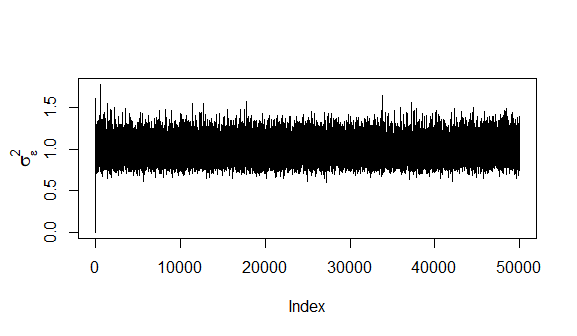}
    \end{subfigure}\hfill
\begin{subfigure}[t]{0.22\textwidth}
    \includegraphics[width=\linewidth]{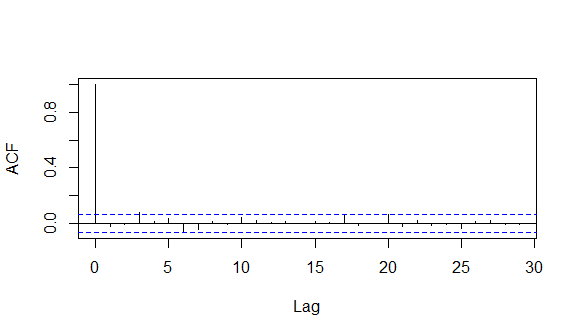}
\end{subfigure}\hfill
\begin{subfigure}[t]{0.22\textwidth}
    \includegraphics[width=\linewidth]{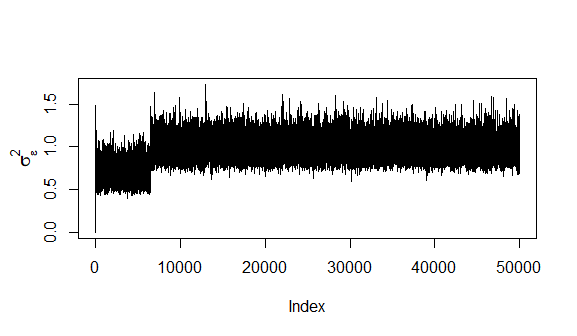}
    \end{subfigure}\hfill
\begin{subfigure}[t]{0.22\textwidth}
    \includegraphics[width=\linewidth]{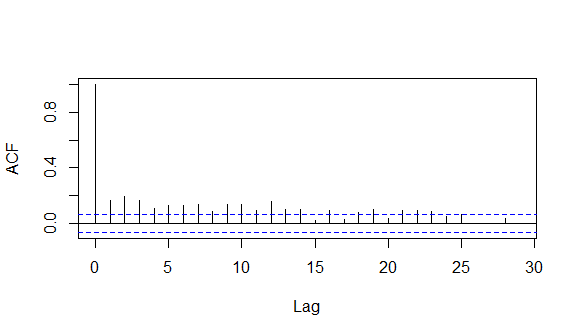}
\end{subfigure}
\caption{For each row, first two plots from the left side is trace plot and autocorrelation plot produced by semi-parametric method and last two plots denotes trace plot and autocorrelation plot produced by parametric method based on the generated samples for $\beta_i$, $i=1,2,3$, $\alpha$ and $\sigma_\epsilon^2$ respectively. Autocorrelation plots are constructed based on the samples at lag size 50.}
\end{figure}

\newpage

\subsection*{Fatigue-Crack Size dataset}

\begin{figure}[htbp]
\begin{subfigure}[t]{0.22\textwidth}
    \includegraphics[width=\linewidth]{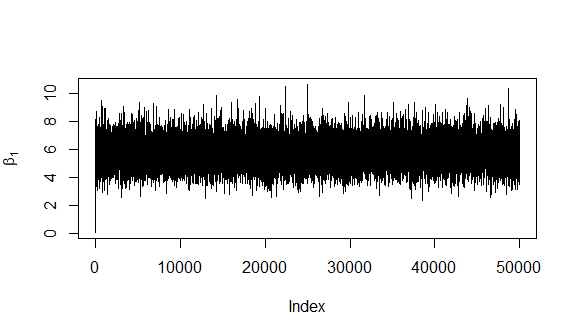}
    \end{subfigure}\hfill
\begin{subfigure}[t]{0.22\textwidth}
    \includegraphics[width=\linewidth]{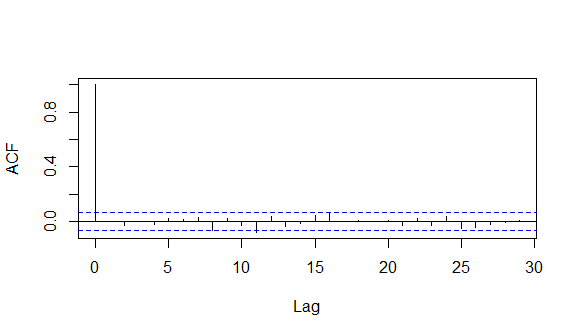}
\end{subfigure}\hfill
\begin{subfigure}[t]{0.22\textwidth}
    \includegraphics[width=\linewidth]{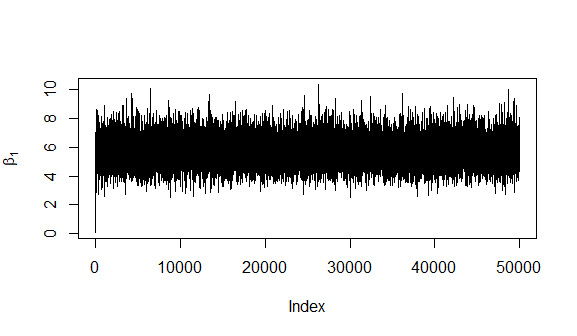}
    \end{subfigure}\hfill
\begin{subfigure}[t]{0.22\textwidth}
    \includegraphics[width=\linewidth]{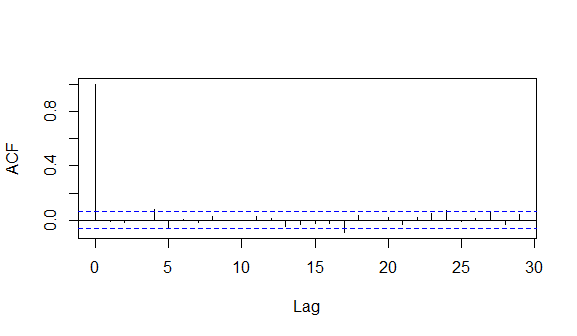}
\end{subfigure}\hfill
\begin{subfigure}[t]{0.22\textwidth}
    \includegraphics[width=\linewidth]{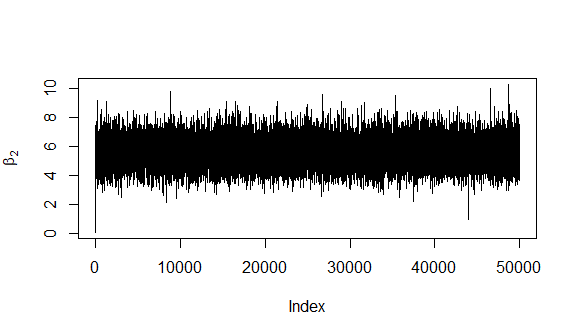}
    \end{subfigure}\hfill
\begin{subfigure}[t]{0.22\textwidth}
    \includegraphics[width=\linewidth]{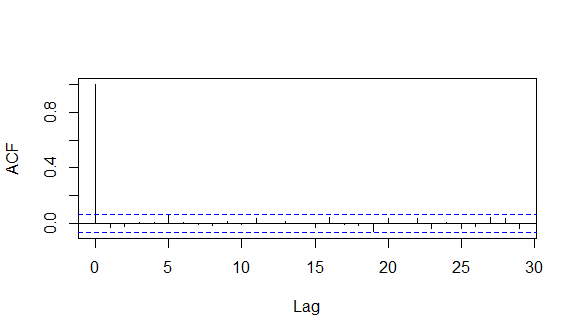}
\end{subfigure}\hfill
\begin{subfigure}[t]{0.22\textwidth}
    \includegraphics[width=\linewidth]{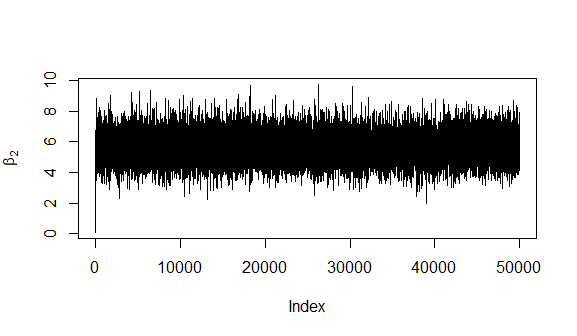}
    \end{subfigure}\hfill
\begin{subfigure}[t]{0.22\textwidth}
    \includegraphics[width=\linewidth]{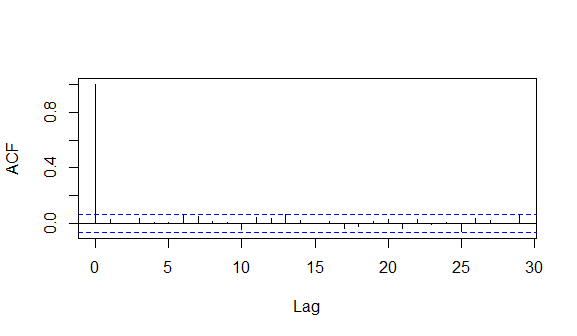}
\end{subfigure}\hfill
\begin{subfigure}[t]{0.22\textwidth}
    \includegraphics[width=\linewidth]{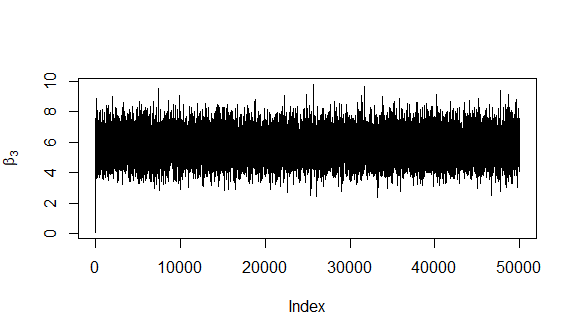}
    \end{subfigure}\hfill
\begin{subfigure}[t]{0.22\textwidth}
    \includegraphics[width=\linewidth]{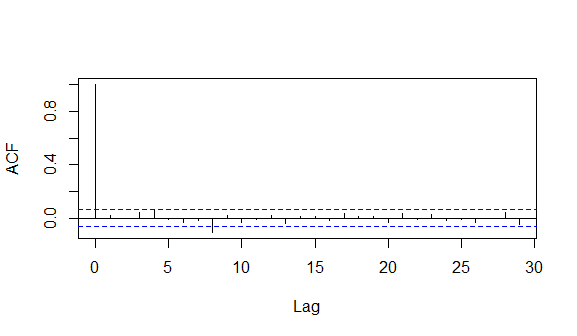}
\end{subfigure}\hfill
\begin{subfigure}[t]{0.22\textwidth}
    \includegraphics[width=\linewidth]{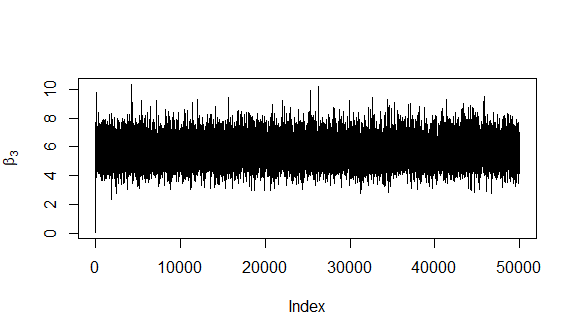}
    \end{subfigure}\hfill
\begin{subfigure}[t]{0.22\textwidth}
    \includegraphics[width=\linewidth]{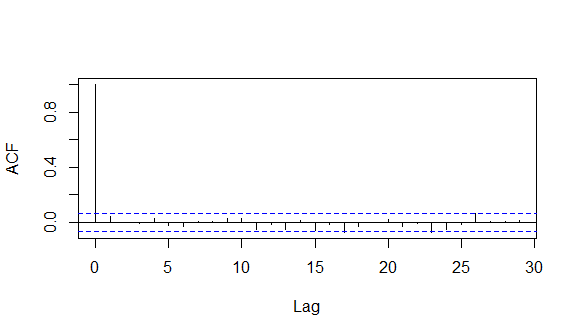}
\end{subfigure}\hfill
\begin{subfigure}[t]{0.22\textwidth}
    \includegraphics[width=\linewidth]{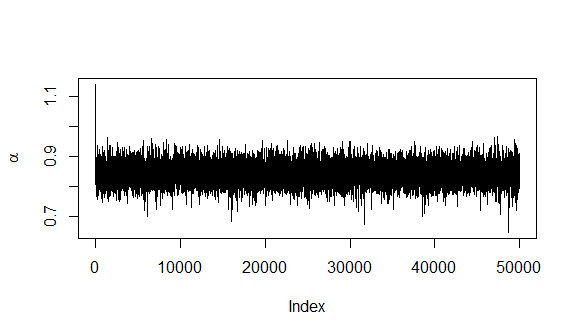}
    \end{subfigure}\hfill
\begin{subfigure}[t]{0.22\textwidth}
    \includegraphics[width=\linewidth]{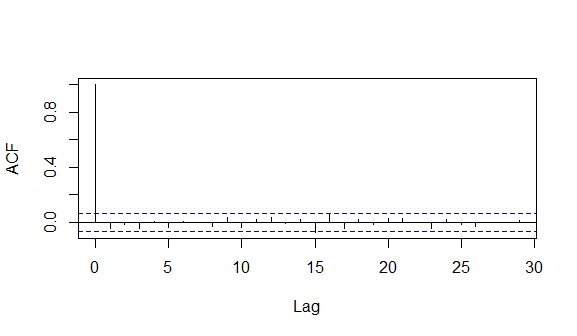}
\end{subfigure}\hfill
\begin{subfigure}[t]{0.22\textwidth}
    \includegraphics[width=\linewidth]{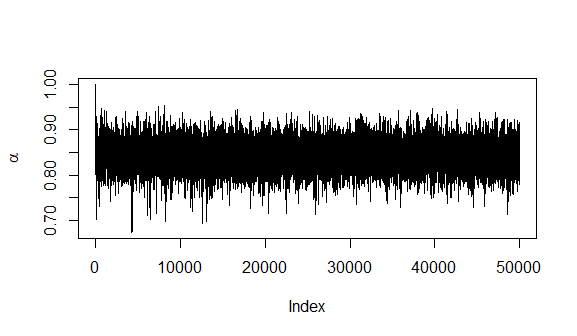}
    \end{subfigure}\hfill
\begin{subfigure}[t]{0.22\textwidth}
    \includegraphics[width=\linewidth]{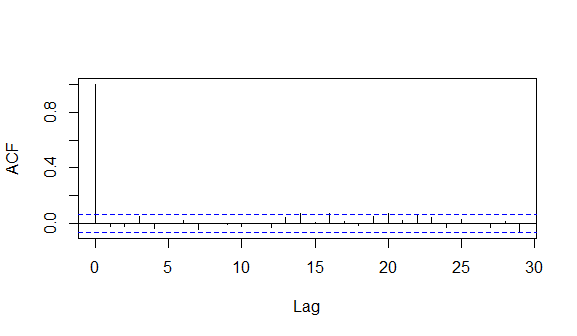}
\end{subfigure}\hfill
\begin{subfigure}[t]{0.22\textwidth}
    \includegraphics[width=\linewidth]{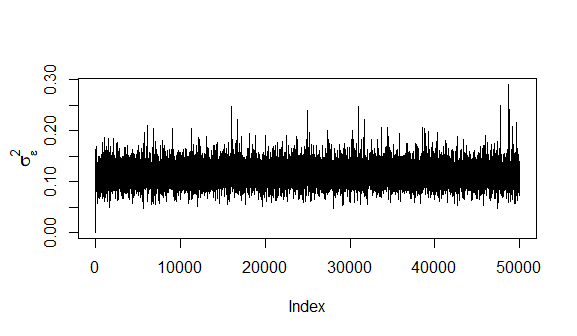}
    \end{subfigure}\hfill
\begin{subfigure}[t]{0.22\textwidth}
    \includegraphics[width=\linewidth]{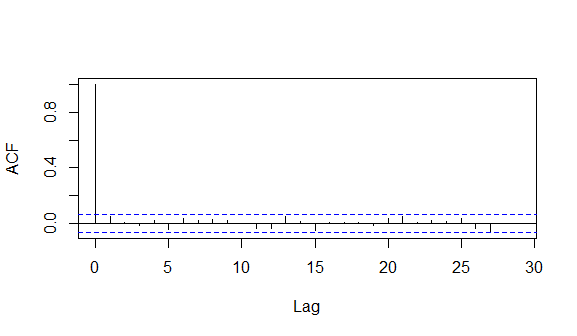}
\end{subfigure}\hfill
\begin{subfigure}[t]{0.22\textwidth}
    \includegraphics[width=\linewidth]{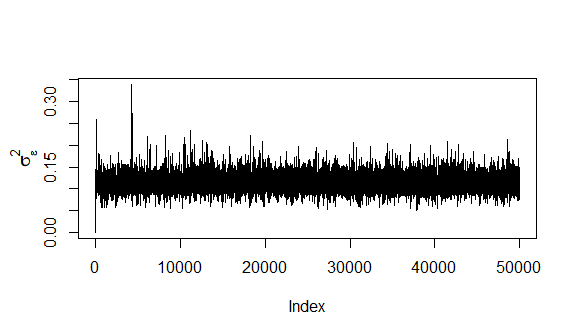}
    \end{subfigure}\hfill
\begin{subfigure}[t]{0.22\textwidth}
    \includegraphics[width=\linewidth]{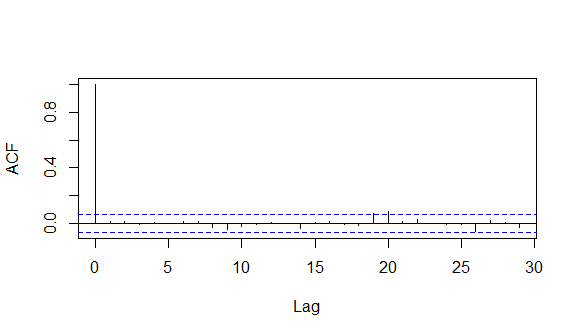}
\end{subfigure}
\caption{For each row, first two plots from the left side is trace plot and autocorrelation plot produced by semi-parametric method and last two plots denotes trace plot and autocorrelation plot produced by parametric method based on the generated samples for $\beta_i$, $i=1,2,3$, $\alpha$ and $\sigma_\epsilon^2$ respectively. Autocorrelation plots are constructed based on the samples at lag size 50.}
\end{figure}

\end{document}